\documentclass[%
 aip,
 amsmath,amssymb,
 reprint,%
]{revtex4-1}

\usepackage{graphicx}% Include figure files
\usepackage{dcolumn}% Align table columns on decimal point
\usepackage{bm}% bold math
\usepackage[utf8]{inputenc}
\usepackage[T1]{fontenc}
\usepackage{mathptmx}
\usepackage{etoolbox}

\usepackage[table,dvipsnames]{xcolor}
\usepackage{tikz,siunitx}

\makeatletter
\def\@email#1#2{%
 \endgroup
 \patchcmd{\titleblock@produce}
  {\frontmatter@RRAPformat}
  {\frontmatter@RRAPformat{\produce@RRAP{*#1\href{mailto:#2}{#2}}}\frontmatter@RRAPformat}
  {}{}
}%
\makeatother
\begin{document}

\preprint{AIP/123-QED}
\title{Model form uncertainty quantification of Reynolds-averaged Navier-Stokes modeling of flows over a SD7003 airfoil}

\author{Minghan Chu}
 \altaffiliation{Mechanical and Materials Engineering Department, Queen's University, Kingston, ON K7L 2V9, Canada.}%Lines break automatically or can be forced with \\
 
\author{Xiaohua Wu}%
\homepage{https://www.rmc-cmr.ca/en/mechanical-and-aerospace-engineering/xiaohua-wu}
 \email{17mc93@queensu.ca.}
\affiliation{ 
Mechanical and Aerospace Engineering, Royal Military College of Canada, Kingston, ON K7K 7B4, Canada.%\\This line break forced with \textbackslash\textbackslash
}%

\author{David E. Rival}
 %\homepage{http://www.Second.institution.edu/~Charlie.Author.}
\affiliation{%
Mechanical and Materials Engineering Department, Queen's University, Kingston, ON K7L 2V9, Canada.%\\This line break forced% with \\
}%

\date{\today}% It is always \today, today,

\begin{abstract}
%Estimation of the model form uncertainty in Reynolds-averaged Naiver-Stokes (RANS) models is imperative to build confidence in RANS predictions. 
It is well known that the Boussinesq turbulent viscosity hypothesis can yield inaccurate predictions when complex flow features are involved, e.g. laminar-turbulent transition. The focus of the study is to explore the capability of a physics-based uncertainty quantification (UQ) approach to quantify the model-form uncertainty in Reynolds-averaged Naiver-Stokes (RANS) simulations of laminar-turbulent transitional flows over an Selig-Donovan (SD) 7003 airfoil. This methodology perturbs the modeled Reynolds stress tensor in the momentum equations; perturbations are injected into the amplitude, eigenvalues and eigenvectors of the anisotropy Reynolds stress tensor undergone an eigen-decomposition. In this study, our analyses focus upon the amplitude perturbation. We observed a monotonic behavior of the magnitude of the predicted uncertainty bounds for different quantities of interest. High-order regressions based on the turbulence kinetic energy discrepancies are used to develop a novel switch marker function $M_{k}$ to introduce perturbations in a non-uniform manner over different regions of the domain based upon prior knowledge of the limitations of the model. Importantly, the compound effect of $M_{k}$ and eigenvalue perturbations show a synergy behavior, e.g., dramatically increased uncertainty bounds to account for the discrepancy in the RANS prediction; and the $M_{k}$ function effectively avoids over-perturbation to the amplitude of the anisotropy Reynolds stress tensor. In this context, regression based amplitude perturbation of the anisotropy Reynolds stress tensor makes a new contribution to the RANS UQ methodology in the simulations of the airfoil transitional flows, which shows very encouraging results.

%It is well known that Boussinesq turbulent-viscosity hypothesis can introduce uncertainty in predictions for complex flow features such as separation, reattachment, and laminar-turbulent transition. This study adopts a recent physics-based uncertainty quantification (UQ) approach to address such model form uncertainty in Reynolds-averaged Naiver-Stokes (RANS) simulations. Thus far, almost all UQ studies have focused on quantifying the model form uncertainty in turbulent flow scenarios. The focus of the study is to advance our understanding of the performance of the UQ approach on two different transitional flow scenarios: a flat plate and a SD7003 airfoil, to close this gap. For the T3A (flat-plate flow) flow, most of the model form uncertainty is concentrated in the laminar-turbulent transition region. For the SD7003 airfoil flow, the eigenvalue perturbations reveal a decrease as well as an increase in the length of the separation bubble. As a consequence, the uncertainty bounds successfully encompass the reattachment point. Likewise, the region of reverse flow that appear in the separation bubble is either suppressed or bolstered by the eigenvalue perturbations. In this context, the UQ methodology is applied to transition and show great results. This is the first successful RANS UQ study for transitional flows. 
\end{abstract}

\maketitle

%\begin{quotation}
%The ``lead paragraph'' is encapsulated with the \LaTeX\ 
%\verb+quotation+ environment and is formatted as a single paragraph before the first section heading. 
%(The \verb+quotation+ environment reverts to its usual meaning after the first sectioning command.) 
%Note that numbered references are allowed in the lead paragraph.
%%
%The lead paragraph will only be found in an article being prepared for the journal \textit{Chaos}.
%\end{quotation}

\section{\label{sec:level1}INTRODUCTION}
Transitional flow regime is very frequently encountered in turbomachines and especially in aircraft engines at relatively low Reynolds numbers. As a consequence, a significant part of the flow on the blade surfaces is under the laminar-turbulent transition process. The boundary development, losses, efficiency, and momentum transfer are greatly affected by the laminar-turbulent transition. Therefore, accurate prediction for the transition process is crucial for the design of efficient as well as reliable aerospace designs \cite{pecnik2007application}.   

RANS simulations remain the most commonly used computational technique for analysis of turbulent flows. There has been considerable effort spent in the past two decades to develop RANS based transition models for engineering applications to predict various kinds of transitional flows \cite{menter2002transition,menter2004correlation,menter2006transition,langtry2009correlation,menter2015one,wei2017modeling,tousi2021active}. Each model has its strengths and weaknesses, and by far the correlation-based transition models by Langtry and Menter \cite{langtry2009correlation,menter2015one} have been widely used in engineering industries, in particular, aerospace industry. Most RANS models have adopted the Boussinesq turbulent viscosity hypothesis, i.e., anisotropy Reynolds stresses are proportional to the mean rate of strain, therefore also referred to as linear eddy viscosity models. It is well known that linear eddy viscosity models are limited due to the restrictions of the Boussinesq turbulent viscosity hypothesis on yielding accurate predictions for complex flow features such as flow with significant streamline curvature, separation, reattachment, and laminar-turbulent transition. Large eddy simulations (LES) or Direct numerical simulations (DNS) provide high-fidelity solution for such problems, but the calculations are often too expensive in computational time and cost, especially for high-Reynolds number flows. Therefore, accounting for the errors and uncertainties in the RANS model predictions provides a means to quantify trust in the predictions, as well as enabling the application of robust and reliability based design optimization. More expensive LES or DNS would only be considered necessary if the model form uncertainty is too large. 

The current study considers a physics-based approach that has been recently introduced by Emory \textit{et al.} \cite{emory2013modeling}, namely eigenspace perturbation method. This framework quantifies the model form uncertainty associated with the linear eddy viscosity model via sequential perturbations in the predicted amplitude (turbulence kinetic energy), shape (eigenvalues), and orientation (eigenvectors) of the anisotropy Reynolds stress tensor. This is an established method for RANS model UQ and has been applied to analyze and estimate the RANS uncertainty in flow through scramjets \cite{emory2011characterizing}, aircraft nozzle jets, turbomachinery, over stream-lined bodies \cite{gorle2019epistemic}, supersonic axisymmetric submerged jet \cite{mishra2017rans}, and canonical cases of turbulent flows over a backward-facing step \cite{iaccarino2017eigenspace,cremades2019reynolds}.  This method has been used for robust design of Organic Rankine Cycle (ORC) turbine cascades \cite{razaaly2019optimization}. In aerospace applications, this method has been used for design optimization under uncertainty\cite{cook2019optimization,mishra2020design,matha2022extending,matha2022assessment}. In civil engineering applications, this method is being used to design urban canopies \cite{garcia2014quantifying}, ensuring the ventilation of enclosed spaces, and used in the wind engineering practice for turbulent bluff body flows \cite{gorle2015quantifying}. This perturbation method for RANS model UQ has been used in conjunction with Machine Learning algorithms to provide precise estimates of RANS model uncertainty in the presence of data \cite{xiao2016quantifying,wu2016bayesian,parish2016paradigm,xiao2017random,wang2017physics,wang2017comprehensive,heyse2021estimating}. The method is also being used for the creation of probabilistic aerodynamic databases, enabling the certification of virtual aircraft designs \cite{mukhopadhaya2020multi,nigam2021toolset}.

All of the aforementioned studies that adopted the eigenspace perturbation framework focused on eigenvalue and eigenvector perturbations but did not consider the turbulence kinetic energy perturbation. According to Mishra and Iaccarino \cite{mishra2019theoretical}, turbulence kinetic energy perturbation varies the coefficient of turbulent viscosity in the Boussinesq turbulent viscosity hypothesis. Currently all eddy viscosity models utilize a predetermined constant value of this coefficient. In reality, the coefficient of turbulent viscosity varies between different turbulent flow scenarios and even between different regions in the same turbulent flow \cite{mishra2019theoretical}.   Therefore, perturbing the amplitude of the anisotropy Reynolds stress tensor not only captures the full ranges of uncertainties introduced by the Boussinesq turbulent viscosity hypothesis, but plays an important role in capturing the true physics of the turbulent flow. However, studies of turbulence kinetic energy perturbation are lacking. The only studies that have been conducted to address the turbulence kinetic energy perturbation are proposed by \cite{gorle2013framework,cremades2019reynolds}. Yet to date, the combined effect of the turbulence kinetic energy and eigenvalue perturbation have not been examined for airfoil flows. It should be noted that introducing uniform perturbations in the entire flow field often lead to overly conservative confidence intervals, because decades of experience in RANS modeling show that the models are not always inaccurate. Consequently, it is reasonable for one to only introduce uncertainties in the regions of the flow where the model is deemed plausibly untrustworthy. Gorl{\'e} \textit{et al.} \cite{gorle2014deviation} first proposed the concept of \textit{ad hoc} ``marker function'' that identifies regions that deviate from parallel shear flow. A recent study of Gorl{\'e} \textit{et al.} \cite{gorle2019epistemic} employed this marker function and applied it to the simulation for a flow over a periodic wavy wall. Emory \textit{et al.} \cite{emory2013modeling} also provided a variety of marker functions aimed at spatially varying the magnitude of the eigenvalue perturbation in a computational domain. Nevertheless, marker function development is still very under-explored and more rigorous discussion and validation of new marker is needed.

There are few methods for implementing the effects of the model form uncertainty on a transitional near-wall flow in a RANS formulation. In this case, the  local-correlation laminar-turbulent transition model of Langtry and Menter \cite{langtry2009correlation} is used to close the mean transport equations. It has been extensively used to predict a wide variety of transitional flows such as natural transition and laminar-turbulent transition. However, there are few studies concerning the model form uncertainty in transition modeling.

Therefore, the objective of this paper is to advance the understanding of the performance of the eigenspace perturbation approach for quantifying the model form uncertainty in RANS simulations of transitional flows over a SD7003 airfoil using the transition model of Langtry and Menter \cite{langtry2009correlation}. Specifically, the objectives of this study are (1) to develop a new regression based marker function $M_{k}$ for the perturbation to the amplitude of the anisotropy Reynolds stress tensor based on the turbulence kinetic energy discrepancy between the RANS and in-house DNS \cite{zhang2021turbulent} datasets; (2) to explore the effect of turbulence kinetic perturbation on various quantities of interests (QoIs) through a sets of uniform perturbations; (3) and to have a thorough understanding of the combined effect of the shape and marker-involved amplitude perturbation to the anisotropy Reynolds stress tensor. A novelty of this study lies in the application of the eigenspace perturbation method to transitional flows, as opposed to fully developed turbulent flows as is done in almost prior investigations.

\section{Methodology}
\subsection{\label{sec:level2}Governing equations}
The flow was assumed to be two-dimensional and incompressible. The RANS formulation of the continuity and momentum equations is as follows:

\begin{equation} \label{p_Continuity}
   \frac{\partial \left\langle U_{i} \right\rangle}{\partial x_{i}}=0,
\end{equation}
%\left\langle u_{i} u_{j}\right\rangle
\begin{equation} \label{p_Momentum}
   \frac{ D \left\langle U_{j}\right\rangle}{\mathrm{Dt}}=-\frac{1}{\rho} \frac{\partial \left\langle P \right\rangle}{\partial x_{j}}+\nu \frac{\partial^{2} {\left\langle U_{j} \right\rangle}}{\partial x_{i} \partial x_{i}}-\frac{\partial \left\langle u_{i} u_{j}\right\rangle}{\partial x_{i}}
\end{equation}

\noindent where $\left\langle \ \right\rangle$ represents time-averaging, $\rho$ is the density, $\left\langle P \right\rangle$ is the time-averaged pressure, and $\nu$ is the kinematic viscosity. The $\left\langle U_{i}\right\rangle$ are the time-averaged velocity components. Reynolds stress terms in Eqs. \ref{p_Continuity} - \ref{p_Momentum}, i.e., $\left\langle u_{i}u_{j}\right\rangle$, are unknowns that need to be approximated using a RANS model. In the results presented in this study for a flow over a SD7003 airfoil, the modified version of shear-stress transport (SST) $k-\omega$ \cite{menter1993zonal,hellsten1998some,menter2001elements,menter2003ten} for transitional flow simulations by Langtry and Menter \cite{langtry2009correlation} is considered. The RANS based transition model \cite{langtry2009correlation} is a linear eddy viscosity model based on the Bossinesq turbulent viscosity hypothesis as follows:

\begin{equation}\label{Eq:noMark_uiuj}
    \left\langle{u_{i} u_{j}}\right\rangle=\frac{2}{3} k \delta_{i j}-2 \nu_{\mathrm{t}} \left\langle S_{i j} \right\rangle,
\end{equation}

\noindent where $k$ is the turbulence kinetic energy, $\delta_{i j}$ is the Kronecker delta, $\nu_\mathrm{t}$ is the turbulent viscosity, and $\left\langle S_{i j} \right\rangle$ is the rate of mean strain tensor. Results obtained from the RANS based transition model bereft of any perturbations are refered to as ``baseline'' solutions. In Eq. \ref{Eq:noMark_uiuj}, the deviatoric anisotropic part is 

\begin{equation}\label{Eqn:Bou_Ani_Tensor}
\begin{aligned}
a_{i j} & \equiv\left\langle u_{i} u_{j}\right\rangle-\frac{2}{3} k \delta_{i j} \\
&=-\nu_{\mathrm{t}}\left(\frac{\partial\left\langle U_{i}\right\rangle}{\partial x_{j}}+\frac{\partial\left\langle U_{j}\right\rangle}{\partial x_{i}}\right) \\
&=-2 \nu_{\mathrm{t}} \left\langle S_{i j} \right\rangle.
\end{aligned}
\end{equation}

The (normalized) anisotropy is defined by 

\begin{equation}\label{Eq:noMark_AnisotropyTensor}
    b_{i j}= \frac{a_{ij}}{2k} = \frac{\big \langle {u_{i} u_{j}} \big \rangle }{2 k}-\frac{\delta_{i j}}{3} = -\frac{\nu_{t} }{k}\big \langle {S_{i j}} \big \rangle. 
\end{equation}

%In the results presented hereafter for the flat plate, three different linear eddy viscosity models were considered: the shear-stress transport (SST) $k - \omega$ \cite{menter1993zonal,hellsten1998some,menter2001elements,menter2003ten}, the modified version of SST $k - \omega$ for transitional flow simulations by Langtry and Menter (SST $k - \omega$ LM) \cite{menter2004correlation,menter2006correlation,langtry2009correlation}, and the $k - \varepsilon$ model \cite{el1983k,launder1983numerical}. By considering three different models, we intend to contrast the uncertainty bounds generated by the transition model (SST $k - \omega$ LM) with the two turbulence models (SST $k - \omega$ and $k - \varepsilon$). Results corresponding to these linear eddy viscosity models bereft of any perturbations are referred to as ``baseline'' solutions.

\subsection{Eigenspace perturbation method}
The Reynolds stress tensor $\left\langle u_{i} u_{j}\right\rangle$ is symmetric positive semi-definite \cite{pope2001turbulent}, thus it can be eigen-decomposed as follows:

\begin{equation} \label{Eq:noMarker_Rij}
    \left\langle u_{i} u_{j}\right\rangle=2 k\left(\frac{\delta_{i j}}{3}+v_{i n}  \hat{b}_{n l} v_{j l}\right),
\end{equation}

\noindent in which $k \equiv {u_{i} u_{i}} / 2$, $v$ represents the matrix of orthonormal eigenvectors, $\hat{b}$ represents the diagonal matrix of eigenvalues ($\lambda_{i}$), which are arranged in a non-increasing order such that $\lambda_{1} \geq \lambda_{2} \geq \lambda_{3}$. The amplitude, the shape and the orientation of $\left\langle u_{i}u_{j} \right\rangle$ are explicitly represented by $k$, $\lambda_{i}$, and $v_{i j}$, respectively. Equations \ref{Eq:noMark_AnisotropyTensor} and \ref{Eq:noMarker_Rij} lead to 

\begin{equation}\label{Eq:noMarker_bij}
    b_{i j}=-\frac{\nu_{t} }{k}\big \langle {S_{i j}} \big \rangle = v_{i n} \hat{b}_{n l} v_{j l}.
\end{equation}

Equation \ref{Eq:noMarker_bij} indicates that the Boussinesq turbulent viscosity hypothesis requires that the shape and orientation of $\left\langle u_{i}u_{j} \right\rangle$ to be determined by $(\nu_{t}/k)\big \langle {S_{i j}} \big \rangle$. This assumption implies the $a_{i j}$ tensor is aligned with the $\big \langle {S_{i j}} \big \rangle$ tensor, which is not true in most circumstances in practice, in particular, complex flows, e.g., strongly swirling flows, flow with significant streamline curvature, and flow with separation and reattachment, and thus a source of the model form uncertainty.

\begin{figure} 
\centerline{\includegraphics[width=3.4in]{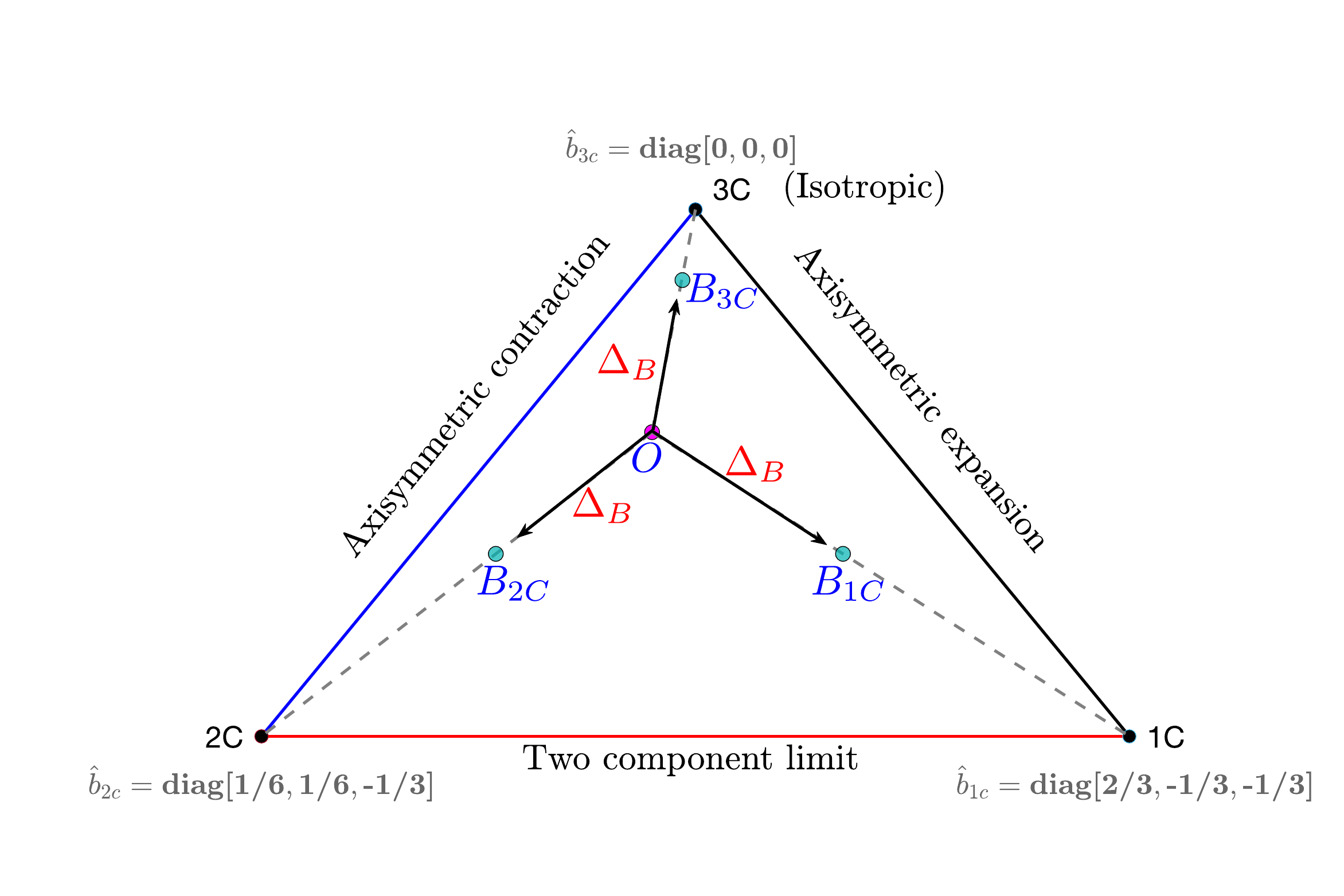}}
\caption{Barycentric map.}
\label{fig:BMap_Sketch.pdf}
\end{figure}

The eigenspace perturbation method was first proposed in \cite{emory2011modeling,gorle2012epistemic}. To quantify errors introduced by the model form uncertainty, perturbation is injected into the eigen-decomposed Reynolds stress defined in Eq. \ref{Eq:noMarker_Rij}. The perturbed Reynolds stresses are defined as

\begin{equation}\label{Eqn_Rij_perturbed}
    \left\langle u_{i} u_{j}\right\rangle^{*}=2 k^{*}\left(\frac{1}{3} \delta_{i j}+v_{i n}^{*} \hat{b}_{n l}^{*} v_{j l}^{*}\right),
\end{equation}

\noindent where $k^{*}$ is the perturbed turbulence kinetic energy, $\hat{b}_{k l}^{*}$ is the diagonal matrix of perturbed eigenvalues, and $v_{i j}^{*}$ is the matrix of perturbed eigenvectors. For eigenvalue perturbations, Pecnik and Iaccarino \cite{emory2011modeling} proposed a perturbation approach, which enforces the realizability constraints on $\left\langle u_{i}u_{j} \right\rangle$ via the barycentric map \cite{banerjee2007presentation}, as shown in Fig. \ref{fig:BMap_Sketch.pdf}, because the map contains all realizable sates of $\left\langle u_{i}u_{j} \right\rangle$. Due to the realizability constraint of the semi-definiteness of $\left\langle u_{i}u_{j} \right\rangle$, there are three extreme states of componentiality of $\left\langle u_{i}u_{j} \right\rangle$: one component limiting state ($1C$), which has one non-zero principal fluctuation, i.e., $\hat{b}_{1c}=\operatorname{diag}[2 / 3,-1 / 3,-1 / 3]$; two component limiting state ($2C$), which has two non-zero principal fluctuations of the same intensity, i.e., $\hat{b}_{2c}=\operatorname{diag}[1 / 6,1 / 6, -1 / 3]$; and three component (isotropic) limiting state (3C), which has three non-zero principal fluctuations of the same intensity, i.e., $\hat{b}_{3c}=\operatorname{diag}[0,0,0]$. In addition, the $\hat{b}_{1c}$, $\hat{b}_{2c}$, and $\hat{b}_{3c}$ limiting states correspond to the three vertices of the barycentric map. Given an arbitrary point $\mathbf{x}$ within the barycentric map, any realizable $\left\langle u_{i}u_{j} \right\rangle$ can be determined by a convex combination of the three vertices $\mathbf{x}_{i c}$ (limiting states) and $\lambda_{l}$ as follows:

\begin{equation}\label{Eq:noMarker_Coordinates_InsideBary}
    \mathbf{x} = \mathbf{x}_{1 \mathrm{c}}\left(\lambda_{1}-\lambda_{2}\right)+\mathbf{x}_{2 \mathrm{c}}\left(2 \lambda_{2}-2 \lambda_{3}\right)+\mathbf{x}_{3 \mathrm{c}}\left(3 \lambda_{3}+1\right).
\end{equation}

In order to define the perturbed eigenvalues $\hat{b}_{i j}^{*}$, first determine the location on the barycentric map for the Reynolds stresses computed by a linear eddy viscosity model and subsequently inject uncertainty by shifting it to a new location on the barycentric map. In Fig. \ref{fig:BMap_Sketch.pdf}, perturbations toward $1c$, $2c$, and $3c$ vertices of the barycentric map shift point $O$ to $B_{1c/2c/3c}$, respectively, which can be written as 

\begin{equation}\label{Eq:noMarker_xstar}
    \mathbf{x_{B(1c/2c/3c)}^{*}}=\mathbf{x_{O}}+\Delta_{B}\left(\mathbf{x}_{1c/2c/3c}-\mathbf{x_{B(1c/2c/3c)}}\right),
\end{equation}

\noindent where $\Delta_{B}$ is the magnitude of perturbation. Once the new location is determined, a new set of eigenvalues $\lambda_{i}$ can be computed from Eq. \ref{Eq:noMarker_Coordinates_InsideBary} and $b_{i j}$ can be reconstructed, which eventually yields $\left\langle u_{i}u_{j} \right\rangle^{*}$. 

%there are relatively few studies for directly perturbing $k$.
As noted earlier in Eq. \ref{Eq:noMarker_bij}, the unperturbed anisotropy Reynolds stress tensor is modeled as $b_{i j}=-\nu_{t} \backslash {k} \big \langle {S_{i j}} \big \rangle = v_{i n} \hat{b}_{n l} v_{j l}$ or, equivalently, $a_{i j} = -2\nu_{t} \big \langle {S_{i j}} \big \rangle =2kv_{i n} \hat{b}_{n l} v_{j l}$. Accordingly, the anisotropy Reynolds stress tensor subject to turbulence kinetic energy perturbation becomes

\begin{equation}\label{Eq:perturb_aij}
     a_{i j}^{*} = -2\nu_{t}^{*} \big \langle {S_{i j}} \big \rangle =2k^{*}v_{i n} \hat{b}_{n l} v_{j l}.
\end{equation}

Because perturbing $k$ does not affect the eigenvalues and eigenvectors of the anisotropy Reynolds stress tensor, the change in the turbulent viscosity hypothesis has to be accounted in the turbulent viscosity coefficient \cite{mishra2019theoretical}. Comparing the unperturbed anisotropy Reynolds stress tensor to Eq. \ref{Eq:perturb_aij}, it is easy to obtain \cite{mishra2019theoretical}

\begin{equation}\label{Eq:k_nut}
    \frac{k^{*}}{k} = \frac{\nu_{T}^{*}}{\nu_{T}}, \quad \text{or equivalently,} \quad \nu_{T}^{*} = \frac{\nu_{T}k^{*}}{k},
\end{equation}

\noindent where $k^{*} = k + \Delta_{k}$. From Eq. \ref{Eq:k_nut}, turbulence kinetic energy perturbation leads to spatial variation of turbulent viscosity coefficient. Specifically, the relation between the turbulent viscosity and the turbulent viscosity coefficient $C_{\mu}$ is given by

\begin{equation} \label{nut_unperturb}
    \nu_{T} = C_{\mu}\frac{k^{2}}{\varepsilon},
\end{equation}

\noindent where $\varepsilon$ is the dissipation rate.  

Thus, the perturbed turbulent viscosity can be expressed as follows:

\begin{equation}\label{nut_perturb}
    \nu_{T}^{*} = C_{\mu}^{*}\frac{{k^{*}}^{2}}{\varepsilon},
\end{equation}
 
\noindent where $C_{\mu}^{*} = C_{\mu} + \Delta_{C_{\mu}}$. Substituting Eqs. \ref{nut_unperturb} and \ref{nut_perturb} into Eq. \ref{Eq:k_nut}, we get \cite{mishra2019theoretical}

\begin{equation} \label{Eq:Mishra_Eq}
    \frac{k}{k^{*}} = \frac{C_{\mu}^{*}}{C_{\mu}}, \quad \text{or equivalently,} \quad \Delta_{C_{\mu}} = -\frac{\Delta_{k}C_{\mu}}{k+\Delta_{k}}.
\end{equation}

In this study, the turbulence kinetic energy discrepancies between the RANS based predictions and the in-house DNS data \cite{zhang2021turbulent} are modeled by high-order regressions. These regressions generate values of $k^{*}$ that vary spatially in the computational domain:

\begin{equation}\label{Eq:Marker_Mk_Method}
    k^{*} = k +\Delta k = kM_{k}, \quad M_{k} \sim f(x,y).
\end{equation}

In Eq. \ref{Eq:Marker_Mk_Method}, $M_{k}$ is a marker function of the $x$ and $y$ coordinate in a computational domain. Additionally, substituting Eq. \ref{Eq:Mishra_Eq} into Eq. \ref{Eq:Marker_Mk_Method} and rearranging, we get:

\begin{equation} \label{Eq:1_M_k}
    \frac{1}{M_{k}} = \frac{C_{\mu}^{*}}{C_{\mu}}.
\end{equation}

Substituting Eq. \ref{Eq:1_M_k} to Eq. \ref{Eq:Mishra_Eq}, the relation between $M_{k}$ and $\Delta_{C_{\mu}}$ can be expressed as follows:

\begin{equation}\label{Eq:Minghan_Eq}
    \Delta_{C_{\mu}} = \frac{C_{\mu}(1-M_{k})}{M_{k}}.
\end{equation}

Therefore, Eq. \ref{Eq:Minghan_Eq} provides the underlying model structure of turbulence kinetic energy perturbation with a marker function involved.

A detailed description for the modeling of $k^{*}$ is presented in Section \ref{RegressionM}. In addition, eigenvector perturbations rotate the eigenvectors of the anisotropy Reynolds stress tensor with respect to the principal axes of the mean rate of strain. Recall that the eigenvectors of the anisotropy Reynolds stress tensor are forced to align along the principal axes of the mean rate of strain due to the limitations of the Boussinesq turbulent viscosity hypothesis \cite{pope2001turbulent}. This again violates the true physics of turbulent flow. Consequently, eigenvector perturbations extend the Boussinesq turbulent viscosity hypothesis to anisotropy turbulent viscosity hypothesis. Unlike eigenvalue perturbations, which are strictly constrained by realizability. Eigenvector perturbations are more difficult to be physically constrained in a local sense. In this study, eigenvector perturbations are omitted for brevity. Therefore, the present study restricts the contribution to the amplitude and shape perturbation to the anisotropy Reynolds stress tensor.

\subsection{Eigenspace perturbation framework in OpenFOAM}
At present the eigenspace perturbation framework is available only in Stanford University's SU2 CFD suite \cite{mishra2019uncertainty} and the TRACE solver of DLR \cite{matha2022assessment}. In spite of its utility to the design and simulation community, there are no tested and validated implementations of this framework available in popular CFD software. OpenFOAM \cite{winkelman1980flowfield} is the most widely used open source CFD software in research and academia. A contribution of this investigation, is the development of a verified and validated implementation of the eigenspace perturbation framework for the OpenFOAM software. Relatively few studies have been conducted to implement the eigenspace perturbation framework in a RANS formulation using OpenFOAM, e.g., see \cite{cremades2019reynolds,hornshoj2021quantifying}. All of these studies employed the MATLAB software compounded with OpenFOAM to decompose and recompose the Reynolds stress tensor. This increases the complexity of using the eigenspace perturbation framework \cite{emory2013modeling} in OpenFOAM, which is prone to errors and violating the spirit of versatility. In addition, C++ is inherently faster than Matlab, which reduces the computational expense. In this study, the eigenspace perturbation framework along with the novel marker functions were completely implemented in C++ in OpenFOAM, which greatly reduces the number of user-defined inputs and allows the users without much knowledge of the fluid mechanics to use the eigenspace perturbation framework in OpenFOAM.

\begin{figure*} 
\centerline{\includegraphics[width=6in]{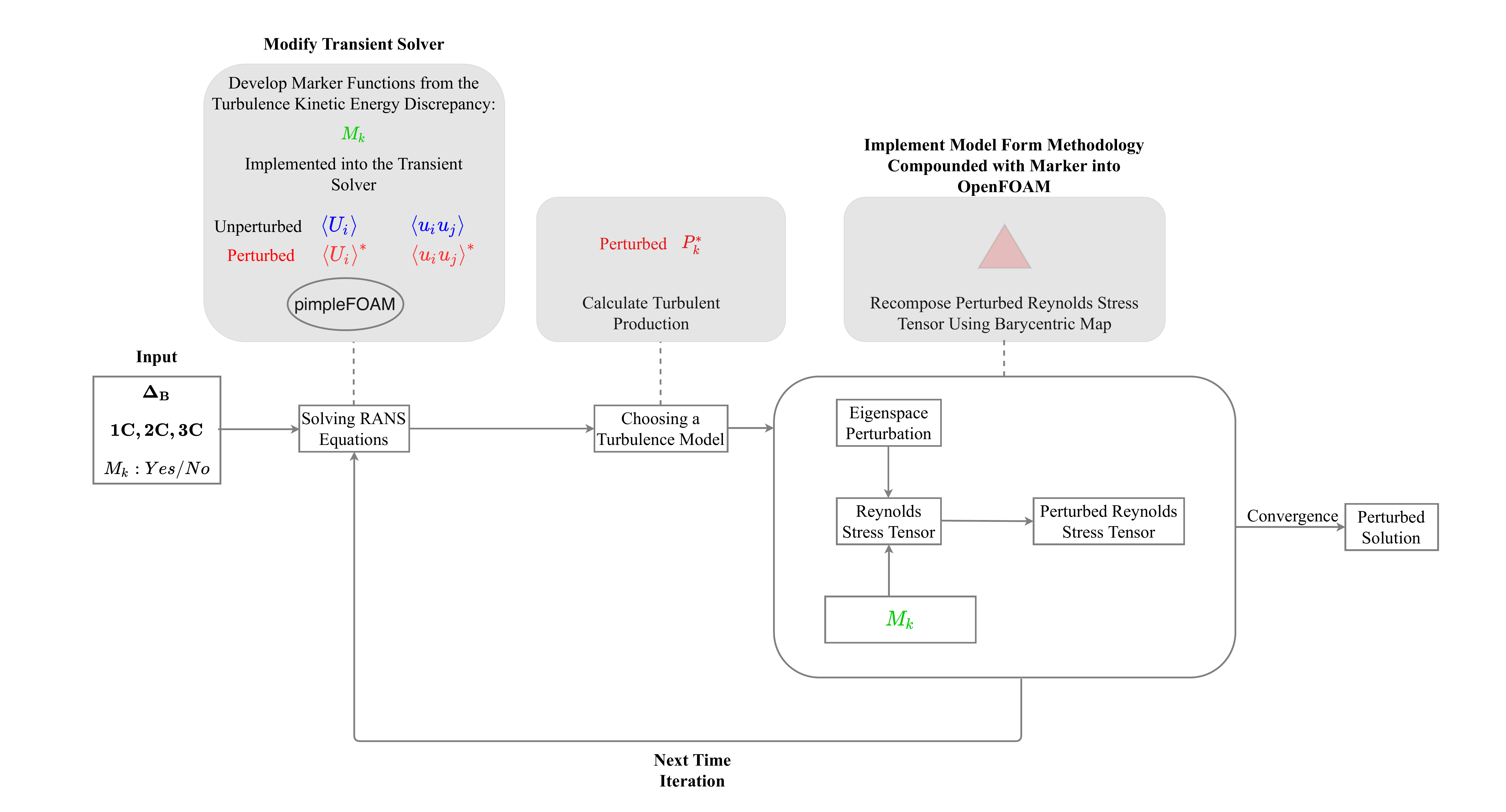}}
\caption{Flow chart showing the implementation of model form framework within OpenFOAM with marker configuration involved.}
\label{fig:FlowChart_Method_Marker.pdf}
\end{figure*}

In the input files (located under the ``constant'' directory in OpenFOAM), user needs to specify what magnitude of $\Delta_{B}$ should be assigned, if $M_{k}$ is needed, and which eigenvalue perturbation ($1c$, $2c$, $3c$) is to be performed. The eigenspace framework conducts the perturbations during the execution of simulations, as illustrated in Fig. \ref{fig:FlowChart_Method_Marker.pdf}. At each control volume (CV), the baseline Reynolds stress tensor is calculated and decomposed into its eigenvalue and eigenvector matrices, which are perturbed using the eigenspace perturbation method as prescribed earlier. If $M_{k}$ is involved, perturbation to the turbulence kinetic energy will be performed. The perturbed eigenvalue and eigenvector matrices are then recomposed into a perturbed Reynolds stress tensor for each CV. These perturbed Reynolds stress matrices together with the perturbed turbulence kinetic energy are then used to compute the perturbed velocity field and the perturbed turbulent production to advance each node to the next time step. At convergence, the Reynolds stress also converges to its perturbed state.

\section{Flow description and numerical method}

\begin{figure*} 
\centerline{\includegraphics[width=5.0in]{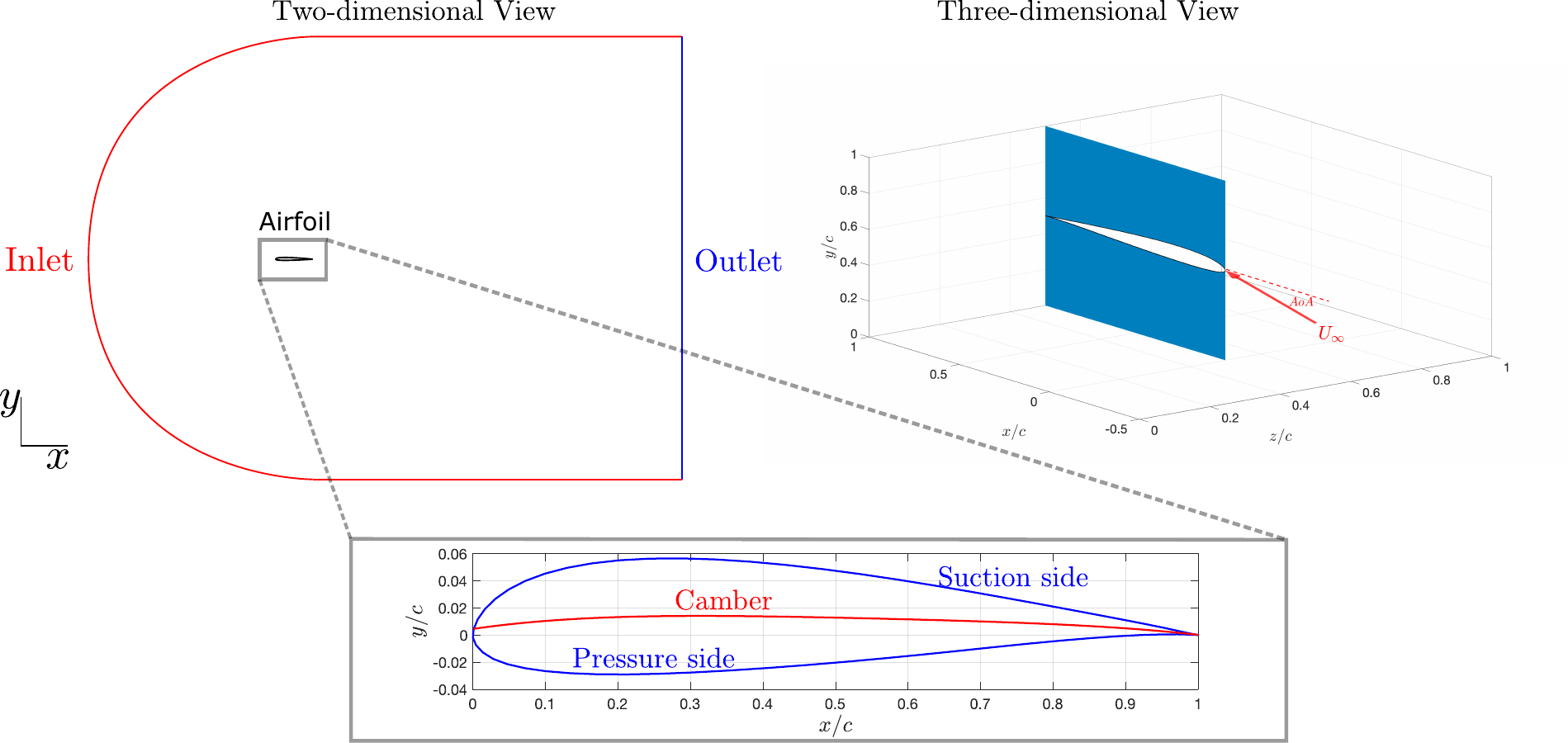}}
  \caption[SD7003 computational domain and boundary conditions: {\color{red} \rule{0.7cm}{0.4mm}} far field, {\color{blue} \rule{0.7cm}{0.4mm}} outflow, {\color{black} \rule{0.7cm}{0.4mm}}, and no-slip walls.]{SD7003 computational domain and boundary conditions: {\color{red} \rule{0.7cm}{0.4mm}} far field, {\color{blue} \rule{0.7cm}{0.4mm}} outflow, {\color{black} \rule{0.7cm}{0.4mm}}, and no-slip walls. Depiction of the suction side, camber, and pressure side of the SD7003 airfoil is displayed in the magnified plot. A three-dimensional version of the computational domain is provided with freestream ($U_{\infty}$) encountering the leading edge at $8^{\circ}$ AoA.}
\label{fig:no_Mk_SD7003domain.pdf}
\end{figure*}

The flow being considered is around an SD7003 airfoil, as shown in Fig. \ref{fig:no_Mk_SD7003domain.pdf}. At the low Reynolds number based on the chord length of $\operatorname{Re}_{c} = 60000$, a laminar separation bubble (LSB) is formed on the suction side of the airfoil. Note that the bubble moves upstream as the angle of attack (AoA) increases \cite{catalano2011rans}. In this study, an $8^{\circ}$ AoA (nearing stall) was considered. Figure \ref{fig:no_Mk_SD7003domain.pdf} schematically shows that the solution domain is a two-dimensional C-topology grid of $389$ (streamwise) $\times$ $280$  (wall-normal) $\times$ $1$ (spanwise) control volumes, which is comparable to the number of control volumes ($768 \times 176$) used in the numerical study of \cite{catalano2011rans}. The magnified view of the two-dimensional SD7003 airfoil labels the camber, suction side and pressure side, as shown in Fig. \ref{fig:no_Mk_SD7003domain.pdf}. The first grid node to the wall was placed at $y^{+} \approx 1.0$ in the turbulent boundary layer, in which more than $20$ CVs were placed. A grid convergence study has been performed to test the influence of the grid resolution on the results. Grid dependency study indicated that higher grid resolution in the near-wall region results in negligible changes in the predicted results: the effect of increasing the number of CVs in the wall-normal direction on the predicted mean velocity and Reynolds shear stress profile was at most $1\%$. Therefore, the simulation results based on the smaller grid ($389 \times 280$) has been used in the present analysis.

The governing Eqs. \ref{p_Continuity} - \ref{p_Momentum} were closed by the RANS based transition model of \cite{langtry2009correlation} using OpenFOAM. The transport equations were discretized on a staggered mesh using finite volume method. The scheme is second order upwind for spatial dicretization, and Gauss linear scheme was used to evaluate the gradients. The PIMPLE algorithm was adopted for pressure-velocity coupling, which is a combination of PISO (Pressure Implicit with Splitting of Operator) \cite{ferziger2002computational} and SIMPLEC (Semi-Implicit Method for Pressure Linked Equations-Consistent)  \cite{van1984enhancements}. It should be noted that PIMPLE algorithm can deal with large time steps where the maximum Courant (C) number may consistently be above $1$. In this study, the maximum value of C was set consistently equal to $0.6$, and OpenFOAM automatically adjusted the time step to achieve the set maximum. In addition, both residuals and distributions of lift and drag coefficients that vary with respect to time ($T$) were used to track convergence status. The solution fields were iterated until convergence, which required residuals of energy and momentum to drop more than four orders of magnitude, and both lift and drag coefficients almost stopped changing with time. This happened at $T \approx 0.3$, which corresponds to a normalized time $T^{*} = T U_{\infty} / c = 6.75$, and similar behavior has been observed by Catalano and Tognaccini \cite{catalano2011rans} in their numerical study for a low-Reynolds number flow over a SD7003 airfoil at $\mathrm{AoA} = 10^{\circ}$. Sampling began at $T = 0.6$ (double the time of convergence) and ended at $T = 1.4$, which required approximately $35000$ iterations for all simulations.

The fluid was assumed to be air, with freestream turbulence intensity of $\operatorname{Tu} = 0.03\%$ and kinematic viscosity of $\nu = 1.5 \times 10^{-5} \mathrm{~m}^{2} / \mathrm{s}$. Ideally, the value of $\operatorname{Tu}$ should be close to zero. 
%which is the smallest possible value of $\operatorname{Tu}$ according to \cite{langtry2009correlation}.
From Fig. \ref{fig:no_Mk_SD7003domain.pdf} at the inlet of the domain, the freestream velocity was set equal to $4.5 \ m/s$, which corresponds to $\operatorname{Re}_{c} = 60000$. The chord length was set equal to $c = 0.2 \ m$. At the outlet, a zero-gradient boundary condition was implemented for $\left\langle U_{i}\right\rangle$ ($\left\langle U \right\rangle$ for $x$ direction, $\left\langle V \right\rangle$ for $y$ direction), $k$, $\omega$ and pressure. At the wall, a no-slip boundary condition was used.

\section{Regression model for amplitude perturbation}
An important and novel focus of this study is the development of a marker function that modulates the degree of perturbations over the entire flow domain. We have explained earlier that this should lead to better calibrated confidence intervals. In this section, high-order polynomial regressions are constructed using MATLAB software in a least-squares sense to fit both the baseline RANS and in-house DNS datasets. Note that these high-order regressions lay the foundation for the development of the new marker functions.

\subsection{Example: a linear regression}\label{RegressionM}
\begin{figure} 
\centerline{\includegraphics[width=3.5in]{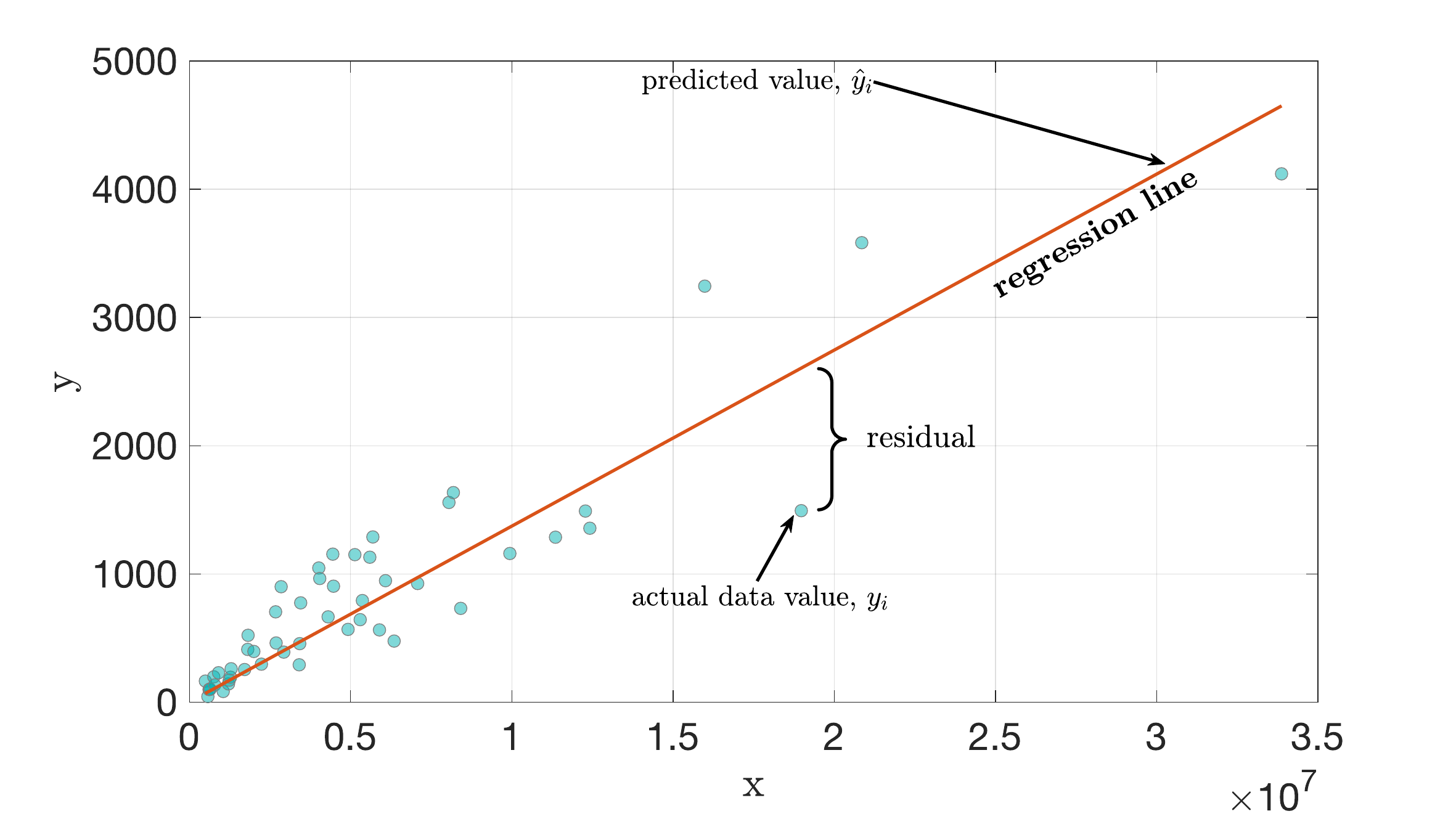}}
\caption{Linear regression relation between $x$ and $y$.}
\label{fig:Linear_Regression_Eg.pdf}
\end{figure}

The $n^{th}$ polynomial regression model that describes the relationship between a dependent $y$ and an independent $x$ can be expressed as

\begin{equation} \label{Eqn:poly_fit}
    y(x)=p_{1} x^{n}+p_{2} x^{n-1}+\ldots+p_{n} x+p_{n+1}
\end{equation}    

where $p = 1,\ldots,n+1$ stands for the coefficients in descending orders, $x$ is the independent variable. Fig. \ref{fig:Linear_Regression_Eg.pdf} illustrates a first-order or linear regression model on a random dataset. The errors $y_{i}-\hat{y}_{i}$ between the predicted values $\hat{y}_{i}$ and the actual data values $y_{i}$ are referred to as residuals. Using MATLAB software, the least-squares method finds the coefficients $p_{i}$ that best fit this datasets by minimizing the sum of squared residuals, i.e.:    

\begin{equation}
    RSS = \sum_{j=1}^{n}\left(y_{j}-\hat{y}_{j}\right)^{2}, \quad j = 1,\ldots,m
\end{equation}

where $RSS$ stands for residual sum of squares, $y_{j}$ is the $j$th actual value of the dependent variable to be predicted, $m$ represents the number of points of the datasets, $\hat{y}_{j}$ is the $j$th predicted value of $y_{j}$. 

\subsubsection{Define untrustworthy regions}\label{Sec:Def_untrust}
\begin{figure*} 
\centerline{\includegraphics[width=5.5in]{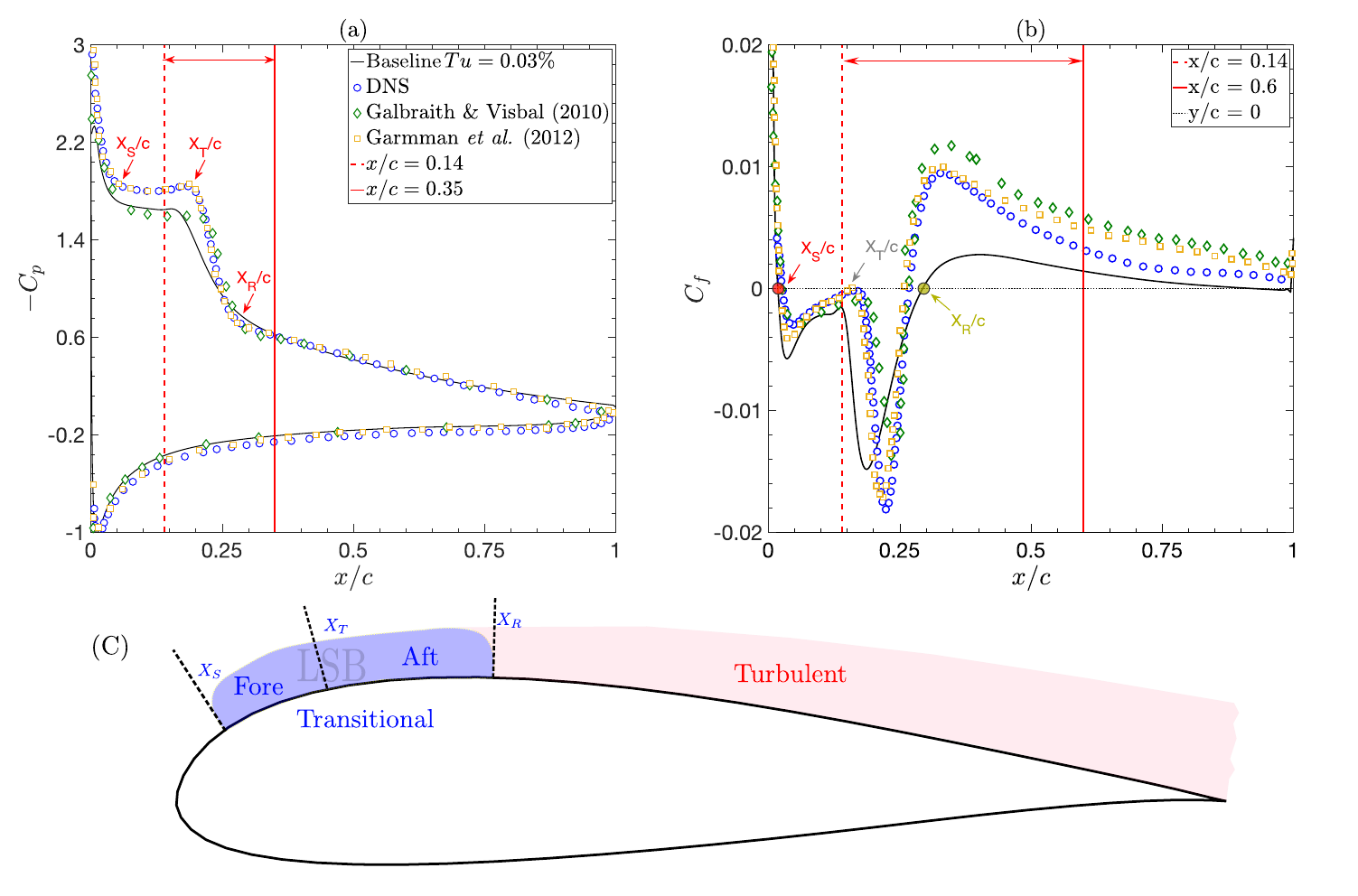}}
\caption[Distribution of (a) pressure coefficient and (b) skin friction coefficient over the SD7003 airfoil at $\operatorname{Re}_{c}=6 \times 10^{4}$ and $\operatorname{AoA} = 8^{\circ}$.]{Distribution of (a) pressure coefficient and (b) skin friction coefficient over the SD7003 airfoil at $\operatorname{Re}_{c}=6 \times 10^{4}$ and $\operatorname{AoA} = 8^{\circ}$. Two headed arrow is added to indicate the untrustworthy region. (c) Schematic of transitional and turbulent regions over a SD7003 airfoil with important transitional parameters highlighted.}
\label{fig:cfcp_Tu003_forMarker.pdf}
\end{figure*}

%Comparison of transition parameters
\begin{table}
\begin{center}
%\captionsetup{width=.58\textwidth}
%\captionsetup{skip=0pt}
\caption{Comparison of transition parameters.}

\label{table:transi_parameters}

\begin{ruledtabular}

\begin{tabular}{c c c c}
%& & \\ % put some space after the caption
%\hline
Method & $X_{S}/c$ &$X_{T}/c$ &$X_{R}/c$\\
\hline
SSTLM (Baseline) \cite{langtry2009correlation} & $0.03$& $0.15$ &$0.29$ \\

In-house DNS \cite{zhang2021turbulent} & $0.02$ &$0.16$&$0.27$\\

LES \cite{garmann2013comparative}& $0.02$ &$0.16$&$0.27$\\

ILES \cite{galbraith2010implicit}& $0.03$ &$0.18$&$0.27$\\

%\hline
\end{tabular}
\end{ruledtabular}
\end{center}
\end{table}

To construct marker functions for $k^{*}$, first and foremost is to identify the regions where the turbulent viscosity hypothesis becomes invalid. This study identifies the regions where the RANS model gives plausible untrustworthy results based on the comparison between the baseline prediction and the in-house DNS data of \cite{zhang2021turbulent}. For the flow over an airfoil geometry, perhaps the local wall shear stress and the local pressure are the most important parameters, whose dimensionless forms become the skin friction coefficient $C_{f}=\tau_{w} / {0.5 \rho U_{\infty}^{2}}$, where $\tau_{w}$ is the wall shear stress, and the pressure coefficient $C_{p}=(p-p_{\infty}) / {0.5 \rho U_{\infty}^{2}}$, where $p$ is the undisturbed static pressure and $p_{\infty}$ is the static pressure in the freestream, respectively. In Figs. \ref{fig:cfcp_Tu003_forMarker.pdf} (a) and (b), the predicted $C_{f}$ and $C_{p}$ are plotted. According to the technique described by Boutilier and Yarusevych \cite{boutilier2012parametric}, Fig.  \ref{fig:cfcp_Tu003_forMarker.pdf} (a) shows three ``kinks'' as representatives of the separation, transition and reattachment points, denoted $X_{S}/c$, $X_{T}/c$ and $X_{R}/c$, respectively. Moreover, the size of the LSB can be determined by finding the $X_{S}/c$ and $X_{R}/c$ points, which can be determined as the zeros of the skin friction coefficient \cite{de2021model}. The two methods showed good agreement with each other, and a summary of these important transition parameters are tabulated in Table \ref{table:transi_parameters}. In this study, the LSB is treated to be composed of a ``fore'' (from $X_{S}/c$ to $X_{T}/c$) and an ``aft'' (from $X_{T}/c$ to $X_{R}/c$) portion for the sake of analysis simplicity, followed by a fully turbulent region, as shown in Fig. \ref{fig:cfcp_Tu003_forMarker.pdf} (c). The in-house DNS \cite{zhang2021turbulent} and implicit LES (ILES)/LES data of \cite{galbraith2010implicit} and \cite{garmann2013comparative} for $C_{f}$ and $C_{p}$ are included for comparison. In the fore portion of the LSB, the predicted $C_{p}$ profile shows relatively good agreement with the ILES data of \cite{galbraith2010implicit}, while a clear discrepancy is observed in the aft portion, where it gives a smaller value of $C_{p}$, i.e. the region indicated by the two headed arrow. This kind of discrepancy was observed by Tousi \textit{et al.} \cite{tousi2021active} in their numerical study as well. Besides, the predicted $C_{p}$ shows good agreement with the reference data for the turbulent region on the suction side, as well as shows good agreement with the reference data for the entire pressure side. On the other hand, a noticeable discrepancy is observed on the $C_{f}$ profile at the negative ``trough'' in the aft portion of the LSB, as well as at the positive ``crest'' in the turbulent boundary layer after the reattachment point $X_{R}$. In Fig. \ref{fig:cfcp_Tu003_forMarker.pdf} (b), a shift of the predicted $C_{f}$ profile in the upstream direction at the trough is observed, and the value of $C_{f}$ is significantly under-predicted at the crest in the region of turbulent boundary layer for $0.3 < x/c < 0.6$. This behavior has been observed by other researchers as well, e.g., see \cite{catalano2011rans,bernardos2019rans,tousi2021active}. Therefore, it can be concluded that the region for $0.14 \leq x/c \leq 0.6$, as indicated by the double headed arrow shown in Fig. \ref{fig:cfcp_Tu003_forMarker.pdf} (b), should be identified as representative for the untrustworthy regions where perturbations should be introduced. 

\begin{figure} 
\centerline{\includegraphics[width=3.5in]{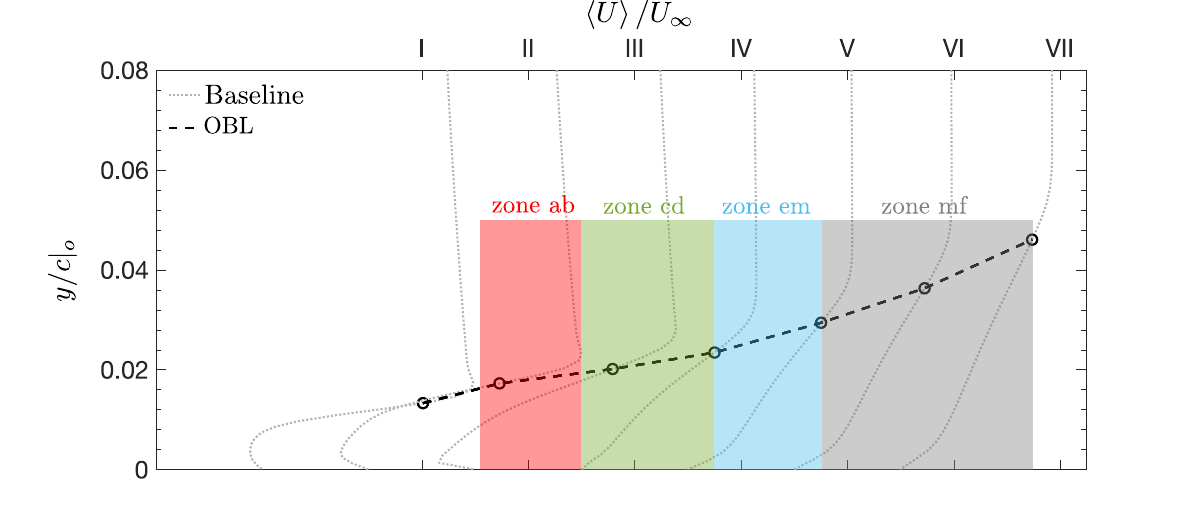}}
 \caption[Injection of uncertainty into the untrustworthy zones: \textcolor{red}{zone $ab$}, \textcolor{OliveGreen}{zone $cd$}, \textcolor{Cyan}{zone $em$} and \textcolor{gray}{zone $mf$}.]{Injection of uncertainty into the untrustworthy zones: \textcolor{red}{zone $ab$}, \textcolor{OliveGreen}{zone $cd$}, \textcolor{Cyan}{zone $em$} and \textcolor{gray}{zone $mf$}. The outer edge of the boundary layer ({\color{black} \protect\tikz[baseline]{\protect\draw[dashed] (0,.5ex)--++(.5,0) ;}} with $\circ$) at seven locations $I = x/c = 0.02$, $II = x/c = 0.03$, $III = x/c = 0.04$, $IV = x/c = 0.06$, $V = x/c = 0.08$, $VI = x/c = 0.10$, $VII = x/c = 0.12$ ($\cdots$) selected on the suction side are provided for reference.}
\label{fig:BL_black}
\end{figure}

This study ensures that the amplitude perturbation is introduced across the entire boundary layer within the untrustworthy region $0.14 \leq x/c \leq 0.6$, which is further divided into the $ab$, $cd$, $em$ and $mf$ zone. In Fig. \ref{fig:BL_black}, the mean velocity profiles at seven locations downstream of the leading edge are used to illustrate the flow development in the streamwise direction, i.e. $I = x/c = 0.1$, $II = x/c = 0.15$, $III = x/c = 0.2$, $IV = x/c = 0.3$, $V = x/c = 0.4$, $VI = x/c = 0.5$, and $VII = x/c = 0.6$. Due to the airfoil curved upper surface, the mean velocity profiles are shifted down to the origin of $y/c$, denoted $y/c|_{o} = (y-y_{w})/c$ for sake of better contrast, where $y_{w}$ is the vertical location of the upper surface of the airfoil. Figure \ref{fig:BL_black} clearly shows that the boundary layer thickness increases as the flow develops in the streamwise direction downstream of the leading edge, i.e. the dash line with open circles indicates the approximate thickness of the outer edge of the boundary layer (OBL). In this study, the regions within which the amplitude perturbation will be introduced are shaded red, green, blue and gray corresponding to the $ab$, $cd$, $em$ and $mf$ zone, respectively, as shown in Fig. \ref{fig:BL_black}. It is clear that all these shaded regions extend well beyond OBL, i.e. $0 < y/c|_{o} < 0.05$, implying the propagation of the amplitude perturbation effect deeper into the outer boundary layer as the flow develops further downstream of the leading edge.

\subsubsection{Polynomial regression for DNS/RANS turbulence kinetic energy datasets}\label{Sec:Poly_reg_k}

\begin{figure} 
\centerline{\includegraphics[width=3.5in]{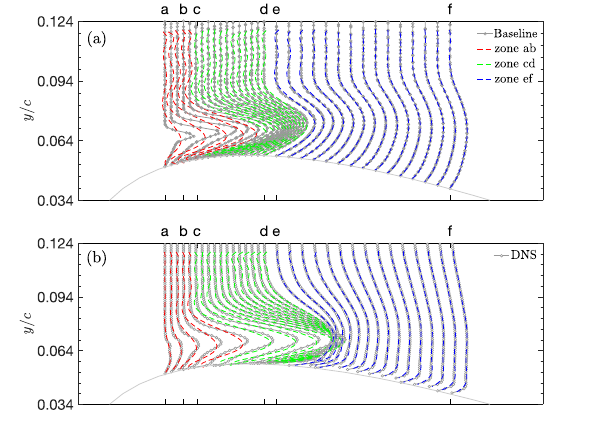}}
\caption{(a) Regressed profile of normalized turbulence kinetic energy for the baseline RANS and (b) in-house DNS datasets (gray profiles) along the suction side of the SD7003 airfoil (geometry depicted by gray line): from left to right are \textcolor{red}{zone $ab$}, \textcolor{Green}{zone $cd$} and \textcolor{Blue}{zone $ef$}.}
\label{fig:Marker_SSTLM_DNS_fit_k_all.pdf}
\end{figure}

In this study, MATLAB software was used to construct a set of least squares higher-order regression lines that are used to fit seventh-order polynomials to both the baseline RANS and in-house DNS datasets, i.e. gray lines with open circles, for turbulence kinetic energy normalized with the freestream velocity squared, $k/U_{\infty}^2$ are shown in Figs. \ref{fig:Marker_SSTLM_DNS_fit_k_all.pdf} (a) and (b). There are 5 locations selected for the $ab$ zone and 12 locations for the $cd$ zone. The regression based $k/U_{\infty}^2$ profiles for the $ab$ and $cd$ zone are colored red and green, respectively, with a uniform spacing of $x/c = 0.01$. As the flow proceeds further downstream, the regression based $k/U_{\infty}^2$ profiles for the $ef$ zone, which comprises a $em$ and $mf$ subzone, are colored blue, within which 15 locations are selected with a uniform spacing of $x/c = 0.02$. Within each zone same number of locations are selected for both the regression based RANS and in-house DNS $k/U_{\infty}^2$ profiles, as documented in Table \ref{table:marker_ranges}. As a result, a total of 32 locations are selected and placed uniformly on the suction side of the airfoil, ranging from the LSB to the fully turbulent flow further downstream. In addition, the locations are more densely packed by imposing a smaller spacing distance within the $ab$ and $cd$ zone, where the LSB evolves and complex flow features start developing. Therefore, a closer investigation into this region is taken. From Figs. \ref{fig:Marker_SSTLM_DNS_fit_k_all.pdf} (a) and (b), the regression based RANS $k/U_{\infty}^2$ profiles in general exhibit a similar behavior as that for in-house DNS, i.e. a gradual increase of the $k/U_{\infty}^2$ profile in the $ab$ and $cd$ zone, followed by a reduction of the profile further downstream in the $ef$ zone. 
%It should be noted that the magnitude of the regression based RANS $k/U_{\infty}^2$ profiles is clearly under-predicted in the $cd$ zone compared to that for in-house DNS, in particular, close to the end of the $cd$ zone  a rather strong increase of the in-house DNS $k/U_{\infty}^2$ profiles penetrating into the profiles further downstream in the $ef$ zone. 

%Table for coefficients
\begin{table*}
\begin{center}

%\captionsetup{width=1\textwidth}
%\captionsetup{skip=0pt}
\caption{Zone ranges for the untrustworthy region.}

\label{table:marker_ranges}

\begin{ruledtabular}
\begin{tabular}{c c c c c c}
%& & \\ % put some space after the caption
%\hline
&\textcolor{red}{zone $ab$} & \textcolor{Green}{zone $cd$} &\textcolor{blue}{zone $em$} &\textcolor{gray}{zone $mf$}\\
\hline
$x/c$ & \textcolor{red}{$0.14 \leq \frac{x}{c}\leq 0.18$}& \textcolor{Green}{$0.18 < \frac{x}{c}\leq 0.3$} &\textcolor{blue}{$0.3 < \frac{x}{c}\leq 0.4$} & \textcolor{gray}{$0.4 < \frac{x}{c}\leq 0.6$}\\

$y/c$ & \textcolor{red}{$y_{w} \leq \frac{y}{c}\leq 0.1$} &\textcolor{Green}{$y_{w} \leq \frac{y}{c}\leq 0.1$}&\textcolor{blue}{$y_{w} \leq \frac{y}{c}\leq 0.1$}&\textcolor{gray}{$y_{w} \leq \frac{y}{c}\leq 0.1$}\\

Number of locations & \textcolor{red}{5} & \textcolor{Green}{12} &\multicolumn{2}{c}{\textcolor{blue}{15}}\\

Spacing of $x/c$ & \multicolumn{2}{c}{\textcolor{black}{0.01}} & \multicolumn{2}{c}{\textcolor{black}{0.02}} \\
%\hline
\end{tabular}
\end{ruledtabular}
\end{center}
\end{table*}

\subsubsection{Spatial discrepancies in $k/U_{\infty}^2$ regressions from DNS/RANS comparison}
\begin{figure} 
\centerline{\includegraphics[width=3.5in]{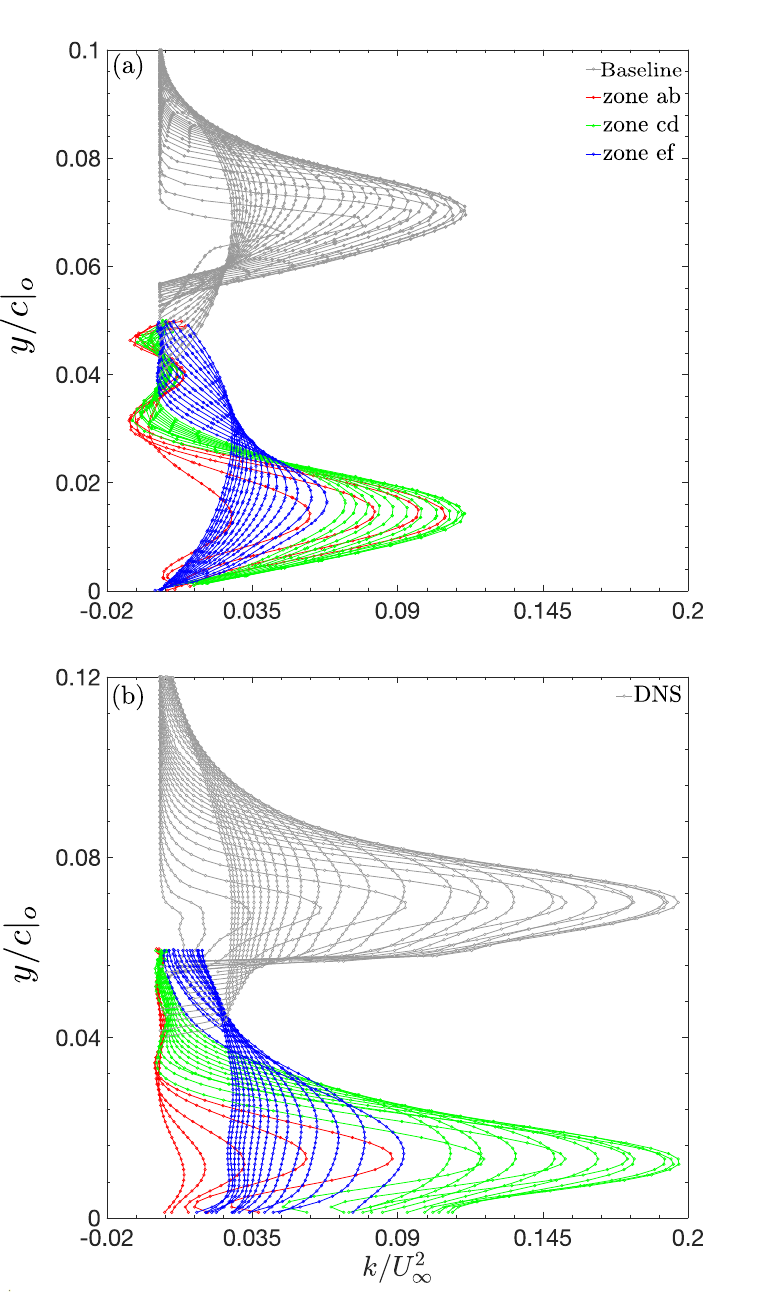}}
\caption[(a) Regression based profile of normalized turbulence kinetic energy for baseline RANS and (b) in-house DNS for \textcolor{red}{zone $ab$}, \textcolor{Green}{zone $cd$} and \textcolor{Blue}{zone $ef$}.]{(a) Regressed profile of normalized turbulence kinetic energy for baseline RANS and (b) in-house DNS for \textcolor{red}{zone $ab$}, \textcolor{Green}{zone $cd$} and \textcolor{Blue}{zone $ef$}. Actual datasets (gray profiles) for baseline RANS and in-house DNS are provided for reference.}
\label{fig:SSTLMDNS_origin_fit_all.pdf}
\end{figure}

In Figs. \ref{fig:SSTLMDNS_origin_fit_all.pdf} (a) and (b), these 32 regression based $k/U_{\infty}^2$ profiles are shifted to the origin of the $x/c$ and $y/c$ axes, respectively, for sake of strong contrast. The baseline predictions and the in-house DNS data are also included for reference, depicted by the gray lines with open circles and shifted to the origin of the $x/c$ axis to be distinguished from these regression based profiles. From Fig. \ref{fig:SSTLMDNS_origin_fit_all.pdf} (a), the regression based RANS $k/U_{\infty}^2$ profiles increase in magnitude as the flow moves further downstream. This is qualitatively similar to that for the in-house DNS profile shown in Fig. \ref{fig:SSTLMDNS_origin_fit_all.pdf} (b). From Figs. \ref{fig:SSTLMDNS_origin_fit_all.pdf} (a) and (b), the regression based RANS $k/U_{\infty}^2$ profiles increase in a somewhat larger magnitude than that for in-house DNS for the $ab$ zone; however, they are significantly reduced in magnitude compared to that for in-house DNS for the $cd$ zone (in the aft portion of the LSB), i.e., the magnitude of reduction is around $50 \%$. For the $ef$ zone, both the regression based RANS and in-house DNS $k/U_{\infty}^2$ profiles show a gradual decrease in magnitude as the flow moves further downstream. Overall, the regression based RANS and in-house DNS $k/U_{\infty}^2$ profiles are similar in magnitude for the $ab$ and $ef$ zones, but the discrepancy is significant in the aft portion of the LSB for the $cd$ zone.

%For the $cd$ zone, the regression based RANS $k/U_{\infty}^2$ profiles show a rather mild increase in magnitude, sitting slightly above the profiles for the $ab$ zone, as shown in Fig. \ref{fig:SSTLMDNS_origin_fit_all.pdf} (a).

\subsubsection{Marker for $k^{*}$}

\begin{figure} 
\centerline{\includegraphics[width=3.5in]{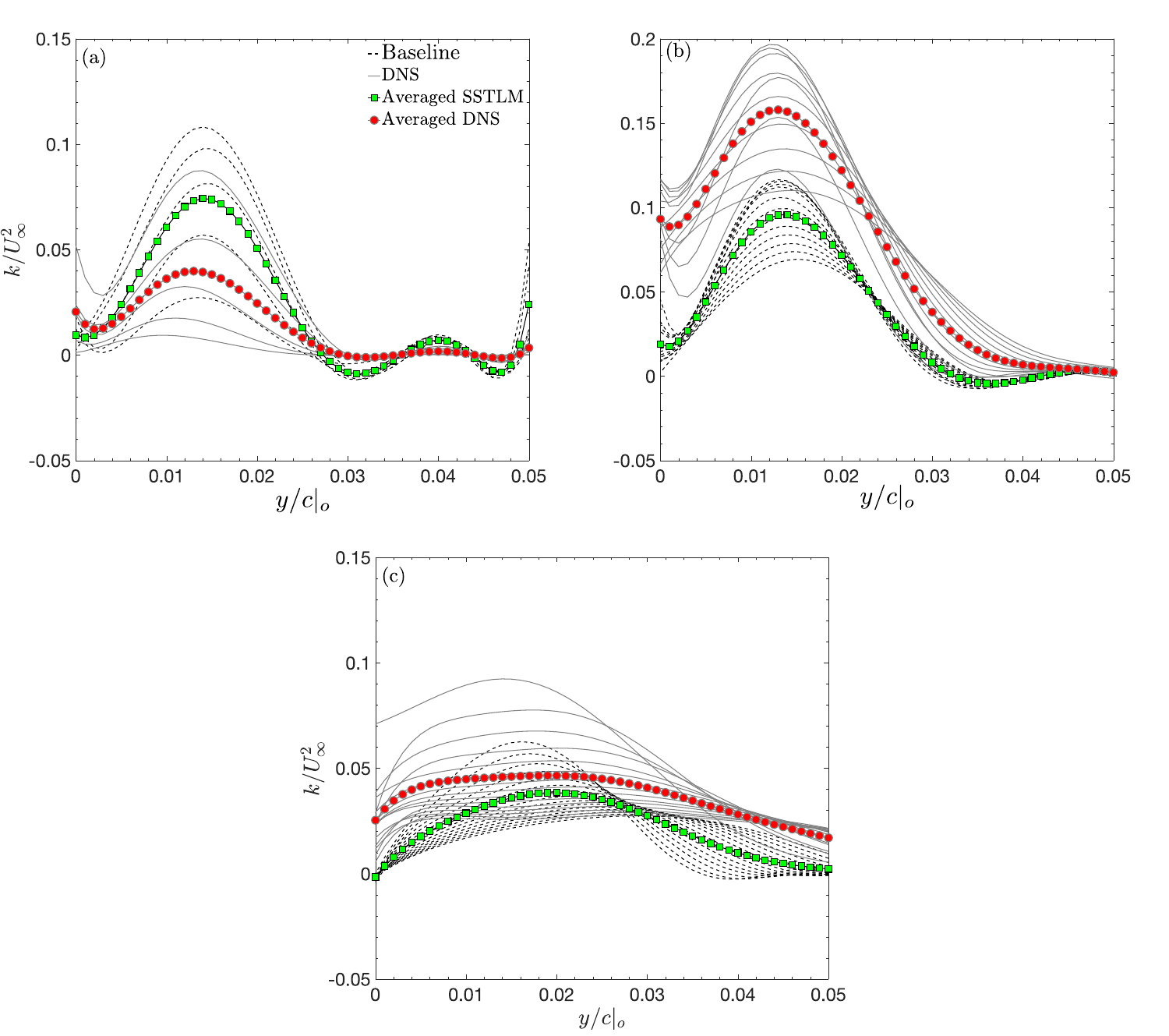}}
\caption[Mean of regression lines for normalized turbulence kinetic energy of both baseline RANS (line with green squares) and in-house DNS (line with red circles). (a) zone ab; (b) zone cd; and (c) zone ef.]{Mean of regression lines for normalized turbulence kinetic energy of both baseline RANS (line with green squares) and in-house DNS (line with red circles). (a) zone ab; (b) zone cd; and (c) zone ef. Also included are profiles of baseline RANS (gray-dashed) and and in-house DNS (gray-solid) for reference.}
\label{fig:avek_RANSDNS_xpls_0pt14To0pt60.pdf}
\end{figure}

\begin{figure*} 
\centerline{\includegraphics[width=5.5in]{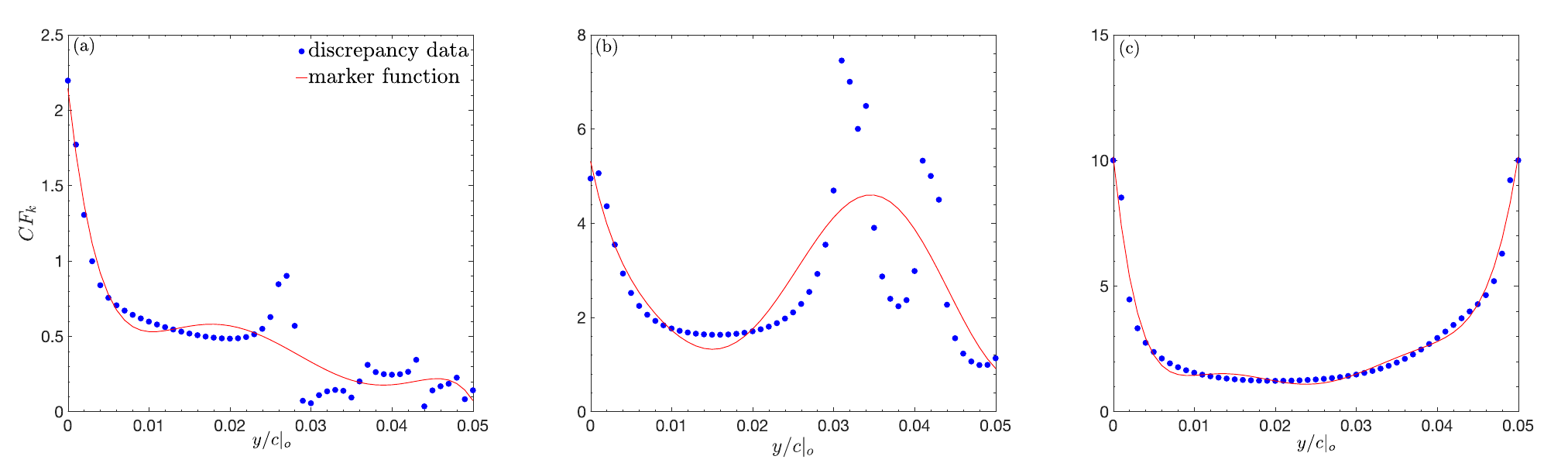}}
\caption{Defining the marker function for (a) zone $ab$, (b) zone $cd$ and (c) zone $ef$ based on the corresponding discrepancy data.}
\label{fig:kcorrection factor_RANSDNS_three.pdf}
\end{figure*}

\begin{figure} 
\centerline{\includegraphics[width=3.5in]{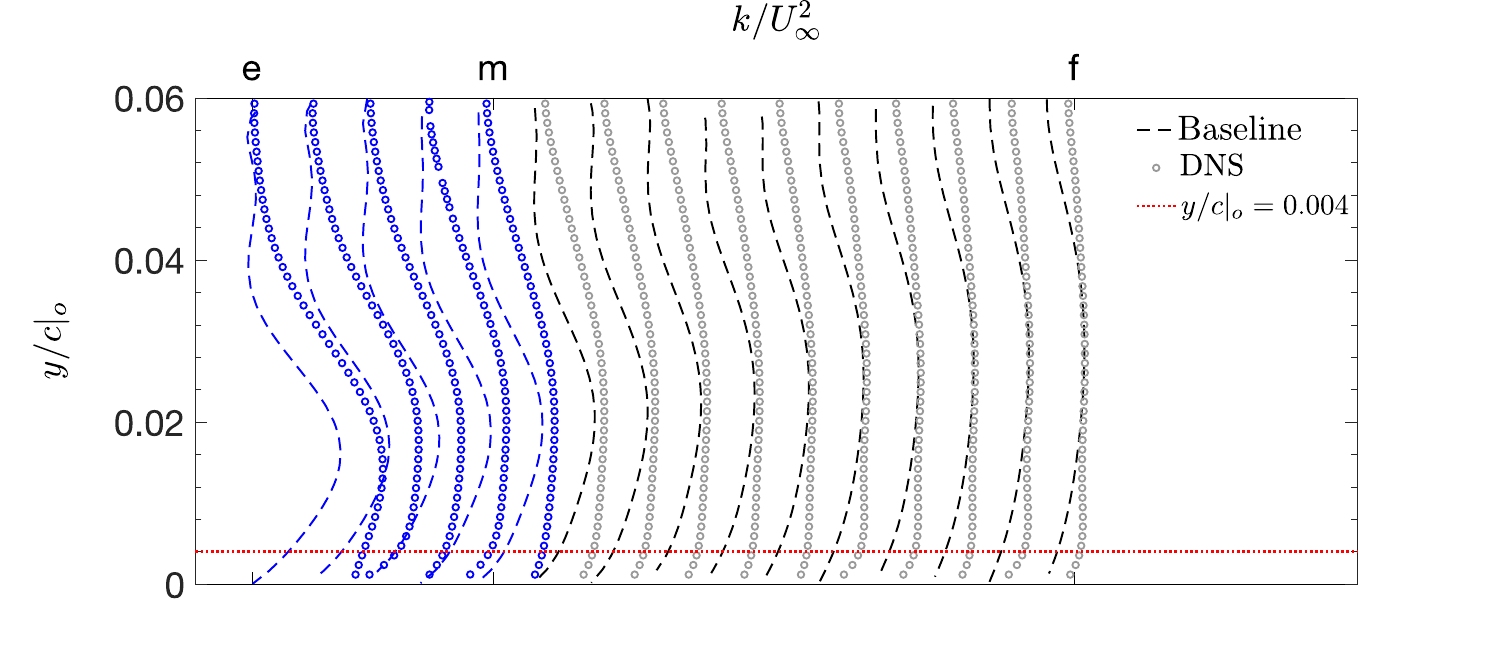}}
\caption{Defining the marker function for subzone $mf$.}
\label{fig:RANSDNS_xpls_0pt3To0pt60_withIndicator.pdf}
\end{figure}

\begin{figure} 
\centerline{\includegraphics[width=3.5in]{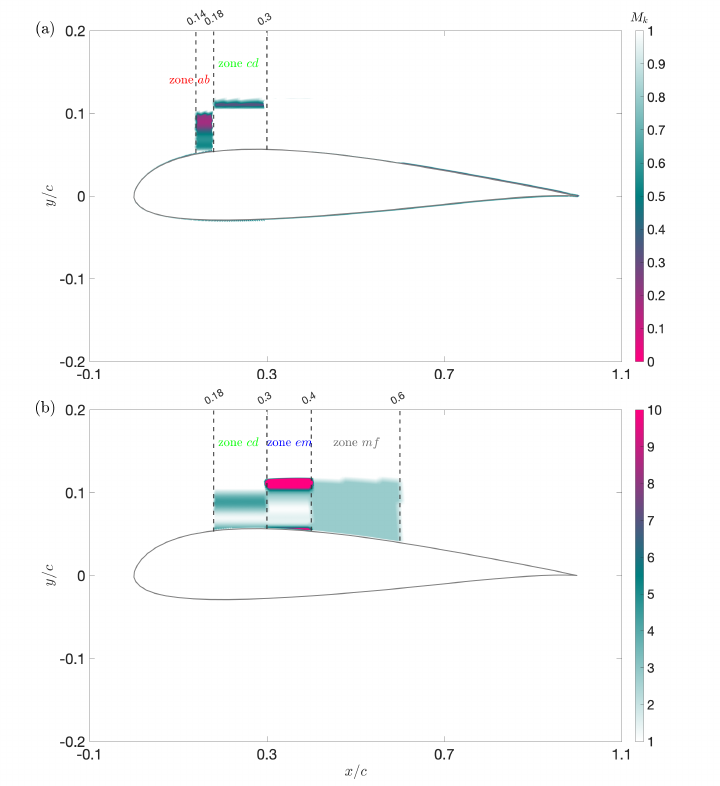}}
\caption[Contours of $M_{k}$ (Eq. \ref{Eqn:Markerfunc}) for (a) $0 < M_{k} < 1$ and (b) $1 < M_{k} < 10$ in an $xy$ plane.]{Contours of $M_{k}$ (Eq. \ref{Eqn:Markerfunc}) for (a) $0 < M_{k} < 1$ and (b) $1 < M_{k} < 10$ in an $xy$ plane. The dashed lines in (a) and (b) denote the actual locations on the suction side of the airfoil, which separate the $ab$, $cd$, $em$ and $mf$ zone.}
\label{fig:RANS_contour_Marker_CF_k.pdf}
\end{figure}

As noted earlier relatively few methods thus far have been developed to construct marker functions, e.g., see \cite{emory2013modeling} and \cite{gorle2014deviation}. Essentially, they can be classified into two categories: (1) spatially varying magnitude of $\Delta_{B}$ and (2) identifying regions that deviate from parallel shear flow. All of these methods essentially use only one explanatory variable to predict the error in RANS model predictions. In this study, a novel method based on least squares high-order regressions is developed to construct a switch marker function for $k^{*}$. This method uses a set of explanatory variables dedicated to the identified untrustworthy zones, which aims at introducing correct level of uncertainty by strictly comparing the RANS predictions for turbulence kinetic energy to the in-house DNS data.

We performed numerous tests and found that increasing the order of polynomial regressions higher than seven no longer gave more accurate results. Consequently, the seventh-order polynomial regression lines for the $k/U_{\infty}^2$ profiles were conducted, as shown in Fig. \ref{fig:avek_RANSDNS_xpls_0pt14To0pt60.pdf}. For each of the $ab$, $cd$ and $ef$ zone, the averaged regression relations for both RANS and in-house DNS are computed using the equation defined as follows: 

%\begin{equation} \label{Eqn:poly_fit}
%    p(x)=p_{1} x^{n}+p_{2} x^{n-1}+\ldots+p_{n} x+p_{n+1}
%\end{equation}    

\begin{equation}\label{Eqn:normk_fit}
    {k}_{RANS/DNS}^{ave}|_{zone \ ab/cd/ef} = \frac{\sum_{i=1}^{n} P_{i}(\frac{y}{c}|_{o})}{n}, 
\end{equation}

where $i$ represents the $i$th location on the suction side of the SD7003 airfoil (there are 32 selected locations), $P_{i}$ represents the polynomial regression at the $i$th location, and $n$ is the number of locations for each zone, as summarized in Table \ref{table:marker_ranges}. 

The regression based $k/U_{\infty}^2$ profiles for the $ab$, $cd$ and $ef$ zone are plotted in Figs. \ref{fig:avek_RANSDNS_xpls_0pt14To0pt60.pdf} (a), (b) and (c), respectively. The two solid lines with filled markers represent the mean of the regression based datasets for both RANS and in-house DNS. In each zone, Figs.  \ref{fig:avek_RANSDNS_xpls_0pt14To0pt60.pdf} (a), (b) and (c) clearly show the discrepancy between these two averaged regression relations. For the $ab$ zone, Fig. \ref{fig:avek_RANSDNS_xpls_0pt14To0pt60.pdf} (a) shows a small discrepancy close to zero at the wall, i.e. $y/c|_{o} = 0$, as well as in the far outer region, i.e. $y/c|_{o} > 0.025$. Besides, the discrepancy tends to increase with $y/c|_{o}$ and peaks around $y/c|_{o} = 0.015$. Within the $cd$ zone shown in Fig. \ref{fig:avek_RANSDNS_xpls_0pt14To0pt60.pdf} (b), there is a large discrepancy at the wall and the discrepancy retains at a nearly consistent level until peaks around $y/c|_{o} = 0.015$, then gradually decreases with $y/c|_{o}$ to approach the value of zero. It is interesting that the discrepancy peaks around $y/c|_{o} = 0.15$ for both the $ab$ and $cd$ zone (aft portion of the LSB). On the other hand, a relatively small discrepancy is observed consistently throughout the entire boundary layer for the $ef$ zone, as shown in Fig. \ref{fig:avek_RANSDNS_xpls_0pt14To0pt60.pdf} (c). This indicates that the RANS based transition model \cite{langtry2009correlation} tends to become more trustworthy in the predictions for the turbulence kinetic energy in the far downstream region than that within/close to the LSB, i.e. the $ab$ and $cd$ zone. From Figs. \ref{fig:avek_RANSDNS_xpls_0pt14To0pt60.pdf} (a), (b) and (c), the discrepancy between the averaged regression relations for RANS and in-house DNS describes the degree of untrustworthiness in the $y/c|_{o}$ direction ranging from the $ab$ zone to the $ef$ zone across the suction side, therefore the discrepancy can be used as an approximation to a marker function. This study defines a correction factor based on the $k/U_{\infty}^2$ discrepancy between the regression based RANS and in-house DNS, which can be written as follows:

\begin{equation}\label{Eqn:CF_k}
    CF_{k} = \left|  \frac{Averaged \ DNS}{Averaged \ SSTLM}  \right| .  
\end{equation}

Equation \ref{Eqn:CF_k} indicates that $CF_{k}$ is constantly positive, which satisfies the physical realizability constraint, i.e. $k^{*} \geq 0$. For each zone, the discrepancy data obtained using Eq. \ref{Eqn:CF_k} are depicted by the blue solid circles, as shown in Figs. \ref{fig:kcorrection factor_RANSDNS_three.pdf} (a), (b) and (c). Using MATLAB software, the marker function for each zone can be constructed by fitting to the corresponding $CF_{k}$ data, i.e., fitting a seventh-order polynomial to the discrepancy data for the $ab$ and $ef$ zone, while fitting a Fourier series to the discrepancy data for the $cd$ zone, as shown in Figs. \ref{fig:kcorrection factor_RANSDNS_three.pdf} (a), (c), and (b), respectively.

Therefore, a switch marker function that introduces local injection of perturbation with respect to the $ab$, $cd$ and $ef$ zone can be written as follows:

%\begin{widetext}
\begin{equation}\label{Eqn:Markerfunc}
    \text {Switch $M_{k}$ }= \begin{cases}\textcolor{red}{a_{0}^{ab}(\frac{y-y_{w}^{ab}}{c})^{7} + a_{1}^{ab}(\frac{y-y_{w}^{ab}}{c})^{6} +...+}\\ \textcolor{red}{a_{5}^{ab}(\frac{y-y_{w}^{ab}}{c})^{2}+a_{6}^{ab}(\frac{y-y_{w}^{ab}}{c}) +a_{7}^{ab}} \quad \textcolor{red}{\text { if }} \textcolor{red}{zone \ ab,} \\ \\

    \textcolor{Green}{a_{0}^{cd}+a_{1}^{cd} \cos \left(w\left(\frac{y-y_{w}^{cd}}{c}\right)\right) +} \\ \textcolor{Green}{b_{1}^{cd} \sin \left(w\left(\frac{y-y_{w}^{cd}}{c}\right)\right)+}\\
    \textcolor{Green}{+ a_{2}^{cd} \cos \left(\left(2w\left(\frac{y-y_{w}^{cd}}{c}\right)\right)\right)+} \quad \textcolor{Green}{\text { if } zone \ cd,} 
    \\ \textcolor{Green}{b_{2}^{cd} \sin \left(\left(2w\left(\frac{y-y_{w}^{cd}}{c}\right)\right)\right)} \\ \\
    
    \textcolor{blue}{a_{0}^{em}(\frac{y-y_{w}^{em}}{c})^{7}+a_{1}^{em}(\frac{y-y_{w}^{em}}{c})^{6}+...+}\\ \textcolor{blue}{a_{5}^{em}(\frac{y-y_{w}^{em}}{c})^{2}+a_{6}^{em}(\frac{y-y_{w}^{em}}{c})+a_{7}^{em}} \quad \textcolor{blue}{\text { if }} \textcolor{blue}{ zone \ em ,}\\ \\
    
    \textcolor{gray}{2.8} \qquad \qquad \textcolor{gray}{\text { if } zone \ mf ,} \end{cases}
\end{equation}
%\end{widetext}

\noindent where a0, a1, a2, a3, a4, a5, a6, a7 represent the polynomial coefficients; a0, a1, b1, a2, b2, w represent the Fourier coefficients. Therefore, the perturbed turbulence kinetic energy, $k^{*}$, is defined as follows:

\begin{equation}\label{Eqn:kstar}
    k^{*} = kM_{k}.  
\end{equation}

It is worth noting that the development of spatial variations in $M_{k}$, are what the turbulence machine learning efforts are focused on. Because when a neural network model is developed to predict the perturbation in the flow, this neural network model will not predict the same perturbation at all points in the flow domain. Instead, it will naturally lead to a non-uniform perturbation. The key differences between my work and the work based on machine learning is two pronged: (1) the choice of the model and (2) the choice of the modeling basis (or the explanatory variables utilized to predict the perturbation). We have used a seventh-order regression, the work based on machine learning uses a random forest or neural network. We have utilized a small set of explanatory variables in $M_{k}$ that is developed based on physics arguments and prior experience. The work based on machine learning utilize a large set of explanatory variables (called features) that is almost $100$ in number and includes invariants of the mean velocity field, scaled distance from the wall.

If a uniform value of $M_{k}$ is used, then Eq. \ref{Eqn:kstar} becomes

\begin{equation}\label{Eqn:Deltak}
    k^{*} = k\Delta_{k},  
\end{equation}

\noindent where $k$ is the perturbed turbulence kinetic energy from the previous time step, and $\Delta_{k}$ represents a uniform value of $M_{k}$. The value of $\Delta_{k}$ must be larger than zero to satisfy physical realizability.  Due to airfoil curved surfaces, $y_{w}$ varies along the suction side. If let $y_{w} = f(x)$ represent the curved upper surface, then its gradient can be calculated by taking the derivative of $f(x)$, e.g., $df(x)/dx$. In Eq. \ref{Eqn:Markerfunc}, the strategy to choose a reasonable magnitude for $y_{w}$ as representative of a zone is to find the minimum value of $y_{w}$, which ensures that the realizability constraint of $M_{k} \geq 0$ is satisfied, and hence $k^{*} \geq 0$. Note that the value of $df(x)/dx$ approaches to the value of zero around $x/c = 0.266$, i.e. sitting within the $cd$ zone. This implies that the minimum value of $y_{w}$ is located closer to the leading edge for $x/c < 0.266$, while the minimum value of $y_{w}$ is located closer to the trailing edge for $x/c > 0.266$.

As noted earlier in Fig. \ref{fig:BL_black}, the $ef$ zone is composed of two subzones, i.e. $em$ and $mf$, as illustrated in Fig. \ref{fig:RANSDNS_xpls_0pt3To0pt60_withIndicator.pdf}. Figure \ref{fig:RANSDNS_xpls_0pt3To0pt60_withIndicator.pdf} enlarges the $ef$ zone, in which the regression based $k/U_{\infty}^2$ profiles for both RANS and in-house DNS are shifted down to the origin of $y$, to highlight the discrepancy in the region for $0.3 < x/c < 0.6$, within which the profiles for the $em$ subzone are painted blue, while the profiles are painted gray for the $mf$ subzone. It is clear that a similar level of discrepancy between the regression based RANS and in-house DNS profile across the $mf$ subzone is observed, as shown in Fig. \ref{fig:RANSDNS_xpls_0pt3To0pt60_withIndicator.pdf}. From Fig. \ref{fig:RANSDNS_xpls_0pt3To0pt60_withIndicator.pdf}, the discrepancy is significant in the vicinity of the wall, i.e at $y/c|_{o} = 0.004$, which corresponds to an approximate value of $2.8$ for $CF_{k}$ in Fig. \ref{fig:kcorrection factor_RANSDNS_three.pdf} (c). For sake of simplicity, the uniform value of $2.8$ for $M_{k}$ is employed for the $mf$ subzone in Eq. \ref{Eqn:Markerfunc}.

We visualize the spatial variation of the magnitude of $M_{k}$ from the contours of $0 < M_{k} < 1$ and $1 < M_{k} < 10$, as shown in Figs. \ref{fig:RANS_contour_Marker_CF_k.pdf} (a) and (b), respectively. From Fig. \ref{fig:RANS_contour_Marker_CF_k.pdf} (a), it is clear that the magnitude of $0 < M_{k} < 1$ is more prevalent in the $ab$ zone, and in the upper portion of the $cd$ zone. In Fig. \ref{fig:RANS_contour_Marker_CF_k.pdf} (a), an overall decreasing trend of $M_{k}$ in magnitude with $y/c$ is observed for the $ab$ zone. On the other hand, the $M_{k}$ magnitude for both the $cd$ and $em$ zone varies with $y/c$ in a fashion consistent with the behavior observed in Figs. \ref{fig:kcorrection factor_RANSDNS_three.pdf} (b) and (c). Moreover, a uniform magnitude of $M_{k}$ is observed for the $ef$ zone, which confirms the uniform magnitude of $2.8$.

\section{Results and discussion}
%Analyzing sensitivity to uniform $M_{k}$, namely, $\Delta_{k}$, is a valid way to gain understanding of the marker effects on various QoIs. This section presents the results subject to the $\Delta_{k}$ perturbations, and explores the propagation effect on different QoIs.

\subsection{Sensitivity to $\Delta_k$}
\subsubsection{Skin friction coefficient}
\begin{figure} 
\centerline{\includegraphics[width=3.5in]{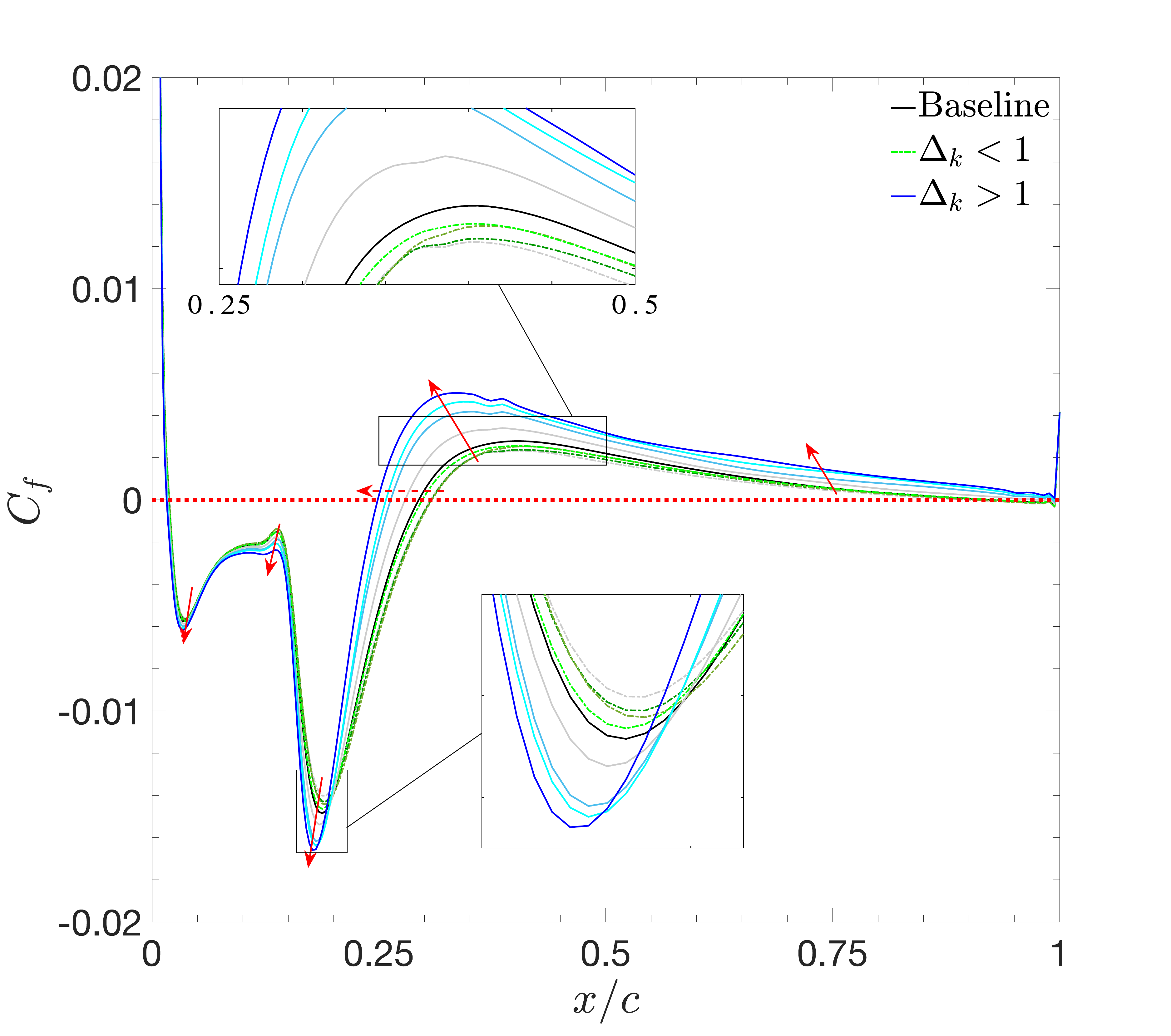}}
\caption[Skin friction coefficient distributions over the suction side of the airfoil with enlarged regions at the trough and the crest. Displayed are $k^{*}$ perturbations with uniform $\Delta_{k}$: $\Delta_{k} < 1$ ($\Delta_{k} = \big\{ 0.1, 0.25, 0.5, 0.75 \big\}$) and $\Delta_{k} > 1$ ($\Delta_{k} = \big\{ 2, 4, 6, 8 \big\}$).]{Skin friction coefficient distributions over the suction side of the airfoil with enlarged regions at the trough and the crest. Displayed are $k^{*}$ perturbations with uniform $\Delta_{k}$: $\Delta_{k} < 1$ ($\Delta_{k} = \big\{ 0.1, 0.25, 0.5, 0.75 \big\}$) and $\Delta_{k} > 1$ ($\Delta_{k} = \big\{ 2, 4, 6, 8 \big\}$); increasing values indicated by lighter to darker hues. Red solid arrows ($\color{red} \longrightarrow$) are provided to indicate increasing magnitude of $C_{f}$ with $\Delta_{k}$; the red dashed arrow ($\color{red} \dashrightarrow$) is provided to indicate the shift in reattachment point with $\Delta_{k}$. Note that the value of $\Delta_{k}$ must be larger than zero to satisfy realizability. The baseline prediction is provided for reference.}
\label{fig:cf_uniformk_line.pdf}
\end{figure}

A set of $C_{f}$ distributions undergoing the $\Delta_{k}$ perturbations are shown in Fig. \ref{fig:cf_uniformk_line.pdf}. The baseline prediction for $C_{f}$ is used as a reference. The increasing magnitude of $\Delta_{k}$ is indicated by lighter to darker hues, as shown in Fig. \ref{fig:cf_uniformk_line.pdf}. In addition, the red solid arrows are added to indicate the trend of $C_{f}$ with increasing $\Delta_{k}$, and the regions that contain a peak negative (trough) and a peak positive (crest) value of $C_{f}$ are enlarged to distinguish the clusters of $C_{f}$ profiles. In Fig. \ref{fig:cf_uniformk_line.pdf}, the magnitude of $C_{f}$ profiles increases with $\Delta_{k}$ perturbations for $\Delta_{k} < 1$ ($\Delta_{k} = \big\{ 0.1, 0.25, 0.5, 0.75 \big\}$) and $\Delta_{k} > 1$ ($\Delta_{k} = \big\{ 2, 4, 6, 8 \big\}$), respectively, at the trough (around $X_{T}$), indicating a monotonic increase. As the flow moves further downstream within the aft portion of the LSB, the magnitude of $C_{f}$ tends to decrease monotonically when the value of $\Delta_{k}$ is increased; as the flow proceeds further downstream of $X_{R}$, a monotonic increase with $\Delta_{k}$ in the magnitude of $C_{f}$ again occurred for $\Delta_{k} < 1$ and $\Delta_{k} > 1$. It should be noted that the $C_{f}$ profiles tend to converge and collapse onto a single curve when $\Delta_{k}$ is increased. The baseline prediction is well enveloped in between the $\Delta_{k} < 1$ and $\Delta_{k} > 1$ perturbations. Compared to the baseline prediction, rather subtle increases in the magnitude of $C_{f}$ for $\Delta_{k} < 1$ is observed, as contrasted with more noticeable increases in $C_{f}$ for $\Delta_{k} > 1$. This indicates that the simulation's response to the injection of $\Delta_{k}$ is more dependent on $\Delta_{k} > 1$ than $\Delta_{k} < 1$. This behavior is highlighted in the enlarged trough and crest region. In addition, the dashed red arrow is added along the line of zeros of $C_{f}$ to indicate the tendency of a shift of $X_{R}$ in the upstream direction when the value of $\Delta_{k}$ is increased for both $\Delta_{k} < 1$ and $\Delta_{k} > 1$. 

Since wall shear stress is a consequence of momentum transfer from the mean flow to the wall surface \cite{monteith2013principles}, the magnitude of mean velocity is closely related to the magnitude of $C_{f}$. In the aft portion of the LSB, the $\Delta_{k} > 1$ perturbations overall yield a smaller magnitude of $C_{f}$, and an increase in the magnitude of mean velocity is expected, while the opposite is true for the $\Delta_{k} < 1$ perturbations.  

\subsubsection{Mean velocity field}
% 0.1, 0.25, 0.5， 4， 6， 8， Baseline
\begin{figure*} 
\centerline{\includegraphics[width=5.5in]{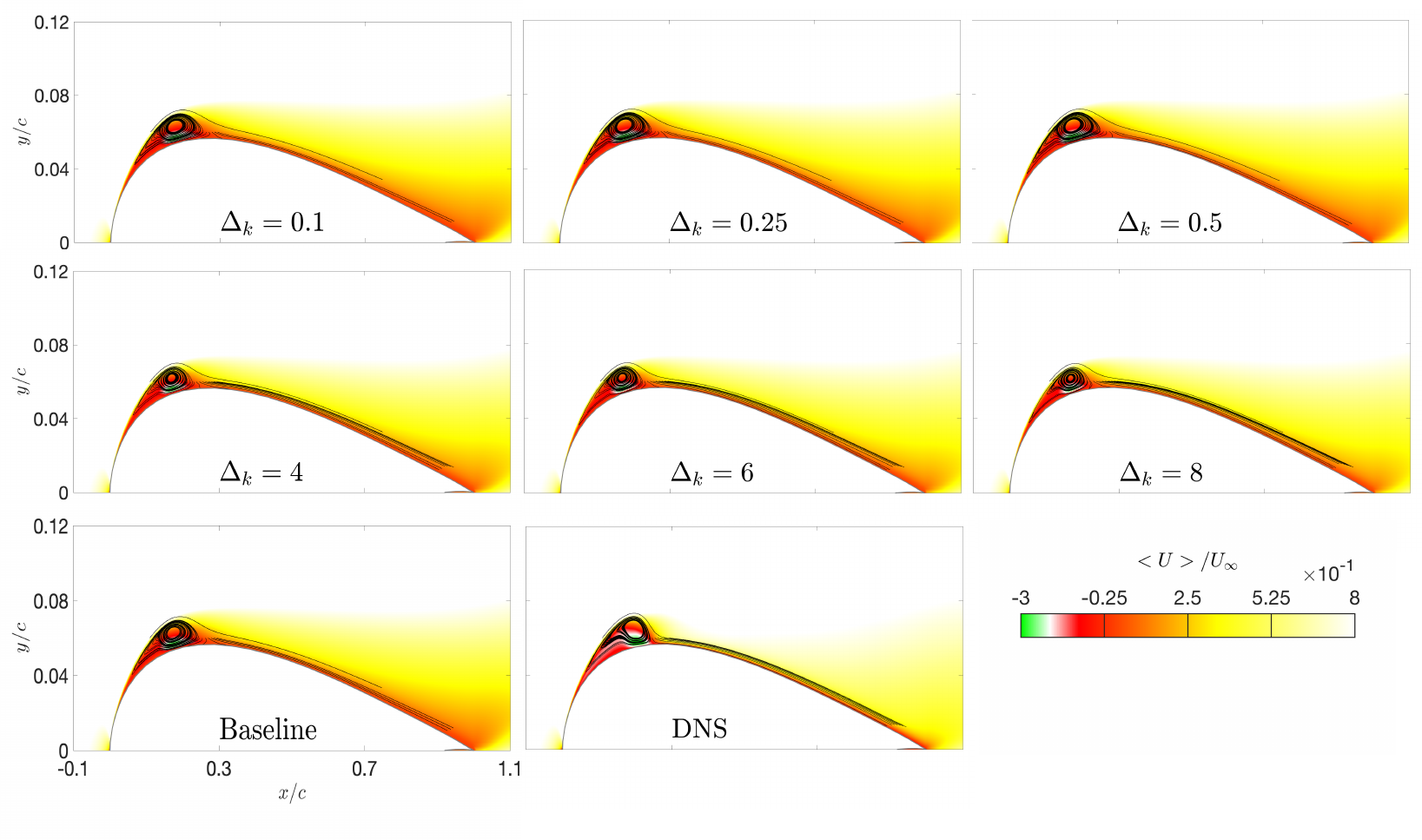}}
\caption[Contours of $\left\langle U \right \rangle/U_{\infty}$ with different values of $\Delta_{k}$: $\Delta_{k} < 1$ ($\Delta_{k} = \big\{ 0.1, 0.25, 0.5 \big\}$) and $\Delta_{k} > 1$ ($\Delta_{k} = \big\{4, 6, 8 \big\}$) in an $xy$ plane.]{Contours of $\left\langle U \right \rangle/U_{\infty}$ with different values of $\Delta_{k}$: $\Delta_{k} < 1$ ($\Delta_{k} = \big\{ 0.1, 0.25, 0.5 \big\}$) and $\Delta_{k} > 1$ ($\Delta_{k} = \big\{4, 6, 8 \big\}$) in an $xy$ plane. Baseline prediction is provided for reference, and in-house DNS data are included for comparison. Streamlines show the size of the LSB on the suction side of the airfoil.}
\label{fig:RANS_contour_streamline_normU_All_uniformk_subplot_foucs.pdf}
\end{figure*}

Contours of the mean velocity normalized with the free stream velocity, $\left\langle U \right \rangle/U_{\infty}$ from the baseline, $\Delta_{k}$ perturbations, and in-house DNS of \cite{zhang2021turbulent} in an $xy$ plane are shown in Fig. \ref{fig:RANS_contour_streamline_normU_All_uniformk_subplot_foucs.pdf}. The streamlines for depicting a large recirculation vortex within the LSB, characterized by the region of reverse flow ($\left\langle U \right \rangle/U_{\infty} < 0$) \cite{rist2002numerical}, are included as well. This large recirculating region contains large-scale events (coherent structures), which are at low-frequency fluctuations due to very-large scale of unsteadiness of the recirculating region itself \cite{kiya1985structure}. As a consequence, the $\left\langle U \right \rangle/U_{\infty}$ contours exhibit a LSB surviving after time-averaging, as shown in Fig. \ref{fig:RANS_contour_streamline_normU_All_uniformk_subplot_foucs.pdf}. This behavior has been observed in the experimental measurements of \cite{zhang2018lagrangian},  RANS analysis of \cite{catalano2011rans}, and ILES/LES data of \cite{galbraith2010implicit} and \cite{garmann2013comparative}. Figure \ref{fig:RANS_contour_streamline_normU_All_uniformk_subplot_foucs.pdf} clearly shows that the baseline prediction for the LSB shows a comparable length to in-house DNS; however, the LSB's height is under-predicted. This inaccurate  prediction for the LSB height alters the effective shape of the airfoil, hence inaccuracy in simulation results \cite{gaster1967structure,spalart2000mechanisms}. This reflects the error in RANS model predictions in the region of the LSB. Compared to the baseline prediction, rather subtle responses to the $\Delta_{k} < 1$ perturbations ($\Delta_{k} = 0.1, 0.25, 0.5$) are observed, which confirms the behavior shown in Fig. \ref{fig:cf_uniformk_line.pdf}. On the other hand, more noticeable changes are observed with the $\Delta_{k} > 1$ perturbations ($\Delta_{k} = 4, 6, 8$), i.e., a clear suppression of the LSB length; in addition, it is clear that the magnitude of mean velocity increases downstream of the LSB within the attached turbulent boundary layer, characterized by the more clustered streamlines compared to the baseline prediction. This confirms the reduction in the magnitude of $C_{f}$ in the aft portion of the LSB, as shown in Fig. \ref{fig:cf_uniformk_line.pdf}. There are two monotonic behaviors: first, the size of the recirculating region deceases monotonically with $\Delta_{k}$ (shallower region of streamlines), showing a tendency of deviating from  the in-house DNS contour; second, the magnitude of $\left\langle U \right \rangle/U_{\infty}$ monotonically increases  with $\Delta_{k}$ in the attached turbulent boundary layer (more densely clustered streamlines), showing a tendency of approaching closer to the in-house DNS contour.

\subsubsection{Reynolds shear stress}
\begin{figure*} 
\centerline{\includegraphics[width=5.5in]{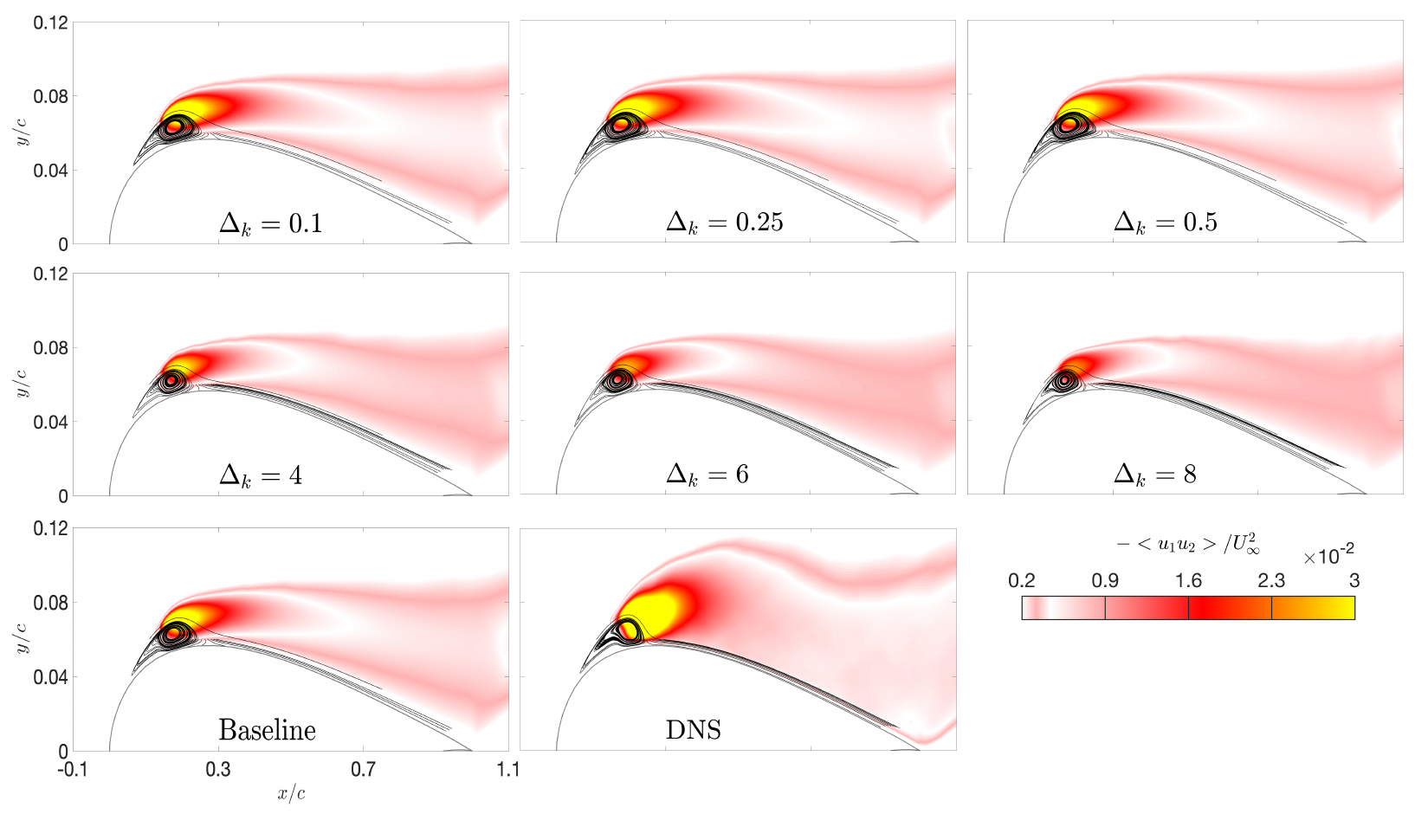}}
\caption[Contours of $-\left\langle u_{1}u_{2} \right \rangle/U_{\infty}^2$ with different values of $\Delta_{k}$: $\Delta_{k} < 1$ ($\Delta_{k} = \big\{ 0.1, 0.25, 0.5 \big\}$) and $\Delta_{k} > 1$ ($\Delta_{k} = \big\{4, 6, 8 \big\}$) in an $xy$ plane.]{Contours of $-\left\langle u_{1}u_{2} \right \rangle/U_{\infty}^2$ with different values of $\Delta_{k}$: $\Delta_{k} < 1$ ($\Delta_{k} = \big\{ 0.1, 0.25, 0.5 \big\}$) and $\Delta_{k} > 1$ ($\Delta_{k} = \big\{4, 6, 8 \big\}$) in an $xy$ plane. Baseline prediction is provided for reference, and in-house DNS data are included for comparison. Streamlines show the size of the LSB on the suction side of the airfoil.}
\label{fig:RANS_contour_streamline_normuv_All_uniformk_subplot_foucs.pdf}
\end{figure*}

Contours of the Reynolds shear stress normalized with the freestream velocity squared, $-\left\langle u_{1}u_{2} \right \rangle/U_{\infty}^2$ from the baseline, $\Delta_{k}$ perturbations, and in-house DNS of \cite{zhang2021turbulent} in an $xy$ plane are presented in Fig. \ref{fig:RANS_contour_streamline_normuv_All_uniformk_subplot_foucs.pdf}. Also included are the streamlines for the depiction of the recirculation vortex region. From Fig. \ref{fig:RANS_contour_streamline_normuv_All_uniformk_subplot_foucs.pdf}, all of the $-\left\langle u_{1}u_{2} \right \rangle/U_{\infty}^2$ contour plots show a magnitude of nearly zero in the region near the leading edge and in the outer region of the flow, and a peak is found within the LSB around $X_{T}$, i.e., the bright yellow region, from which the magnitude of $-\left\langle u_{1}u_{2} \right \rangle/U_{\infty}^2$ reduces as the flow moves further downstream. A similar behavior was also observed by Zhang and Rival \cite{zhang2018lagrangian} in their experimental measurements. Overall, the baseline prediction for Reynolds shear stress gives a smaller value than the in-house DNS data, especially in the LSB. It should be noted that the contour plots of $-\left\langle u_{1}u_{2} \right \rangle/U_{\infty}^2$ and $\left\langle U \right \rangle/U_{\infty}$ show a similar trend: a lack of sensitivity to the $\Delta_{k} < 1$ ($\Delta_{k} = 0.1, 0.25, 0.5$) perturbations, while a rather strong sensitivity to the $\Delta_{k} > 1$ ($\Delta_{k} = 4, 6, 8$) perturbations in both the transitional and turbulent region. In general, the Reynolds shear stress contour exhibits a larger response to $\Delta_{k}$ compared to the mean velocity contour. From Fig. \ref{fig:RANS_contour_streamline_normuv_All_uniformk_subplot_foucs.pdf}, the $\Delta_{k} < 1$ perturbations give a somewhat larger value of $-\left\langle u_{1}u_{2} \right \rangle/U_{\infty}^2$ than the baseline prediction, while the $\Delta_{k} > 1$ perturbations do the opposite. In addition, the $\Delta_{k} < 1$ perturbations slightly reduce the magnitude of $-\left\langle u_{1}u_{2} \right \rangle/U_{\infty}^2$ when the value of $\Delta_{k}$ is increased, as opposed to the $\Delta_{k} > 1$ perturbations, which greatly reduces the magnitude of $-\left\langle u_{1}u_{2} \right \rangle/U_{\infty}^2$. In addition, it is clear that the peak value for $-\left\langle u_{1}u_{2} \right \rangle/U_{\infty}^2$ gradually becomes smaller as the value of $\Delta_{k}$ is increased, in particular for $\Delta_{k} > 1$. This is accompanied with a suppression of the recirculating region and hence a decrease in the turbulence kinetic energy \cite{lengani2014pod}.  In Fig. \ref{fig:RANS_contour_streamline_normuv_All_uniformk_subplot_foucs.pdf}, the $\Delta_{k} > 1$ perturbations tend to approach closer to the in-house DNS data in the turbulent boundary layer, while the $\Delta_{k} < 1$ perturbations tend to result in a closer agreement with the in-house DNS data in the LSB. According to Davide \textit{et al.} \cite{lengani2014pod}, the overall turbulence kinetic energy can be decomposed into the large-scale coherent (Kelvin-Helmholtz induced) and stochastic (turbulence-induced) contributions. With the total energy in the mean flow remained constant, the $\Delta_{k}$ perturbations in a sense redistribute the Reynolds-shear-stress momentum transfer between turbulence and mean flow. 

%Recall that the Reynolds shear stress is dedicated to one of the contributions to the turbulent production \cite{pope2001turbulent}, e.g., i.e., $-\left\langle u_{1} u_{2}\right\rangle\left({\partial \left\langle U \right\rangle}/{\partial y}+{\partial \left\langle V \right\rangle}/{\partial x}\right)$, which, accordingly, causes an impact on the turbulence kinetic energy\cite{pope2001turbulent,durbin2011statistical}. We note that the turbulence kinetic energy contours showed similar observations to the Reynolds shear stress, therefore omitted for brevity. 

%uv 0.1, 0.25, 0.5， 4， 6， 8， Baseline

\subsection{Comparison between uniform $\Delta_{k}$ and $M_{k}$}

\subsubsection{Skin friction coefficient and pressure coefficient}
\begin{figure} 
\centerline{\includegraphics[width=3.5in]{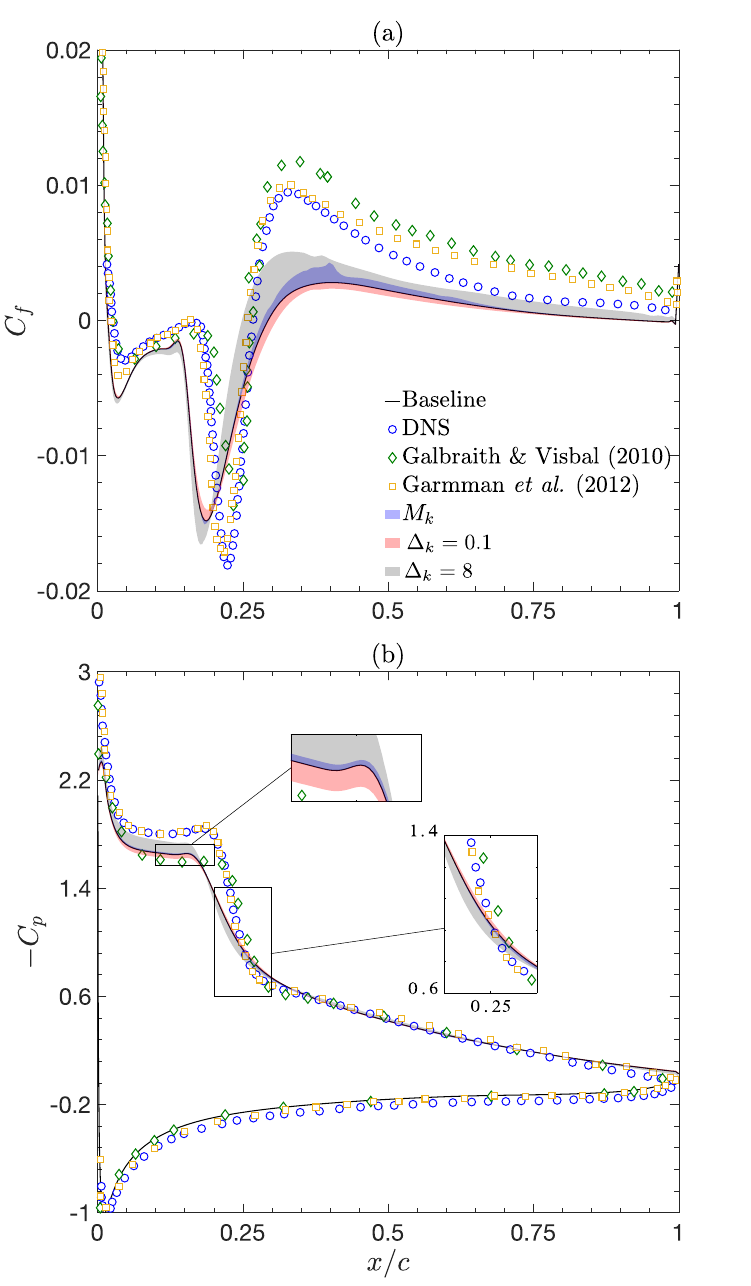}}
\caption[(a) Profile of skin friction coefficient and (b) pressure coefficient with enlarged regions at the flat spot and the kink.]{(a) Profile of skin friction coefficient and (b) pressure coefficient with enlarged regions at the flat spot and the kink. Displayed are envelopes for uniform $k^{*}$ perturbations: $\Delta_{k} = 0.1$ (red envelope), $\Delta_{k} = 8$ (gray envelope), and $M_{k}$ (blue envelope). The baseline prediction is provided for reference. $\circ$ in-house DNS data \cite{zhang2021turbulent}.}
\label{fig:cfcp_uniformkVsOnlyMarker.pdf}
\end{figure}

Distributions of the skin friction coefficient and the pressure coefficient, $C_{f}$ and $C_{p}$, are shown in Figs. \ref{fig:cfcp_uniformkVsOnlyMarker.pdf} (a) and (b). The in-house DNS \cite{zhang2021turbulent} and ILES/LES data of \cite{galbraith2010implicit} and \cite{garmann2013comparative} are included for comparison. In Figs. \ref{fig:cfcp_uniformkVsOnlyMarker.pdf} (a) and (b), an enveloping behavior with respect to the baseline prediction is observed. Figure \ref{fig:cfcp_uniformkVsOnlyMarker.pdf} (a) shows that the effect of $M_{k}$ is more prevalent in the aft portion of the LSB $0.25 < x/c < 0.28$, as well as in the region downstream of the LSB $0.28 < x/c < 0.6$.  This reflects the effect of spatial variability in $M_{k}$. In addition, the uncertainty bound generated from the $M_{k}$ perturbation sits within the gray envelope of $\Delta_{k} = 8$. Figure \ref{fig:cfcp_uniformkVsOnlyMarker.pdf} (a) clearly shows that the uncertainty bound generated from the $M_{k}$ perturbation is well encompassed by the uniform $\Delta_{k} = 0.1$ and $\Delta_{k} = 8$ perturbations. It is interesting to note that the $\Delta_{k} = 8$ perturbation overall tends to approach closer to the in-house DNS \cite{zhang2021turbulent} and ILES/LES data of \cite{galbraith2010implicit} and \cite{garmann2013comparative} than the $\Delta_{k} = 0.1$ perturbation does. At the trough shown in Fig. \ref{fig:cfcp_uniformkVsOnlyMarker.pdf} (a), the $\Delta_{k} = 8$ perturbation gives a larger magnitude of $C_{f}$, sitting below the baseline prediction and showing a clear tendency to approach closer to the in-house DNS \cite{zhang2021turbulent} and LES \cite{garmann2013comparative} data. In addition, Fig. \ref{fig:cfcp_uniformkVsOnlyMarker.pdf} (a) clearly shows that the reattachment point is well encompassed by the $\Delta_{k} = 8$ perturbation. Further downstream of reattachment point, the uncertainty bound generated from both the $\Delta_{k} = 8$ and $M_{k}$ perturbations show a tendency to approach closer to the in-house DNS \cite{zhang2021turbulent} and the ILES/LES data of \cite{galbraith2010implicit} and \cite{garmann2013comparative}, while the $\Delta_{k} = 0.1$ perturbation under-predicts the baseline prediction and deviates from the reference data. 

At the flat spot and the kink ($X_{R}$) followed by a steep drop on the $C_{p}$ profile, the uncertainty bound generated from the $M_{k}$ perturbation is encompassed by the $\Delta_{k} = 8$ perturbation, as shown in the enlarged regions in Fig \ref{fig:cfcp_uniformkVsOnlyMarker.pdf} (b). In addition, a tendency for the $\Delta_{k} = 8$ and $M_{k}$ perturbations to approach closer to the in-house DNS \cite{zhang2021turbulent} and LES data of \cite{garmann2013comparative} is observed at the flat spot and the kink. On the other hand, it is interesting that the $\Delta_{k} = 0.1$ perturbation shows a tendency of approaching toward the ILES data of \cite{galbraith2010implicit} in the enlarged regions shown in Fig. \ref{fig:cfcp_uniformkVsOnlyMarker.pdf} (b). In the region downstream of the kink and along the entire pressure side, both $\Delta_{k}=0.1$ and $\Delta_{k}=8$ perturbations are almost negligible in magnitude, i.e., a collapse onto the baseline prediction. It should be noted that the baseline prediction overall shows good agreement with the in-house DNS data \cite{zhang2021turbulent} and ILES/LES data of \cite{galbraith2010implicit} and \cite{garmann2013comparative}, especially good agreement with the ILES/LES data of \cite{galbraith2010implicit} and \cite{garmann2013comparative} for the pressure side. This indicates a low level of the model form uncertainty in the predictions for $C_{p}$ for these regions.

\subsubsection{Mean velocity field}
\begin{figure} 
\centerline{\includegraphics[width=4.0in]{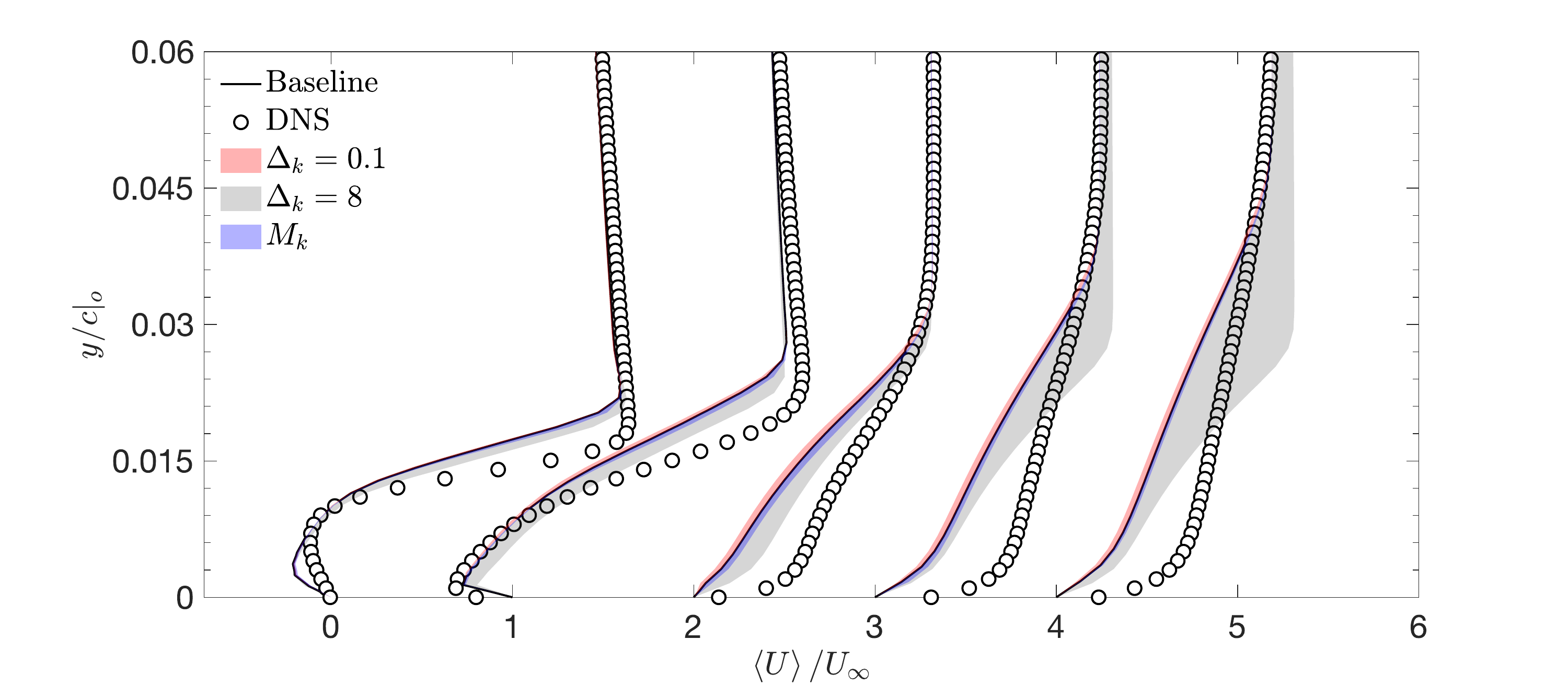}}
\caption[Streamwise mean velocity profiles in the aft portion of the LSB ($x/c = 0.15$ and $0.2$) and in the attached turbulent boundary layer ($x/c = 0.3, 0.4$ and $0.5$).]{Streamwise mean velocity profiles in the aft portion of the LSB ($x/c = 0.15$ and $0.2$) and in the attached TBL ($x/c = 0.3, 0.4$ and $0.5$). From left to right are $x/c = 0.15, 0.2, 0.3, 0.4$ and $0.5$, respectively. Displayed are envelopes for two extreme $\Delta_{k}$ perturbations considered in this study: $\Delta_{k} = 0.1$ (red envelope),  $\Delta_{k} = 8$ (gray envelope), and $M_{k}$ (blue envelope). The baseline prediction is provided for reference. $\circ$ in-house DNS data \cite{zhang2021turbulent}.}
\label{fig:UQ_uniformkvsOnlyMarker_U_RepeatOn_Tu0027.pdf}
\end{figure}

The $\left\langle U \right \rangle/U_{\infty}$ profiles across the entire boundary layer on the suction side of the airfoil are plotted in Fig. \ref{fig:UQ_uniformkvsOnlyMarker_U_RepeatOn_Tu0027.pdf}. Overall, the baseline prediction at each location is encompassed by the $\Delta_{k} = 0.1$ and $\Delta_{k} = 8$ perturbations, exhibiting an enveloping behavior, as shown in Fig. \ref{fig:UQ_uniformkvsOnlyMarker_U_RepeatOn_Tu0027.pdf}. The baseline prediction for the $\left\langle U \right \rangle/U_{\infty}$ profile at $x/c = 0.15$ ($X_{T}$) matches the in-house DNS profile of \cite{zhang2021turbulent}, except in the regions $y/c|_{o} < 0.007$ (next to the wall) and $y/c|_{o} > 0.011$ (upper portion of the boundary layer), where it gives slightly smaller values of $\left\langle U \right \rangle/U_{\infty}$, as shown in Fig. \ref{fig:UQ_uniformkvsOnlyMarker_U_RepeatOn_Tu0027.pdf}. At $x/c = 0.2$ (in the aft portion of the LSB), the baseline prediction for the $\left\langle U \right \rangle/U_{\infty}$ profile shows good agreement with the in-house DNS profile in the region of reverse flow $y/c|_{o} < 0.011$, with a somewhat reduction in the predicted $\left\langle U \right \rangle/U_{\infty}$ profile in the upper portion of the boundary layer $0.011 < y/c|_{o} < 0.027$. For the attached turbulent boundary layer, the baseline predictions for the $\left\langle U \right \rangle/U_{\infty}$ profiles at $x/c = 0.3$, $x/c = 0.4$ and $x/c = 0.5$ give smaller values of $\left\langle U \right \rangle/U_{\infty}$ compared to the in-house DNS profiles, and the discrepancies are comparable with each other.

Figure \ref{fig:UQ_uniformkvsOnlyMarker_U_RepeatOn_Tu0027.pdf} shows that the $\Delta_{k} = 0.1$ perturbation under-predicts the baseline prediction, and the simulation's response to the $\Delta_{k} = 0.1$ perturbation is negligibly small within both the transitional and turbulent boundary layer. This well confirms the behavior of the $\Delta_{k} = 0.1$ perturbation in the prediction for $C_{f}$, as shown in Fig. \ref{fig:cfcp_uniformkVsOnlyMarker.pdf} (a). As the flow proceeds downstream from $x/c = 0.15$ to $x/c = 0.3$, the uncertainty bounds generated from the $\Delta_{k} = 0.1$ perturbations gradually increase in size, although the increase is rather subtle. This confirms the slightly increased $C_{f}$ in magnitude compared to the baseline prediction for the aft portion of the LSB. As the flow moves further downstream from $x/c = 0.4$ to $x/c = 0.5$ (in the attached turbulent boundary layer), the $\Delta_{k} = 0.1$ perturbation gradually reduces the size of the uncertainty bounds at a decreasing rate, reflecting the damping effect of the positive values of $C_{f}$ on the mean flow. 

On the other hand, the $\Delta_{k} = 8$ perturbation over-predicts the baseline prediction, exhibiting rather noticeable uncertainty bounds, as shown in Fig. \ref{fig:UQ_uniformkvsOnlyMarker_U_RepeatOn_Tu0027.pdf}. As the flow proceeds from $x/c = 0.15$ to $x/c = 0.3$, it is interesting to note that the uncertainty bounds generated from the $\Delta_{k} = 8$ perturbations increase blatantly in size, showing a tendency of approaching closer to the in-house DNS data. In addition, the effect of the $\Delta_{k} = 8$ perturbation tends to become more prevalent in the near-wall region, which well confirms the significantly reduced $C_{f}$ in magnitude compared to the baseline prediction for $0.15 < x/c < 0.3$, as shown in Fig. \ref{fig:cfcp_uniformkVsOnlyMarker.pdf} (a). As the flow proceeds further downstream from $x/c = 0.4$ to $x/c = 0.5$, the uncertainty bounds become larger in the upper section of the mean velocity profiles, while remain at a relatively small magnitude in the near-wall region due to the large positive values of $C_{f}$ at the crest shown in Fig. \ref{fig:cfcp_uniformkVsOnlyMarker.pdf} (a), reflecting the weakening propagation of the effect of the positive $C_{f}$ values deeper into the outer boundary layer. 

Unlike the $\Delta_{k} = 0.1$ and $\Delta_{k} = 8$ perturbation, $M_{k}$ identifies the untrustworthy regions in which uncertainty will be injected. In Fig. \ref{fig:UQ_uniformkvsOnlyMarker_U_RepeatOn_Tu0027.pdf}, the uncertainty bounds generated from the $M_{k}$ perturbations in general over-predict the baseline prediction, and sit within the uncertainty bounds generated from the $\Delta_{k} = 8$ perturbations. It should be noted that the sole effect of the $M_{k}$ perturbation on the predicted mean velocity profile is rather small. In section \ref{Sec:compound}, the $M_{k}$ perturbation is compounded with the eigenvalue perturbation ($1c$, $2c$, $3c$) to construct more effective uncertainty bounds.

\subsubsection{Reynolds shear stress}
\begin{figure} 
\centerline{\includegraphics[width=3.7in]{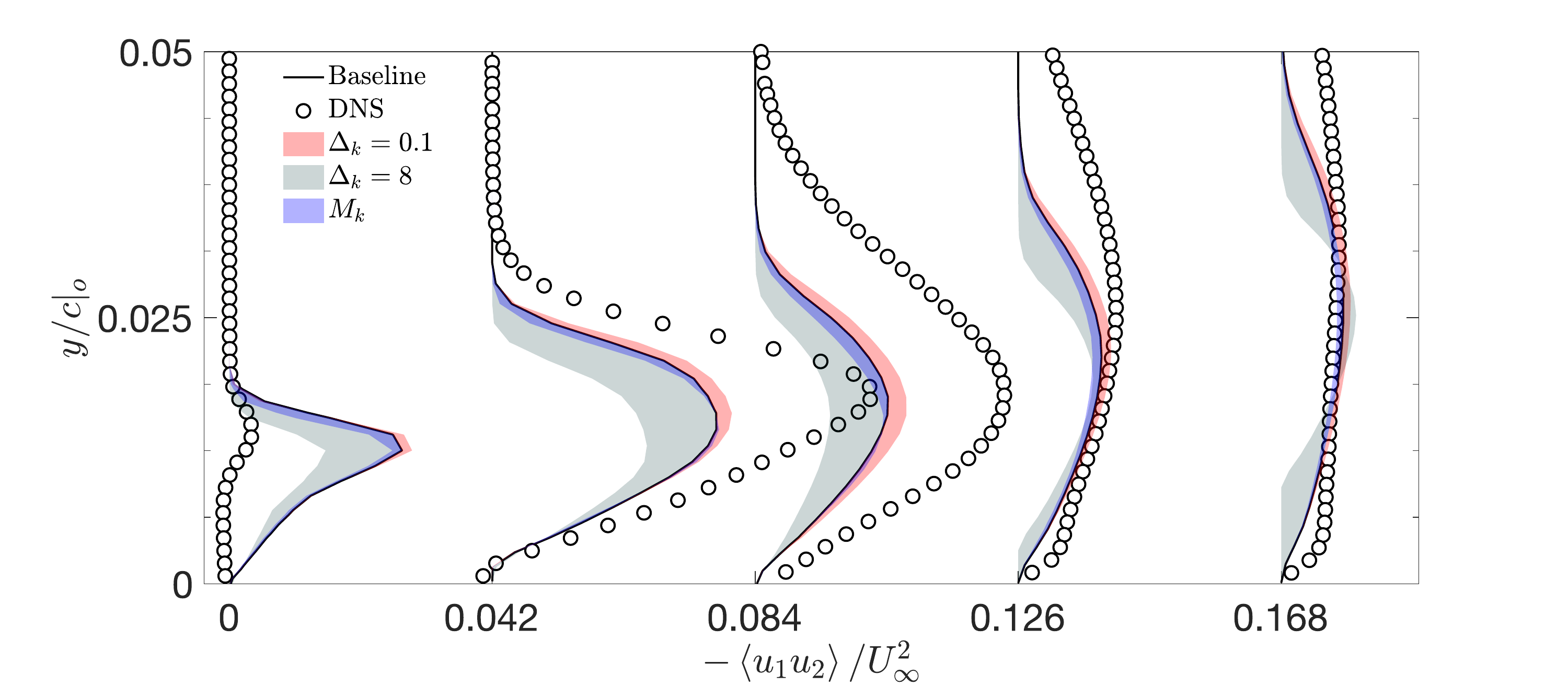}}
\caption[Reynolds shear stress profiles in the aft portion of the LSB ($x/c = 0.15$ and $0.2$) and in the attached turbulent boundary layer ($x/c = 0.3, 0.4$ and $0.5$).]{Reynolds shear stress profiles in the aft portion of the LSB ($x/c = 0.15$ and $0.2$) and in the attached TBL ($x/c = 0.3, 0.4$ and $0.5$). From left to right are $x/c = 0.15, 0.2, 0.3, 0.4$ and $0.5$, respectively. Displayed are envelopes for two extreme $\Delta_{k}$ perturbations considered in this study: $\Delta_{k} =0.1$ (red envelope),  $\Delta_{k} = 8$ (gray envelope), and $M_{k}$ (blue envelope). The baseline prediction is provided for reference. $\circ$ in-house DNS data \cite{zhang2021turbulent}.}
\label{fig:UQ_uniformk_uv_RepeatOn_Tu0027.pdf}
\end{figure}

The predicted profiles for the Reynolds shear stress normalized with the freestream velocity squared, $-\left\langle u_{1}u_{2} \right \rangle/U_{\infty}^2$ are shown in Fig. \ref{fig:UQ_uniformk_uv_RepeatOn_Tu0027.pdf}. Undergoing the $\Delta_{k} = 0.1$ and $\Delta_{k} = 8$ perturbations, an enveloping behavior with respect to the baseline prediction can be observed. Figure \ref{fig:UQ_uniformk_uv_RepeatOn_Tu0027.pdf} shows that the baseline prediction for the $-\left\langle u_{1}u_{2} \right \rangle/U_{\infty}^2$ profile at $x/c = 0.15$ significantly over-predicts the in-house DNS profile, implying a higher level of momentum transfer due to the Reynolds shear stress. In the aft portion of the LSB and downstream of the LSB near the reattachemnt point ($X_{R}$), the predictions for the $-\left\langle u_{1}u_{2} \right \rangle/U_{\infty}^2$ profiles at $x/c = 0.2$ and $x/c = 0.3$ exhibit a shape of parabolic arch, revealing a same effect as the in-house DNS data, i.e., a strong increase in the Reynolds shear stress profile around the peak of the parabolic arch. The magnitude of the increase is much greater for the in-house DNS data probably due to the larger height of the LSB produced than the baseline prediction. Further downstream of the LSB, the baseline predictions for the $-\left\langle u_{1}u_{2} \right \rangle/U_{\infty}^2$ profiles at $x/c = 0.4$ and $x/c = 0.5$ (in the attached turbulent boundary layer) show relatively good agreement with the in-house DNS data, although some discrepancies exist in the regions next to the wall and in the upper section of the Reynolds shear stress profiles. 
 
In Fig. \ref{fig:UQ_uniformk_uv_RepeatOn_Tu0027.pdf}, the $\Delta_{k} = 0.1$ perturbation increases the magnitude of the $-\left\langle u_{1}u_{2} \right \rangle/U_{\infty}^2$ profile compared to the baseline prediction. In the aft portion of the LSB ($x/c = 0.15$, $x/c = 0.2$ and $x/c = 0.3$), the $\Delta_{k} = 0.1$ perturbations retain the shape of parabolic arch, with a peak value around the maximum height of the arch gradually reducing in magnitude to zero from the peak in the opposite directions toward the wall and the OBL, respectively. In addition, the $\Delta_{k} = 0.1$ perturbations increase the momentum transfer due to the Reynolds shear stress compared to the baseline prediction. Consequently, the $\Delta_{k} = 0.1$ perturbations tend to approach closer to the in-house DNS data except at $x/c = 0.15$, where a deviation from the in-house DNS data is observed. As the flow proceeds further downstream within the attached turbulent boundary layer ($x/c = 0.4$ and $x/c = 0.5$), the effect of the $\Delta_{k} = 0.1$ perturbation gradually deteriorates with $x/c$, with some of the in-house DNS data being encompassed. 

On the other hand, the $\Delta_{k} = 8$ perturbation decreases the magnitude of the $-\left\langle u_{1}u_{2} \right \rangle/U_{\infty}^2$ profile compared to the baseline prediction, with the size of the uncertainty bound significantly larger than that for the $\Delta_{k} = 0.1$ perturbation, reflecting the simulation's much stronger response to the $\Delta_{k} = 8$ perturbation. Likewise, a shape of parabolic arch and a similar behavior to the $\Delta_{k} = 0.1$ perturbations are observed for the $\Delta_{k} = 8$ perturbations in the aft portion of the LSB ($x/c = 0.15$, $x/c = 0.2$ and $x/c = 0.3$) as well: peaking around the maximum height of the parabolic arch and gradually decreasing in magnitude toward the wall and OBL. As a result, the $\Delta_{k} = 8$ perturbations reduce the momentum transfer due to the Reynolds shear stress to a great extend around the peak of the parabolic arch. This shows a tendency for the $\Delta_{k} = 8$ perturbations to deviate from the in-house DNS data except at $x/c = 0.15$, where the $\Delta_{k} = 8$ perturbations tend to approach closer to the in-house DNS profile that much lag behind the baseline prediction. Within the attached turbulent boundary layer ($x/c = 0.4$ and $x/c = 0.5$), an important observation for the $\Delta_{k} = 8$ perturbation is that the uncertainty bounds retain a value of zero not only at the wall but also extend for some distance above the wall, which violates the ``rule'' that all Reynolds stresses decrease to zero at the wall surface due to the no-slip wall condition \cite{versteeg2007introduction}. This marks the behavior of ``over perturbation'' with $\Delta_{k} = 8$, and is not physically realizable. Since few studies have been conducted to determine the upper bound of $k^{*}$, this study here sheds light on a possible way of determining the upper bound of $k^{*}$ using the result of Reynolds shear stress. Therefore, the maximum magnitude of $\Delta_{k}$ must ensure that Reynolds stresses must behave in a physically-realizable manner in the near-wall region. 

The $M_{k}$ perturbation in general under-predicts the baseline prediction across the suction side, in general sitting within the gray envelope, with a subtle movement to the red envelope being discerned in the lower section of the Reynolds shear stress profiles for $x/c =  0.2$ and $x/c = 0.3$. Within the attached turbulent boundary layer ($x/c = 0.4$ and $x/c = 0.5$), the uncertainty bounds generated from the $M_{k}$ perturbation remain constantly below the baseline prediction, which is consistent with the uniform magnitude of $\Delta_{k} = 2.8$. It should be noted that the simulation's response to the perturbed Reynolds shear stress profile is in general stronger than that for the perturbed mean velocity profile. This indicates that the level of sensitivity to the $\Delta_{k}$ perturbation varies with different QoIs being observed. In Fig. \ref{fig:UQ_uniformk_uv_RepeatOn_Tu0027.pdf}, the $M_{k}$ function successfully avoids over-perturbations through strictly comparing to the available high-fidelity data, ensuring that only the physical realistic perturbations are considered. 
%Since Reynolds shear stress plays an important role in contributing to the turbulent production (hence turbulence kinetic energy) in linear eddy viscosity models \cite{pope2001turbulent,versteeg2007introduction}, we conjecture that the turbulent production should behave in a similar manner to the Reynolds shear stress. 

\subsection{Combining $M_{k}$ with $1c$, $2c$, and $3c$}\label{Sec:compound}
  
\subsubsection{Skin friction coefficient}
\begin{figure} 
\centerline{\includegraphics[width=3.5in]{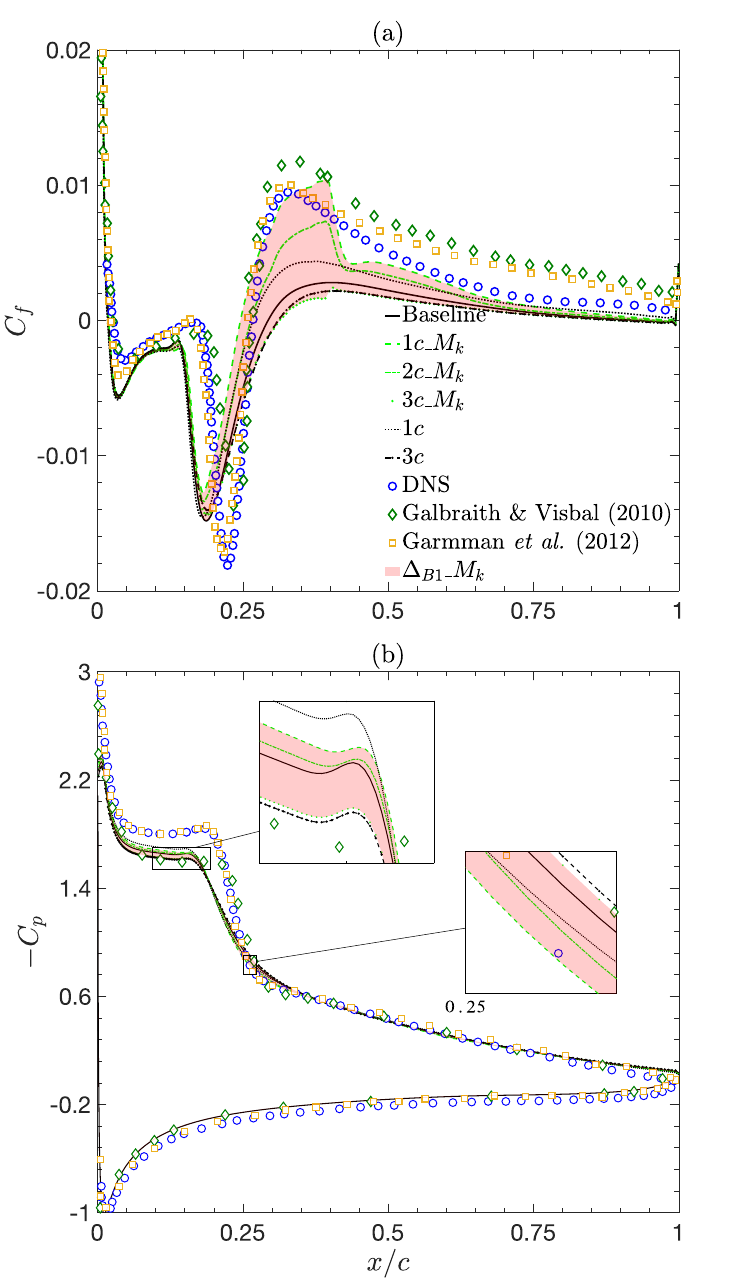}}
\caption[(a) Profile of skin friction coefficient and (b) pressure coefficient with enlarged regions at the flat spot and the kink followed by a sharp drop of $C_{p}$.]{(a) Profile of skin friction coefficient and (b) pressure coefficient with enlarged regions at the flat spot and the kink followed by a sharp drop of $C_{p}$. Displayed are uncertainty bounds for $1c\_M_{k}$, $2c\_M_{k}$ and $3c\_M_{k}$ perturbations (red envelope). $\Delta_{B1}$ stands for $\Delta_B = 1.0$. Profiles of baseline prediction and eigenvalue perturbations ($1c$ and $3c$) are provided for reference. $\circ$ in-house DNS data \cite{zhang2021turbulent}.}
\label{fig:cfcp_marker.pdf}
\end{figure}

Distributions of the skin friction coefficient and the pressure coefficient, $C_{f}$ and $C_{p}$, are shown in Figs. \ref{fig:cfcp_marker.pdf} (a) and (b), respectively. Also included are the in-house DNS \cite{zhang2021turbulent} and ILES/LES data of \cite{galbraith2010implicit} and \cite{garmann2013comparative} for comparison. Integrating the $M_{k}$ perturbation with the eigenvalue perturbation ($1c$, $2c$ and $3c$) using Eqs. \ref{Eqn_Rij_perturbed}, \ref{Eqn:Markerfunc} and \ref{Eqn:kstar} yields compound effect, namely, $1c\_M_{k}$, $2c\_M_{k}$ and $3c\_M_{k}$. Also included are the eigenvalue perturbation ($1c$ and $3c$) as a reference for the $1c\_M_{k}$ $2c\_M_{k}$ and $3c\_M_{k}$ perturbation. In the aft portion of the LSB, Fig. \ref{fig:cfcp_marker.pdf} (a) clearly shows that the $1c\_M_{k}$ perturbation decreases the magnitude of $C_{f}$ more than the $2c\_M_{k}$ perturbation does compared to the baseline prediction, while the $3c\_M_{k}$ perturbation results in the uncertainty bound that almost overlaps the one generated from the $3c$ perturbation, indicating simulation's low sensitivity to the $3c\_M_{k}$ perturbation. In addition, both uncertainty bounds generated from the $3c\_M_{k}$ and $3c$ perturbations in general sit slightly below the baseline prediction except at the trough around $x/c = 0.2$ (in the aft portion of the LSB), where they sit somewhat above the baseline prediction. As a consequence, an enveloping behavior with respect to the baseline prediction is observed. On the other hand, the uncertainty bounds generated from the $1c\_M_{k}$ and $2c\_M_{k}$ perturbations lie significantly above the baseline prediction, encompassing the reference data for $X_{R}$, as well as the steep rise followed by $X_{R}$. Interestingly, it is clear that this promising increase associated with the $1c\_M_{k}$ and $2c\_M_{k}$ perturbations is not a simple sum of the $M_{k}$ and $1c$/$2c$ uncertainty bounds up, but a ``synergy'' has developed. Moreover, the synergy behavior associated with the $1c\_M_{k}$ perturbation results in the encompassing of the gap between the baseline prediction and the reference data in the aft portion of the LSB as well as at the crest. Besides, it is interesting to note that the uncertainty bounds generated from $1c\_M_{k}$ and $2c\_M_{k}$ perturbations tend to retain the shape of the $C_{f}$ profile at the crest for $0.3 < x/c < 0.4$, with the $1c\_M_{k}$ perturbation effectively encompassing the in-house DNS data of \cite{zhang2021turbulent}. This confirms the effect of spatial variability in $M_{k}$. As the flow proceeds further downstream $0.4 < x/c < 0.6$, a rapid collapse of the uncertainty bounds generated from the $1c\_M_{k}$ and $2c\_M_{k}$ perturbations is observed. This confirms the uniform magnitude of $M_{k}$ used in the region for $0.4 < x/c < 0.6$. On the other hand, the $3c\_M_{k}$ and $3c$ perturbations become almost indistinguishable from each other, lying somewhat below the baseline prediction across the entire suction side, except for a slight decrease found associated with the $3c\_M_{k}$ perturbation in the region for $0.35 < x/c < 0.4$.

In Fig. \ref{fig:cfcp_marker.pdf} (b), at the flat spot the $1c\_M_{k}$ perturbation increases the magnitude of $C_{p}$ more than the $2c\_M_{k}$ perturbation does compared to the baseline prediction. Both $1c\_M_{k}$ and $2c\_M_{k}$ perturbations show a tendency to approach toward the in-house DNS \cite{zhang2021turbulent} and LES data of \cite{garmann2013comparative}, and sit within the uncertainty bound generated from the $1c$ perturbation. Interestingly, there is no discernible synergy behavior appearing at the flat spot, the uncertainty bound generated from the $1c\_M_{k}$ and $2c\_M_{k}$ perturbations tend to reduce somewhat in size compared to the $1c$ and $2c$ perturbations instead. On the other hand, the uncertainty bounds generated from the $3c\_M_{k}$ and $3c$ perturbations become almost indistinguishable at the flat spot, sitting slightly below the baseline prediction in a trend of approaching toward the ILES data of \cite{galbraith2010implicit}. At the kink around $X_{R}$, the uncertainty bounds generated from the $1c\_M_{k}$ and $2c\_M_{k}$ perturbations under-predict the baseline prediction and tend to approach closer to the reference data, while the uncertainty bound for the $3c\_M_{k}$ perturbation over-predicts the baseline prediction and retains the behavior of collapsing onto the $3c$ perturbation, showing a trend of deviating from the reference data. As a consequence, a discernible enveloping behavior with respect to the baseline prediction is observed at both the flat spot and the kink around $X_{R}$, where most uncertainty is generated, as shown in Fig. \ref{fig:cfcp_marker.pdf} (b). In addition, the collapsing behavior of the $3c\_M_{k}$ and $3c$ perturbations at the flat spot and the kink indicates the simulation's low sensitivity to the $3c\_M_{k}$ perturbation. Compared to $C_{f}$, it is clear that $Cp$ overall is less sensitive to all kinds of perturbations. This is because the wall pressure is determined by the freestream, which is only modified minutely  by the eigenvalue perturbations \cite{emory2013modeling}. In addition, this reflects that the degree of response to the $\Delta_{k}$ perturbation varies with which QoIs being observed. On the pressure side, the simulation's response to all kinds of perturbations are rather small, indicating a low level of model form uncertainty and hence high trustworthiness in the baseline prediction for $C_{p}$. 

It should be noted that because $3c$ perturbation retains the isotropic nature of the turbulent viscosity model, it yields limited influence on the perturbed results \cite{mishra2019theoretical}. This is well reflected in the smaller size of the uncertainty bound generated from the $3c$ perturbation compared to the $1c$ and $2c$ perturbations. Such inefficacy of $3c$ perturbation has been observed by Emory \textit{et al.} \cite{emory2013modeling} as well. Importantly, this inefficacy persists when being compounded with $M_{k}$, which might partly explain the collapsing behavior of the $3c\_M_{k}$ profile onto the $3c$ profile as can be observed in Figs. \ref{fig:cfcp_marker.pdf} (a) and (b). Moreover, this collapsing behavior not only happens to the results of $C_{f}$ and $C_{p}$ but also happens to the mean velocity profile and the turbulence quantities, as can be observed in the following sections.

\subsubsection{Mean velocity field}
%contourf for U: 1cMarker 2cMarker 3cMarker OnlyMk and DNS
\begin{figure*} 
\centerline{\includegraphics[width=5.5in]{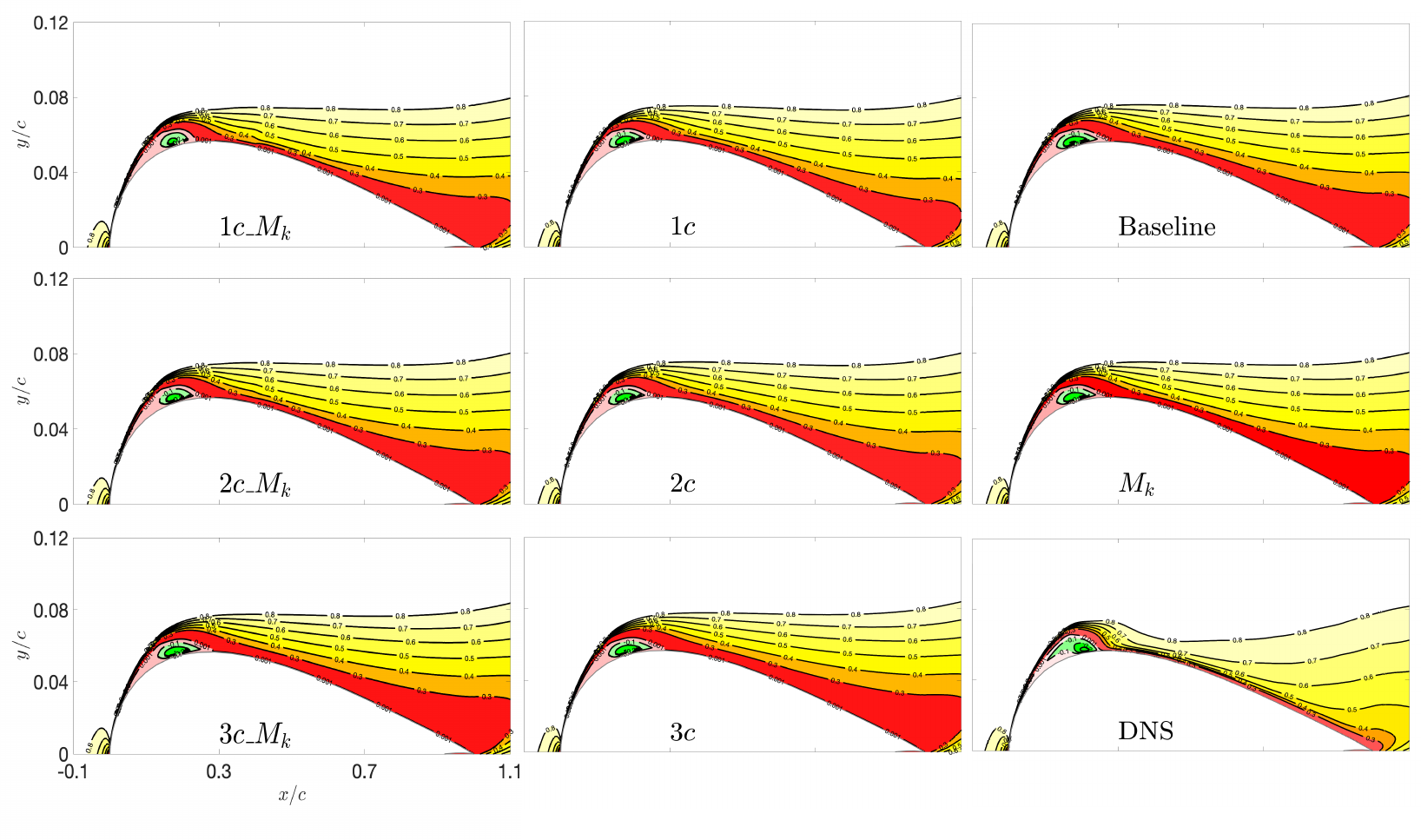}}
\caption[Contours of $\left\langle U \right \rangle/U_{\infty}$ with $1c\_M_{k}$, $2c\_M_{k}$, $3c\_M_{k}$, $1c$, $2c$, $3c$ and $M_{k}$ perturbations in an $xy$ plane.]{Contours of $\left\langle U \right \rangle/U_{\infty}$ with $1c\_M_{k}$, $2c\_M_{k}$, $3c\_M_{k}$, $1c$, $2c$, $3c$ and $M_{k}$ perturbations in an $xy$ plane. Isolines of the mean streamwise velocity are superimposed on the contour plots. The contour of baseline prediction is provided for reference, and the contour of in-house DNS data \cite{zhang2021turbulent} is included for comparison.}
\label{fig:RANS_contourf_normU_All_subplot_outlook.pdf}
\end{figure*}

\begin{figure} 
\centerline{\includegraphics[width=3.7in]{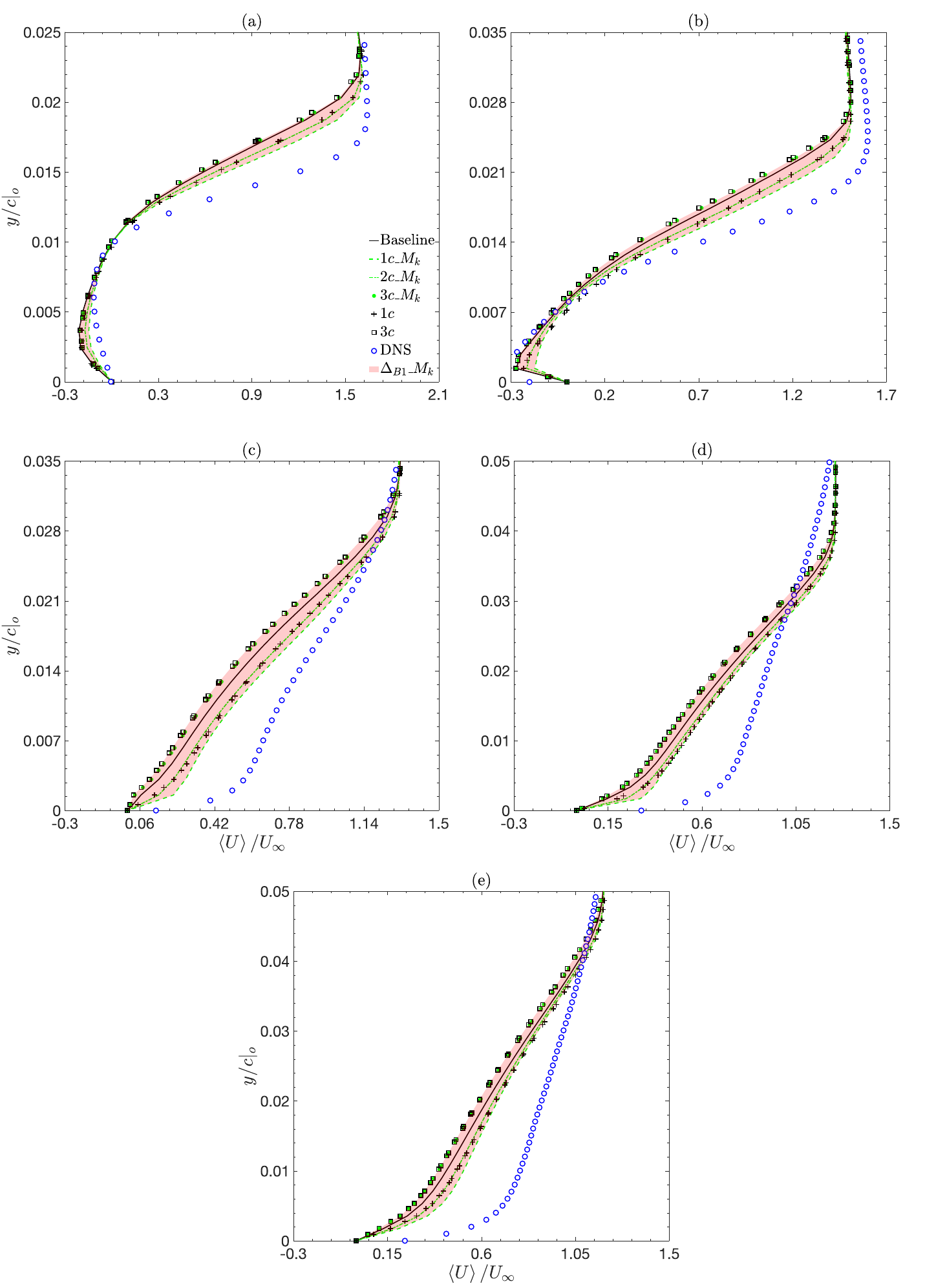}}
\caption[Profile of $\left\langle U \right \rangle/U_{\infty}$ in the aft portion of the LSB for (a) $x/c = 0.15$ and (b) $x/c = 0.2$; and profile of $\left\langle U \right \rangle/U_{\infty}$ in the attached TBL for (c) $x/c = 0.3$, (d) $x/c = 0.4$ and (e) $x/c = 0.5$. Displayed are uncertainty bounds for $1c\_M_{k}$, $2c\_M_{k}$ and $3c\_M_{k}$ perturbations (red envelope).]{Profile of $\left\langle U \right \rangle/U_{\infty}$ in the aft portion of the LSB for (a) $x/c = 0.15$ and (b) $x/c = 0.2$; and profile of $\left\langle U \right \rangle/U_{\infty}$ in the attached TBL for (c) $x/c = 0.3$, (d) $x/c = 0.4$ and (e) $x/c = 0.5$. Displayed are uncertainty bounds for $1c\_M_{k}$, $2c\_M_{k}$ and $3c\_M_{k}$ perturbations (red envelope). $\Delta_{B1}$ stands for $\Delta_B = 1.0$. Profiles of baseline prediction and eigenvalue perturbations ($1c$ and $3c$) are provided for reference. $\circ$ in-house DNS data \cite{zhang2021turbulent}.}
\label{fig:markerfunc_U_five.pdf}
\end{figure}

\begin{figure*} 
\centerline{\includegraphics[width=5.5in]{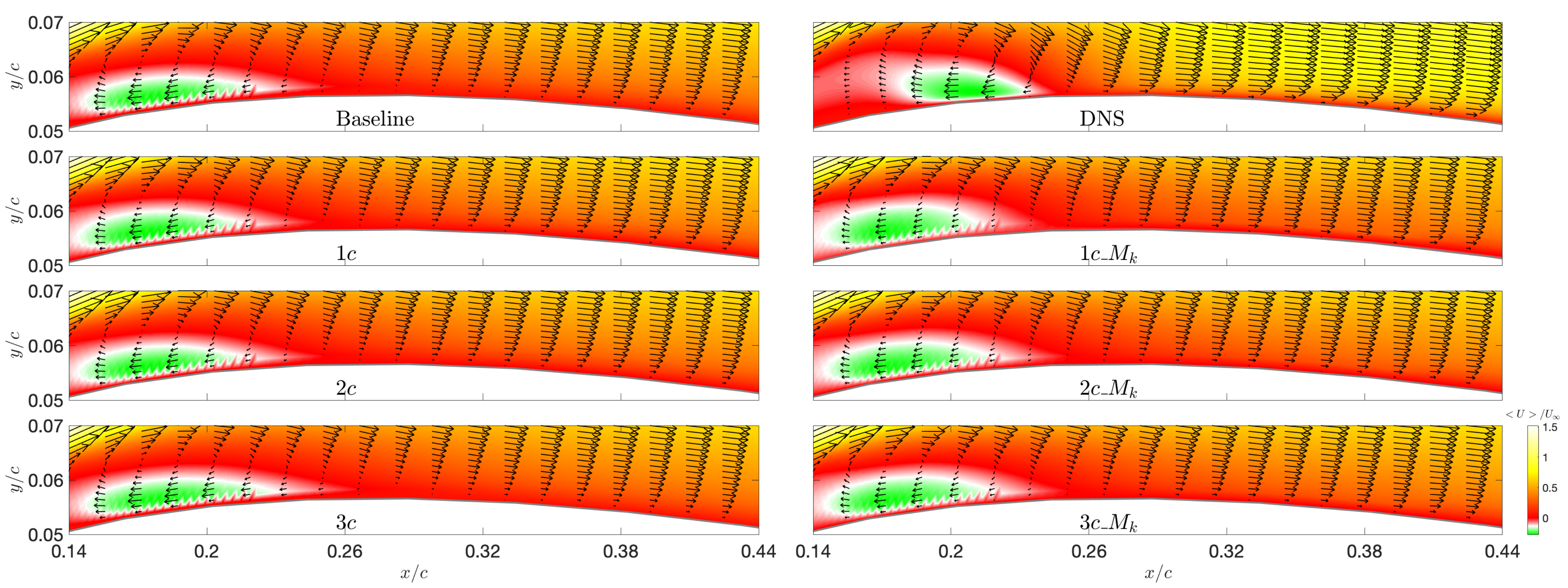}}
\caption[Contours of normalized mean velocity $\left\langle U \right \rangle/U_{\infty}$ with in-plane velocity vectors superimposed on the contours in an $xy$ plane.]{Contours of normalized mean velocity $\left\langle U \right \rangle/U_{\infty}$ with in-plane velocity vectors superimposed on the contours in an $xy$ plane. The region in the vicinity of the wall is enlarged to highlight the flow behavior in the LSB, as well as in the turbulent region right downstream of the LSB. A focus on a section of the airfoil suction side is considered: $0.14 < x/c < 0.44$.}
\label{fig:arrows_1c2c3c_Base_Marker_DNS_Marker.pdf}
\end{figure*}

Contours of the mean velocity normalized by the freestream velocity, $\left\langle U \right \rangle/U_\infty$ from the baseline, $1c\_M_{k}$, $2c\_M_{k}$ and $3c\_M_{k}$ perturbations, eigenvalue perturbations ($1c$, $2c$ and $3c$), $M_{k}$ perturbation and in-house DNS of \cite{zhang2021turbulent} in an $xy$ plane are shown in Fig. \ref{fig:RANS_contourf_normU_All_subplot_outlook.pdf}. From Fig. \ref{fig:RANS_contourf_normU_All_subplot_outlook.pdf}, all of the $\left\langle U \right \rangle/U_\infty$ contours show a recirculating region, i.e., the eye-like green region, where the negative value of velocity is present. In addition, the mean velocity contour generated from the $M_{k}$ perturbation results in a shorter LSB and a slightly increased $\left\langle U \right \rangle/U_\infty$ in magnitude in the region downstream of the reattachment point $0.3 < x/c < 0.6$, in which the untrustworthy zones are identified. This indicates that the $M_{k}$ perturbation tends to suppress the LSB compared to the baseline prediction. This reduces the turbulence kinetic energy contained in the large-scale coherent structures \cite{lengani2014pod}, implying the increased mean-flow energy in the vicinity of the LSB $0.3 < x/c < 0.6$, therefore increased magnitude of $\left\langle U \right \rangle/U_\infty$ in this region, as shown in Fig. \ref{fig:RANS_contourf_normU_All_subplot_outlook.pdf}. For the $1c$ and $2c$ eigenvalue perturbations, the contours show a reduction in the length of the LSB compared to the baseline prediction, which results in an overall increase in the mean-flow magnitude further downstream of the reattachment point, while the $3c$ perturbation does the opposite. In addition, it is clear that the $1c\_M_{k}$ and $2c\_M_{k}$ perturbations further increase the magnitude of $\left\langle U \right \rangle/U_\infty$ than the $1c$ and $2c$ perturbations do compared to the baseline prediction, which confirms the greatly reduced $C_{f}$ in magnitude in the aft portion of the LSB, while the $3c\_M_{k}$ perturbation remains at nearly same magnitude as that for the $3c$ perturbation, which confirms the collapse of the $C_{f}$ profiles generated from the $3c\_M_{k}$ and $3c$ perturbations, as shown in Fig. \ref{fig:cfcp_marker.pdf} (a). This indicates a weak compound effect of the $3c\_M_{k}$ perturbation. Compared to the baseline prediction, both $1c\_M_{k}$ and $2c\_M_{k}$ perturbations shorten the region of reverse flow (deviating from the in-house DNS data), while increase the mean-flow magnitude in the attached turbulent boundary layer (approaching closer to the in-house DNS data); on the other hand, the $3c\_M_{k}$ perturbation bolsters the region of reverse flow, showing a tendency of approaching closer to the in-house DNS data, while shows a reduction in the magnitude of $\left\langle U \right \rangle/U_\infty$ in the attached turbulent boundary layer, causing a deviation from the in-house DNS data.

%\textbf{you need to add couple references indicating that the similar trend: 1c overpredict the base line has been observed by other %researchers, e.g. SU2 paper, openfoam paper, and emory paper}
The predictions for the mean velocity profile normalized by $U_{\infty}$, i.e., $\left\langle U \right \rangle/U_{\infty}$, are presented in Figs. \ref{fig:markerfunc_U_five.pdf} (a) - (e). The in-house DNS data of \cite{zhang2021turbulent} is included for comparison. In Figs. \ref{fig:markerfunc_U_five.pdf} (a) - (e), the $1c$ and $3c$ eigenvalue perturbations are used as a reference for the $1c\_M_{k}$, $2c\_M_{k}$ and $3c\_M_{k}$ perturbation. From Figs. \ref{fig:markerfunc_U_five.pdf} (a) - (e), an enveloping behavior with respect to the baseline prediction is observed, i.e., the $1c\_M_{k}$ and $2c\_M_{k}$ perturbations leading the baseline prediction, while the $3c\_M_{k}$ perturbation lagging behind. A similar behavior of the $1c$ and $3c$ perturbations with respect to the baseline prediction was also observed by Luis \textit{et al.} \cite{cremades2019reynolds} in their numerical study for a turbulent flow over a backward-facing step. In addition, the $3c\_M_{k}$ profile tends to collapse onto the $3c$ profile, reflecting the simulation's low sensitivity to the $3c\_M_{k}$ perturbation, which is consistent with the behavior shown in Fig. \ref{fig:RANS_contourf_normU_All_subplot_outlook.pdf}. At $x/c  = 0.15$ ($X_{T}$), the uncertainty bound generated from the $1c\_M_{k}$ perturbation increases the magnitude of the mean velocity profile more than the $2c\_M_{k}$ perturbation does in both the region of reverse flow ($U/U_{\infty} < 0$) for $0 < y/c|_{o} < 0.007$ and the upper portion of the boundary layer for $0.011 < y/c|_{o} < 0.023$, showing a tendency to approach closer to the in-house DNS data. On the other hand, the $3c\_M_{k}$, $3c$ and baseline profiles show a collapse, indicating a type of similarity. This might be partly explained by the almost same values of $C_{f}$ found in the $3c\_M_{k}$, $3c$ and baseline profiles around $X_{T}$, as shown in Fig. \ref{fig:cfcp_marker.pdf} (a). It should be noted that the uncertainty bounds generated from the $1c\_M_{k}$, $2c\_M_{k}$, and $3c\_M_{k}$ perturbations and the baseline prediction are negligibly small for $0.007 < y/c|_{o} < 0.011$, i.e., all collapsing onto a single curve, which shows good agreement with the in-house DNS data, as shown in Fig. \ref{fig:markerfunc_U_five.pdf} (a). This reveals a low level of the model form uncertainty and hence relatively high trustworthiness in this region. As the flow moves further downstream to $x/c = 0.2$ (the aft portion of the LSB), the effect of the $1c\_M_{k}$ and $2c\_M_{k}$ perturbations has permeated the entire boundary layer, showing a tendency of approaching closer to the in-house DNS data in the upper portion of the boundary layer. Moreover, the uncertainty bounds generated from the $1c\_M_{k}$ and $2c\_M_{k}$ perturbations over-predict the baseline prediction but lie somewhat above the $1c$ perturbation in the region for $0.007 < y/c|_{o} < 0.011$, overall showing closer agreement with the in-house DNS data, as shown in Fig. \ref{fig:markerfunc_U_five.pdf} (b). Again, this reflects the effect of spatial variability in $M_{k}$. Note that the baseline prediction overall shows good agreement with the in-house DNS data in the region of reverse flow at $x/c = 0.2$, at which the baseline prediction, $3c$, and $3c\_M_{k}$ perturbations collapse onto a single curve. The almost identical magnitude of $C_{f}$ retained by the $3c\_M_{k}$, $3c$ and baseline predictions for $x/c = 0.2$, as shown in Fig. \ref{fig:cfcp_marker.pdf} (a), might partly explain this type of similarity. Away from the wall, a collapsing behavior is again observed for the $3c\_M_{k}$ and $3c$ perturbations, with a slight offset from the baseline prediction. Note that the $1c$ perturbation is well encompassed by $1c\_M_{k}$ in the aft portion of the LSB, as shown in Figs. \ref{fig:markerfunc_U_five.pdf} (a) - (b). This confirms the larger values of $C_{f}$ for the $1c\_M_{k}$ perturbation in this region. At $x/c = 0.3$ (downstream of the LSB near \textbf{$X_{R}$}), $x/c = 0.4$ and $x/c = 0.5$ (in the reattached turbulent boundary layer), the uncertainty bounds generated from the $1c\_M_{k}$ and $2c\_M_{k}$ perturbations in general tend to approach closer to the in-house DNS data, with the perturbation effect gradually deteriorating further downstream. This is consistent with the gradual reduction in the positive values of $C_{f}$ as the flow moves further downstream of $X_{R}$, as shown in Fig. \ref{fig:cfcp_marker.pdf} (a). Also the difference between the $1c$ and $1c\_M_{k}$ perturbation becomes smaller, which confirms the comparable magnitude of $C_{f}$ in the region of the attached turbulent boundary layer, as shown in Fig. \ref{fig:cfcp_marker.pdf} (a). On the other hand, a collapse is also observed for the $3c\_M_{k}$ and $3c$ perturbations at $x/c = 0.4$ and $x/c = 0.5$, which confirms the almost identical values of $C_{f}$ retained by the $3c\_M_{k}$ and $3c$ perturbations shown in Fig. \ref{fig:cfcp_marker.pdf} (a). It is interesting to note that $1c$ and $2c$ perturbations respond favorably to $M_{k}$, while $3c$ perturbation remains almost immune to it.  

Figure \ref{fig:arrows_1c2c3c_Base_Marker_DNS_Marker.pdf} shows contours of mean velocity, with the region of reverse flow being enlarged. The region of reverse flow is evidenced by the velocity vectors added in the LSB. The baseline prediction is provided for reference. Also included is the in-house DNS data of \cite{zhang2021turbulent} for comparison. Compared to the in-house DNS data, Fig. \ref{fig:arrows_1c2c3c_Base_Marker_DNS_Marker.pdf} clearly shows that the baseline prediction shifts the region of reverse flow in the upstream direction. Within the LSB (green region), the velocity vectors for the $1c$ and $2c$ perturbations clearly indicate a subdued reverse-flow field, resulting in a shorter LSB and hence a better agreement with the DNS data, and the opposite is true for the $3c$ perturbation. For the attached turbulent boundary layer, the velocity vectors indicate an overall increase in the mean velocity field for the $1c$ and $2c$ perturbations, showing a tendency of approaching to the DNS mean flow field, while the $3c$ perturbation shows an overall reduction in the mean velocity field. Integrating $M_{k}$ into the $1c$, $2c$, and $3c$ perturbation tends to suppress the size of the LSB, but increases the mean flow field downstream of the LSB. Among these perturbations, $1c\_M_{k}$ increases the mean flow field more than $2c\_M_{k}$ in the attached turbulent boundary layer, to the largest extend contributing to a closer approach to the in-house DNS data.   

\subsubsection{Reynolds shear stress}
%contourf and streamlines for 1c2c3cOnlyMarkerk and DNS
\begin{figure*} 
\centerline{\includegraphics[width=5.5in]{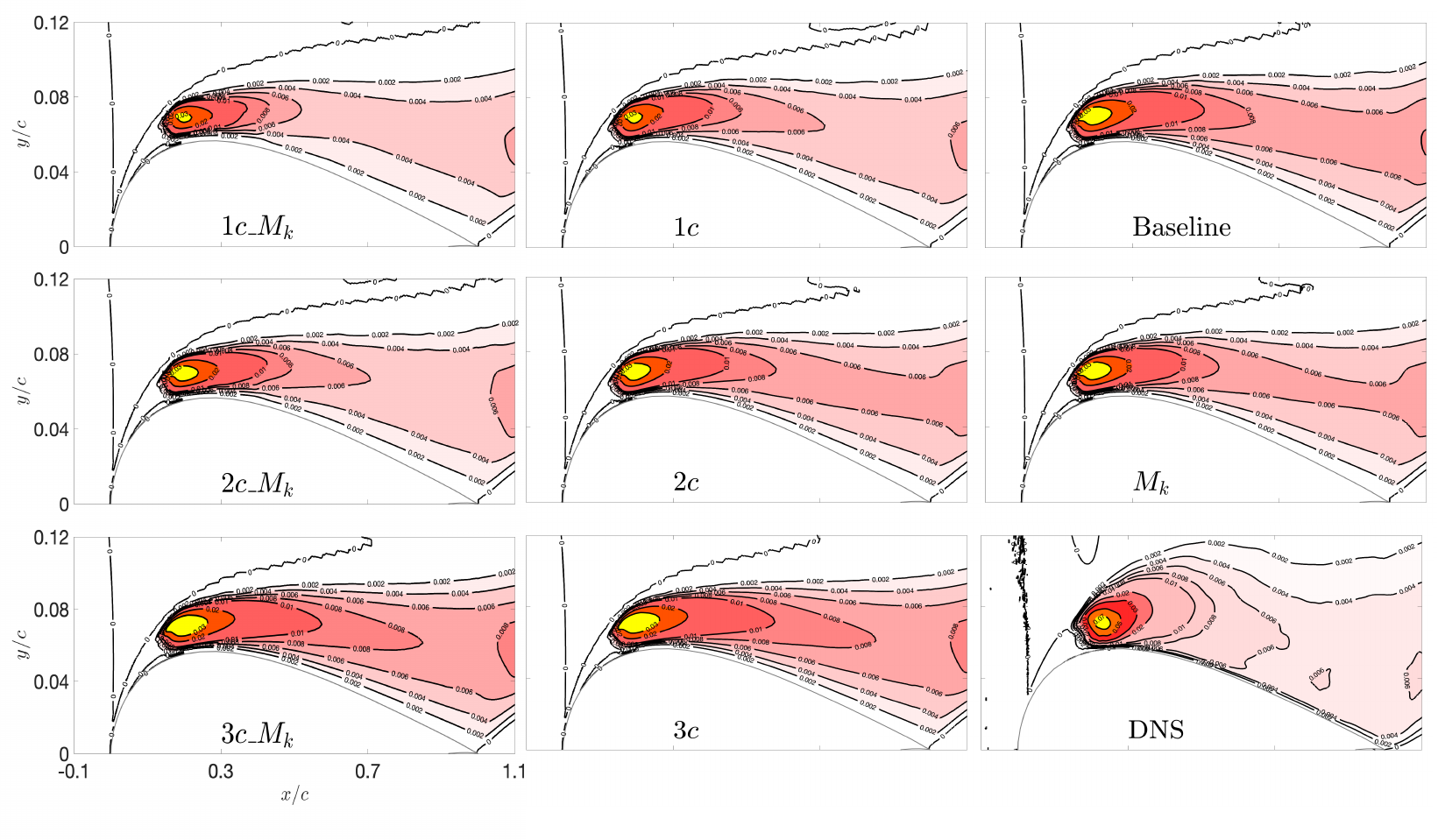}}
\caption[Contours of $-\left\langle u_{1}u_{2} \right \rangle/U_{\infty}^2$ with $1c\_M_{k}$, $2c\_M_{k}$, $3c\_M_{k}$, $1c$, $2c$, $3c$ and $M_{k}$ perturbations in an $xy$ plane.]{Contours of $-\left\langle u_{1}u_{2} \right \rangle/U_{\infty}^2$ with $1c\_M_{k}$, $2c\_M_{k}$, $3c\_M_{k}$, $1c$, $2c$, $3c$ and $M_{k}$ perturbations in an $xy$ plane. Isolines of the Reynolds shear stress are superimposed on the contour plots. The contour of baseline prediction is provided for reference, and the contour of in-house DNS data \cite{zhang2021turbulent} is included for comparison.}
\label{fig:RANS_contourf_normuv_All_1c2c3cOnlyMarkerfunc_subplot_foucs.pdf}
\end{figure*}

%marker for uv x_c = 0.15, 0.2, 0.4, 0.5
\begin{figure} 
\centerline{\includegraphics[width=3.7in]{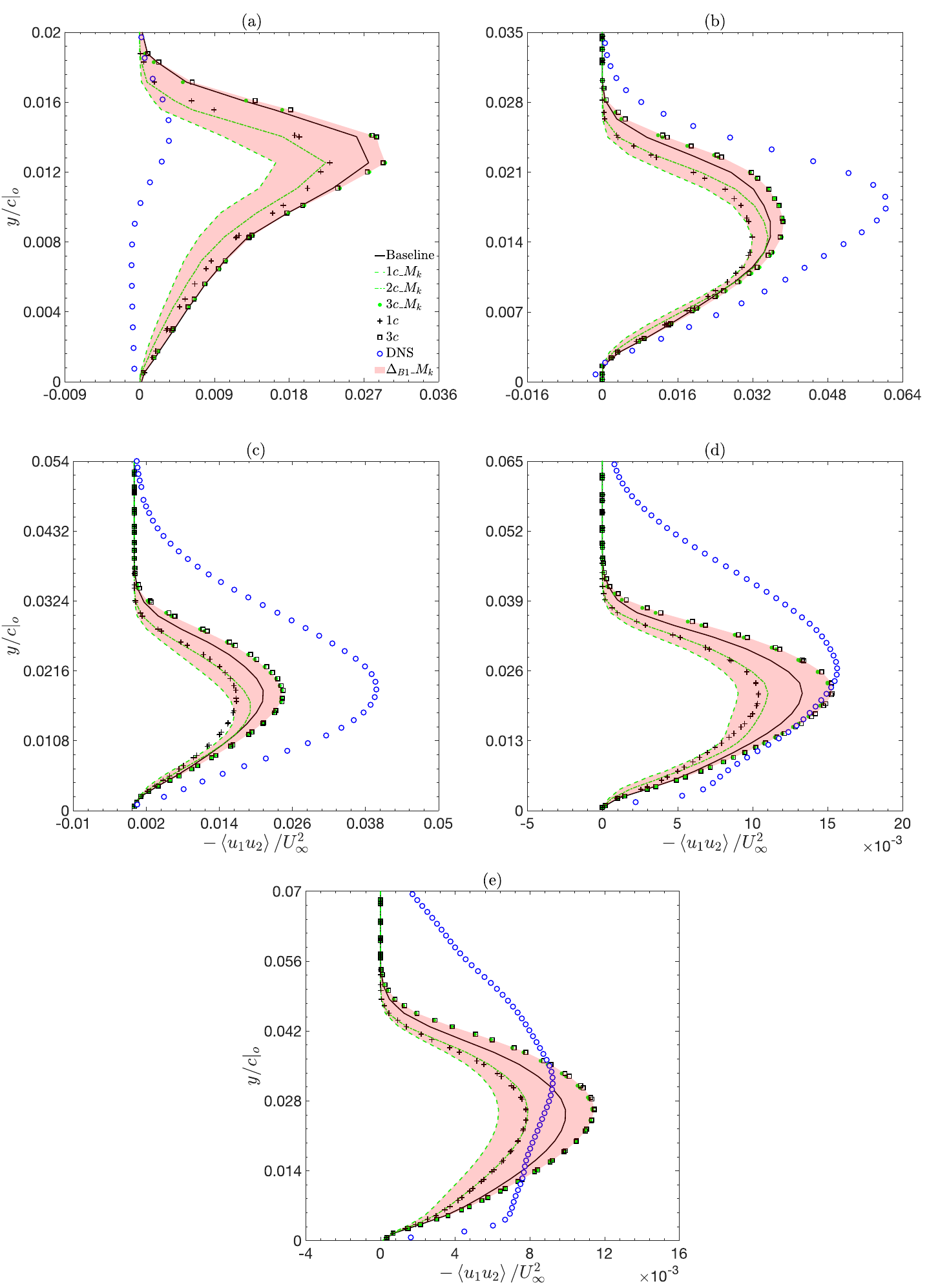}}
\caption[Profile of $-\left\langle u_{1}u_{2} \right\rangle/U_{\infty}^2$ in the aft portion of the LSB for (a) $x/c = 0.15$ and (b) $x/c = 0.2$; and profile of $-\left\langle u_{1}u_{2} \right\rangle/U_{\infty}^2$ in the attached TBL for (c) $x/c = 0.3$, (d) $x/c = 0.4$ and (e) $x/c = 0.5$. Displayed are uncertainty bounds for $1c\_M_{k}$, $2c\_M_{k}$ and $3c\_M_{k}$ perturbations (red envelope).]{Profile of $-\left\langle u_{1}u_{2} \right\rangle/U_{\infty}^2$ in the aft portion of the LSB for (a) $x/c = 0.15$ and (b) $x/c = 0.2$; and profile of $-\left\langle u_{1}u_{2} \right\rangle/U_{\infty}^2$ in the attached TBL for (c) $x/c = 0.3$, (d) $x/c = 0.4$ and (e) $x/c = 0.5$. Displayed are uncertainty bounds for $1c\_M_{k}$, $2c\_M_{k}$ and $3c\_M_{k}$ perturbations (red envelope). $\Delta_{B1}$ stands for $\Delta_B = 1.0$. Profiles of baseline prediction and eigenvalue perturbations ($1c$ and $3c$) are provided for reference. $\circ$ in-house DNS data \cite{zhang2021turbulent}.}
\label{fig:markerfunc_uv_five.pdf}
\end{figure}

Contours of the Reynolds shear stress normalized by the freestream velocity squared,  $-\left\langle u_{1}u_{2} \right \rangle/U_{\infty}^2$ from the baseline, $1c\_M_{k}$, $2c\_M_{k}$ and $3c\_M_{k}$ perturbations, eigenvalue perturbations ($1c$, $2c$, $3c$), $M_{k}$ perturbation in an $xy$ plane are shown in Fig. \ref{fig:RANS_contourf_normuv_All_1c2c3cOnlyMarkerfunc_subplot_foucs.pdf}. Also included is the in-house DNS data of \cite{zhang2021turbulent} for comparison. In Fig. \ref{fig:RANS_contourf_normuv_All_1c2c3cOnlyMarkerfunc_subplot_foucs.pdf}, all of the $-\left\langle u_{1}u_{2} \right \rangle/U_{\infty}^2$ contour plots show a peak around $X_{T}$, i.e., the bright yellow region, downstream from the peak a gradual reduction in the magnitude of $-\left\langle u_{1}u_{2} \right \rangle/U_{\infty}^2$ is observed. From Fig. \ref{fig:RANS_contourf_normuv_All_1c2c3cOnlyMarkerfunc_subplot_foucs.pdf}, the $M_{k}$ perturbation overall reduces the magnitude of $-\left\langle u_{1}u_{2} \right \rangle/U_{\infty}^2$ in both the transitional and turbulent region compared to the baseline prediction. From Fig. \ref{fig:RANS_contourf_normuv_All_1c2c3cOnlyMarkerfunc_subplot_foucs.pdf}, it is clear that the $1c\_M_{k}$ and $2c\_M_{k}$ perturbations further reduce the magnitude of $-\left\langle u_{1}u_{2} \right \rangle/U_{\infty}^2$ compared to the $1c$ and $2c$ perturbations, while the $3c\_M_{k}$ perturbation remains at a nearly same magnitude as that of the $3c$ perturbation. In addition, the $1c\_M_{k}$ and $2c\_M_{k}$ perturbations under-predict the baseline prediction, and in general tend to approach closer to the in-house DNS data in the attached turbulent boundary layer, although an under-prediction for $-\left\langle u_{1}u_{2} \right \rangle/U_{\infty}^2$ is observed in the LSB; on the other hand, the $3c\_M_{k}$ perturbation over-predicts the baseline prediction, showing better agreement with the in-house DNS data within the LSB. 
%\textbf{you need to add couple references indicating that the similar trend: 1c overpredict the base line has been observed by other %researchers, e.g. SU2 paper, openfoam paper, and emory paper}

The predicted profiles for the Reynolds shear stress normalized by the freestream velocity squared, $-\left\langle u_{1}u_{2} \right \rangle/U_{\infty}^2$, are shown in Figs. \ref{fig:markerfunc_uv_five.pdf} (a) - (e). Also included are the in-house DNS data of \cite{zhang2021turbulent} for comparison, as well as the $1c$ and $3c$ eigenvalue perturbations used as a reference for the ${1c}\_M_{k}$, ${2c}\_M_{k}$ and ${3c}\_M_{k}$ perturbations. Figures \ref{fig:markerfunc_uv_five.pdf} (a) - (e) show that the baseline prediction is well enveloped by the uncertainty bounds generated from the ${1c}\_M_{k}$, ${2c}\_M_{k}$ and ${3c}\_M_{k}$ perturbations. In addition, the ${1c}\_M_{k}$ and ${2c}\_M_{k}$ perturbations reduce the magnitude of the predicted $-\left\langle u_{1}u_{2} \right \rangle/U_{\infty}^2$ profiles compared to the baseline prediction, while the ${3c}\_M_{k}$ perturbation does the opposite. A similar behavior of the $1c$ and $3c$ perturbations with respect to the baseline prediction was also observed by Luis \textit{et al.} \cite{cremades2019reynolds} in their numerical study for a turbulent flow over a backward-facing step. Figures \ref{fig:markerfunc_uv_five.pdf} (a) - (e) show that the simulation's sensitivity to the $3c\_M_{k}$ perturbation is rather low, with the ${3c}\_M_{k}$ profile nearly collapsing onto the $3c$ profile. A similar behavior is also observed in Fig. \ref{fig:markerfunc_U_five.pdf} (a) - (e). At $x/c  = 0.15$ ($X_{T}$), Fig. \ref{fig:markerfunc_uv_five.pdf} (a) shows that the $1c\_M_{k}$ perturbation results in a rather strong reduction in the magnitude of $-\left\langle u_{1}u_{2} \right \rangle/U_{\infty}^2$ across the entire boundary layer compared to the baseline prediction, showing a tendency of approaching closer to the in-house DNS data, and a ``synergy behavior'' is observed. On the other hand, the $3c\_M_{k}$ perturbation that tends to deviate from the in-house DNS data, indicating a weak response to the $M_{k}$ perturbation. In Figs. \ref{fig:markerfunc_uv_five.pdf} (b) - (e), the baseline predictions and in-house DNS data are similar in shape: the convexity of the profile strongly increases in the vicinity of the wall, and then relaxes as the distance from the wall increases, with some discrepancies that overall mark the under-prediction of the momentum transfer due to the Reynolds shear stress within both the transitional and turbulent boundary layer. Therefore, there is an important observation: the synergy behavior seems only active for the $1c\_M_{k}$ and $2c\_M_{k}$ perturbations. As the flow moves further downstream to $x/c = 0.2$ (the aft portion of the LSB) and $x/c = 0.3$ (downstream of the LSB near $X_{R}$), the $1c\_M_{k}$ perturbation decreases the predicted Reynolds shear stress more than the $1c$ perturbation does in the outer portion of the boundary layer, indicating a deviation from the in-house DNS data, while shows a somewhat reduction and no discernible change (a collapse onto the $1c$ profile) at $x/c = 0.2$ and $x/c = 0.3$, respectively, in the near-wall region, as shown in Figs. \ref{fig:markerfunc_uv_five.pdf} (b) and (c). This reflects the spatial variability in $M_{k}$. On the other hand, the $-\left\langle u_{1}u_{2} \right \rangle/U_{\infty}^2$ profiles generated from the $3c\_M_{k}$ perturbations nearly collapse onto that for the $3c$ perturbations at $x/c = 0.2$ and $x/c = 0.3$; consequently, the $-\left\langle u_{1}u_{2} \right \rangle/U_{\infty}^2$ profiles generated from the $3c\_M_{k}$ perturbations tend to approach closer to the in-house DNS data at $x/c = 0.2$ and $x/c = 0.3$, as shown in Figs. \ref{fig:markerfunc_uv_five.pdf} (b) and (c). At $x/c = 0.4$ and $x/c = 0.5$ (in the attached turbulent boundary layer), a synergy behavior is again observed for the $1c\_{M_{k}}$ perturbations across the entire boundary layer, which enhances the deviation from the in-house DNS data at $x/c = 0.4$, while encompasses the in-house DNS data at $x/c = 0.5$. On the other hand, a collapse is again observed for the $3c\_M_{k}$ and $3c$ perturbations. Besides, the uncertainty bounds generated from the $2c\_M_{k}$ and $3c\_M_{k}$ perturbations successfully encompass the in-house DNS data in the lower portion of the attached turbulent boundary layer at $x/c = 0.4$ and $x/c = 0.5$, although there is a small discrepancy present in the region next to the wall. 

\section{Conclusions}
The goal of the present study was to advance our understanding of a physics-based methodology to quantify transition model-form uncertainty in RANS predictions of unsteady flow over a SD7003 airfoil. The method is based on the framework proposed in the study of \cite{emory2013modeling}, which introduces perturbations to a decomposition of the Reynolds stress tensor, namely, the amplitude and the eigenvalue of the anisotropy Reynolds stress tensor. In this study, the methodology was completely implemented in C++ in OpenFOAM. Based on the baseline predictions for $C_{f}$ and $C_{p}$, we presented analyses to locate the untrustworthy region, which is further divided into four zones to cover both the LSB and turbulent flow region further downstream. A novel regression based marker function was developed to inject an accurate level of the amplitude perturbation into the identified untrustworthy region. 

We presented analyses to understand the effect of the uniform amplitude perturbation to the skin friction coefficient, mean velocity, and Reynolds shear stress. Importantly, we observed a monotonic behavior of the magnitude of the predicted bounds with $\Delta_{k}$ perturbations, in particular more noticeable bounds for $\Delta_{k} > 1$: a clear shift of the reattachment point in the upstream direction, a noticeable suppression of the length of the LSB, and a greatly reduced magnitude of Reynolds shear stress in the LSB region; for the turbulent flow region further downstream of the LSB, results for both the mean velocity and the Reynolds shear stress showed better agreement with the in-house DNS data of \cite{zhang2021turbulent}. Such monotonic behavior is imperative for the development of a marker function that aims to predict plausible bounds for QoIs.  

The predicted bounds generated from the marker function $M_{k}$ was contrasted with the uniform amplitude perturbations $\Delta_{k} = 0.1$ and $\Delta_{k} = 8$ for different QoIs. Results for the QoIs clearly showed the spatial variability in $M_{k}$, and the bounds generated from $M_{k}$ in general sat within the bounds generated from $\Delta_{k} = 8$. The $\Delta_{k} = 8$ perturbations showed a clear tendency to approach closer to the reference data \cite{galbraith2010implicit,garmann2013comparative,zhang2021turbulent} for $C_{f}$ and $C_{p}$, and well encompassed the reattachment point in the predicted bounds. Overall, the $\Delta_{k} = 0.1$ perturbation was the opposite of the behavior of $\Delta_{k} = 8$: deviating from the reference data and showing rather small bounds. On the pressure side, the $C_{p}$ profile for $\Delta_{k} = 0.1$, baseline prediction, and $\Delta_{k} = 8$ showed a collapse, which indicated a low model form uncertainty. Importantly, the over-perturbation behavior associated with the predicted Reynols shear stress profile undergoing the $\Delta_{k} = 8$ perturbation could facilitate the approximating of the upper-bound of the amplitude perturbation.

When compounding $M_{k}$ with the eigenvalue perturbations $1c$, $2c$, the predicted bounds for $C_{f}$ was dramatically increased to encompass the reattachment point and the reference data of \cite{galbraith2010implicit,garmann2013comparative,zhang2021turbulent} at the crest, which showed a synergy behavior and consistently sat above the baseline prediction. Overall, the uncertainty bounds retained the shape of the baseline prediction for $C_{f}$, which confirmed the effect of spatial variability in $M_{k}$. The predicted $1c\_M_{k}$ and $2c\_M_{k}$ bounds for $C_{p}$ sat above the baseline prediction at the flat spot, which did not exhibit a synergy behavior, but reduced in magnitude compared to the $1c$ and $2c$ perturbations instead. The opposite was true at the kink (or the reattachment point) of the $C_{p}$ distribution, where the $1c\_M_{k}$ and $2c\_M_{k}$ perturbations under-predicted the baseline prediction. The $3c$ and $3c\_M_{k}$ bounds for both $C_{f}$ and $C_{p}$ showed a collapse, which deviated slightly away from the baseline prediction.  the perturbed mean velocity profile approached a lot closer to the in-house DNS data near the reattachment point.

When the contours of the mean velocity were plotted in an $xy$ plane, the $1c$ and $2c$ perturbations suppressed the LSB compared to the baseline prediction, which increased the magnitude of the mean flow. This behavior was enhanced by compounding with $M_{k}$: $1c\_M_{k}$ and $2c\_M_{k}$ further increased the magnitude of the mean flow in the attached turbulent boundary layer through a more suppression of the LSB. This behavior is qualitatively similar to that observed in the in-house DNS contour \cite{zhang2021turbulent}. Again, the $3c\_M_{k}$ remained at nearly same magnitude as that for the $3c$ perturbation, which bolstered the region of reverse flow to approach closer to the in-house DNS data \cite{zhang2021turbulent}. When the predictions for the mean velocity profile were plotted in coordinates shifted vertically, the predicted bounds generated from the $1c\_M_{k}$ and $2c\_M_{k}$ perturbations in general led ahead the baseline prediction, while the $3c\_M_{k}$ perturbation lagged behind it, which showed an enveloping behavior with respect to the baseline prediction. This behavior is qualitatively similar to the $1c$ and $3c$ perturbations observed by Luis \textit{et al.} \cite{cremades2019reynolds}. At the transition point $X_{T}$, all of the perturbations and the baseline prediction showed a collapse for $0.007 < y/c|_{o} < 0.011$, and showed a good agreement with the in-house DNS data \cite{zhang2021turbulent}. As the flow moves further downstream of $X_{R}$, the $1c\_M_{k}$ and $2c\_M_{k}$ perturbations showed a tendency to approach closer to the in-house DNS data, while the effect of perturbation gradually deteriorated due to gradual reduction in the positive values of $C_{f}$. Overall, the compound effect of $3c\_M_{k}$ was weak, which indicated the immunity of the $3c$ perturbation to the marker function. With the velocity vectors added to the mean velocity contour, a clear visualization again, confirmed the effect of all of the perturbations in the region of reverse flow and the attached turbulent boundary layer.   

The dimensionless Reynolds shear stress contours in an $xy$ plane were also analyzed. The $1c\_M_{k}$ and $2c\_M_{k}$ perturbations under-predicted the baseline prediction, which showed a tendency to approach closer to the in-house DNS data in the region downstream of the LSB. While, the $3c\_M_{k}$ perturbation over-predicted the baseline prediction, and showed good agreement with the in-house DNS data \cite{zhang2021turbulent} in the region of the LSB. When the predictions for the dimensionless Reynolds shear stress $-\left\langle u_{1}u_{2} \right \rangle/U_{\infty}^2$ profile were plotted, the $1c\_M_{k}$ and $2c\_M_{k}$ perturbations reduced the magnitude of the $-\left\langle u_{1}u_{2} \right \rangle/U_{\infty}^2$ profiles compared to the baseline prediction, while the $3c\_M_{k}$ perturbation did the opposite, which resulted in an enveloping behavior. This behavior is qualitatively similar to that observed by Luis \textit{et al.} At the transition point, the $1c\_M_{k}$ perturbation greatly reduced the magnitude of the $-\left\langle u_{1}u_{2} \right \rangle/U_{\infty}^2$ profile, which marked a synergy behavior. An important observation was that the synergy behavior seems only active for the $1c\_M_{k}$ and $2c\_M_{k}$ perturbations. 

Overall, the marker function $M_{k}$ was effective in the eigenspace perturbation framework in constructing uncertainty bounds for both mean velocity and turbulence properties. Future work will focus on the development of different types of marker functions based on a variety of transitional flow scenarios. Eigenvector perturbations to the Reynolds stress tensor should also be conducted to complete the full range of the model form uncertainty in the Boussinesq turbulent viscosity models. Also a wider range of RANS based transition models will be tested using the eigenspace perturbation framework with marker involved.

\begin{acknowledgments}
The support of the Natural Sciences and Engineering Research Council (NSERC) of Canada for the research program of Professor Xiaohua Wu and Professor David E. Rival is gratefully acknowledged.  
\end{acknowledgments}

\section*{Data Availability Statement}
The data that support the findings of this study are available from the corresponding author upon reasonable request. 

\section*{References}
\bibliography{aipsamp}% Produces the bibliography via BibTeX.

\end{document}